\documentclass[reprint, floatfix,amsmath,amssymb, eqsecnum, superscriptaddress,nofootinbib,groupedaddress,prx,aps,longbibloiography]{revtex4-1}

\usepackage{hyperref}
\hypersetup{
     colorlinks   = true,
     linkcolor    = blue,
     urlcolor     = blue,
     citecolor    = red
}

\usepackage{graphicx}
\usepackage{bm}
\usepackage{dcolumn}
\usepackage{amssymb}
\usepackage{amsmath}
\usepackage{slashed}
\usepackage{amsfonts}
\usepackage{epsfig}
\usepackage{stackrel}
\usepackage{paralist}
\usepackage[rgb]{xcolor}
\usepackage{todonotes}
\usepackage{pdfcomment}
\usepackage{epstopdf}

\usepackage{units}
\usepackage{amsthm}
\usepackage{amsmath}
\usepackage{amssymb}

\usepackage{centernot}

\usepackage{multirow}

\usepackage[caption=false]{subfig}

\definecolor{NewColor}{rgb}{1,0,0}
\graphicspath{{./Figs/}}

\newcommand{\fsl}[1]{{\centernot{#1}}}

\global\long\def\ket#1{\left|#1\right\rangle }
\global\long\def\bra#1{\left\langle #1\right|}

\begin{document}

\title{Probing topological superconductors with emergent gravity}
\author{Omri Golan} \email{omri.golan@weizmann.ac.il.}
\author{Ady Stern}  
\affiliation{Department of Condensed Matter Physics, Weizmann
Institute of Science, Rehovot 76100, Israel}
\date{\today}

\begin{abstract}

Topological superconductors are characterized by topological invariants that describe the number and nature of their robust boundary modes. These invariants must also have observable consequences in the \textit{bulk} of the system, akin to the quantized bulk Hall conductivity in the quantum Hall effect, but such consequences are made elusive by the spontaneous breaking of $U(1)$ symmetry in the superconductor. Here we focus on 2+1 dimensional spin-less $p$-wave superconductors and show that \textit{emergent gravity} serves as a natural \textit{bulk} probe for their topological invariant. This emergent gravity is due to the same attractive interaction between fermions that leads to superconductivity, and is therefore built into topological superconductors. The bulk response of a topological superconductor to the emergent gravitational field is encoded in a gravitational Chern-Simons term, and is related to the existence of robust boundary modes via energy-momentum conservation, or gravitational anomaly inflow. The gravitational Chern-Simons term implies a universal relation between variations in the superconducting order parameter and the energy-momentum currents and densities that they induce. The spontaneous breaking of $U(1)$ symmetry in the superconductor leads to additional bulk responses, encoded in a gravitational \textit{pseudo} Chern-Simons term. Although not of topological nature, these carry surprising similarities to the topological responses of the gravitational Chern-Simons term. We show how these two types of responses can be disentangled.

\end{abstract}

\maketitle

\begingroup
\hypersetup{linkcolor=blue}

\tableofcontents

\endgroup

\section{Introduction }

In this paper we study spin-less $p$-wave superconductors (SC) in
2+1 dimensions. These are superconductors that can be thought of as
microscopically comprised of charge 1 spin-less fermions with an attractive
two body interaction. The interaction is such that it can efficiently
be described as an interaction of the fermions with a charge 2 spin 1 boson, which
is the superconducting order parameter, as in BCS theory. This boson
represents a condensate of Cooper pairs, where the fermions in a pair
have relative orbital angular momentum 1, as opposed to $s$-wave
SC, where the relative orbital angular momentum is 0. $p$-wave pairing
has been experimentally observed in thin films of superfluid He-3
 \cite{vollhardt2013superfluid}, and there are many solid state
candidates \cite{sato2016topological}. Another notable candidate
is the $5/2$ fractional quantum Hall state which has been proposed
to be a $p$-wave SC of composite fermions \cite{moore1991nonabelions,read2000paired}. 


Within mean field theory, $p$-wave SC are known to realize gapped topological phases \cite{read2000paired}, or symmetry protected topological phases (SPT) \cite{wang2017symmetric}. The most notable known manifestation of the existence of distinct topological
phases, is the formation of chiral Majorana (Majorana-Weyl) spinors
on spatial domain walls between different phases, and Majorana bound
states, or zero modes, in the cores of vortices \cite{read2000paired}.
These Majorana bound states exhibit non abelian braiding statistics,
and may therefore be used as building blocks for a topological
quantum computer \cite{kitaev2003fault,nayak2008non}. This is the main drive behind the intense research of $p$-wave SC in
recent years.


The different topological phases of the $p$-wave SC are characterized by an integer valued topological invariant, which is the Chern number $\nu$. An important physical manifestation of the Chern number is the net chirality $C$ of the chiral Majorana spinors on the boundary between a $p$-wave SC and vacuum. More generally, there are Majorana spinors with net chirality $C=\Delta\nu$ on spatial boundaries between different topological phases, where the Chern number jumps by $\Delta\nu$  \cite{read2000paired,stone2004edge,volovik2009universe}. The equation $C=\Delta\nu$ is referred to as \textit{bulk-boundary correspondence}.

Although the bulk of a topological superconductor is expected to manifest the topological invariant, in a way similar to the quantized Hall conductivity of a quantum Hall state, the spontaneous breaking of $U(1)$ symmetry in superconductors makes this manifestation elusive. In this paper we address two fundamental questions in this context:

Question 1: What is the physical manifestation of the Chern number in the \textit{bulk}, i.e, what is the \textit{topological bulk response}?

Question 2: What is the physical principle behind bulk-boundary correspondence? 

Question 1 is of both conceptual and practical importance. Answering it provides a definition for the Chern number in terms of physical bulk observables, which may be used in experiment to probe the topological phase diagram of a $p$-wave SC. Question 2 is of conceptual importance. It asks for the physical obstruction to the existence of edge states without a topological bulk. As explained below, the two questions are intimately related. 

In order to clarify the above questions, and the type of answers we are after, it will be useful to briefly review the closely related integer quantum Hall effect (IQHE), where the answers to both of our questions are known, in the language of the \textit{anomaly inflow} mechanism. 

\begin{figure}[!th]

\includegraphics[width=\linewidth]{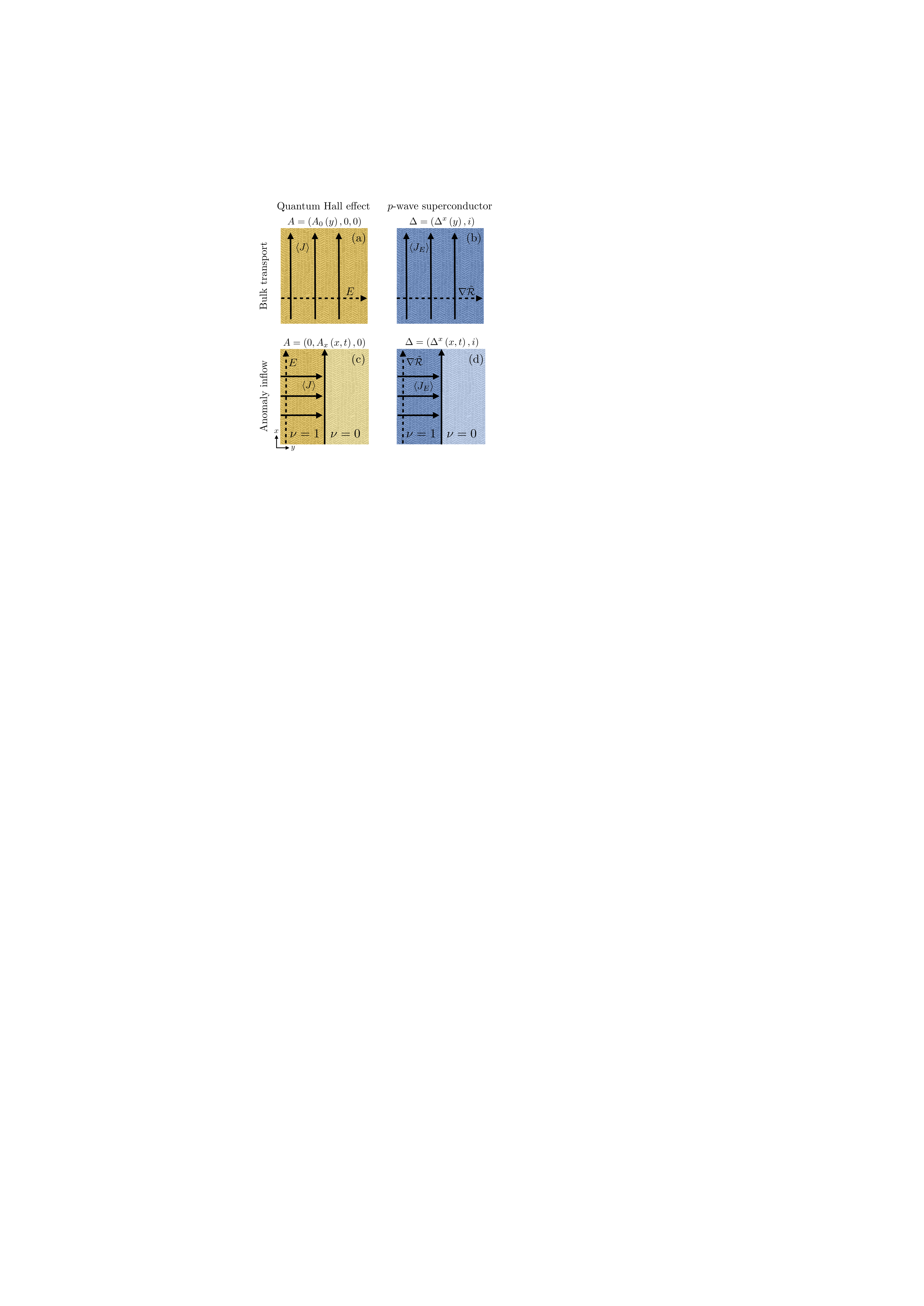}

\caption{A comparison of topological electromagnetic effects in the integer
quantum Hall effect (IQHE), and their energy-momentum analogs in the
$p$-wave superconductor (SC). (a) In the IQHE
there is a perpendicular electric current $\left\langle J\right\rangle $
in response to an applied electric field $E$, with a quantized Hall
conductivity, proportional to the Chern number, as encoded in a $U(1)$ Chern-Simons term.
(b) In the $p$-wave SC,
an energy current $\left\langle J_{E}\right\rangle$ flows
in response to a space dependent order parameter $\Delta$, as encoded in a gravitational Chern-Simons term. Derivatives
of the curvature $\tilde{\mathcal{R}}$ associated with $\Delta$
play the role of the electric field in the IQHE, and $\left\langle J_{E}\right\rangle $
is perpendicular to the curvature gradient $\nabla\tilde{\mathcal{R}}$.
The ratio between the magnitudes of $\left\langle J_{E}\right\rangle $
and $\nabla\tilde{\mathcal{R}}$ is quantized, and proportional to
the Chern number. As described in the text, the spontaneous breaking of $U(1)$ symmetry in the $p$-wave SC
allows also a gravitational \textit{pseudo} Chern-Simons term, which encodes closely related bulk responses, which are not topological in nature. (c) The quantized Hall conductivity implies the existence of a
chiral boundary fermion with a $U\left(1\right)$ anomaly, which can
be described as a Weyl fermion at low energy. (d) The analogous response
in the $p$-wave SC implies the existence of a boundary chiral Majorana
fermion with a gravitational anomaly, which can be described as a
Majorana-Weyl fermion at low energy.\label{fig:A-comparison-of}}
\end{figure}

Like the $p$-wave SC, the IQHE is 2+1 dimensional and is characterized by the same Chern number $\nu$, despite the difference in symmetries\footnote{Depending on convention, one either says that the $p$-wave SC does not have any symmetries and that the IQHE has
$U\left(1\right)$ symmetry \cite{kitaev2009periodic}, or that the
$p$-wave SC has particle-hole symmetry (symmetry class D) while the
IQHE has no symmetries (symmetry class A) \cite{schnyder2008classification,ryu2010topological}.}. For the IQHE, the answer to Question 1 is that $\nu/2\pi$ is the quantized
Hall conductivity in units of $e^{2}/\hbar$ \cite{thouless1982quantized,avron1983homotopy,golterman1993chern,qi2008topological,mera2017topological}, illustrated in Fig.\ref{fig:A-comparison-of}(a). Equivalently, the effective action for a background $U\left(1\right)$ gauge field $A$ contains a $U\left(1\right)$ Chern-Simons (CS) term $\frac{\nu}{4\pi}\int A\text{d}A$, where we have set $e=1=\hbar$. The answer to Question 2 is that charge conservation, or $U(1)$ symmetry, is the physical principle behind bulk-boundary correspondence, as depicted in Fig.\ref{fig:A-comparison-of}(c). In the IQHE boundaries carry chiral (Weyl) spinors with net chirality $C=\Delta\nu$, which have a $U(1)$ \textit{anomaly} \cite{manes1985algebraic,bertlmann2000anomalies,kaplan1992method,shamir1993chiral,chandrasekharan1994anomaly}. A physical implication of the anomaly is that the expectation value of the boundary current $j^{\alpha}$ ($\alpha=t,x$) is not conserved in the presence of an electric field parallel to the boundary, $\partial_{\alpha}\left\langle j^{\alpha}\right\rangle=\frac{C}{2\pi}E_{x}$\footnote{This is the \textit{covariant} $U(1)$ anomaly}. A 1+1 dimensional system that microscopically conserves charge cannot be described by Weyl spinors with $C\neq0$, because the anomaly implies an unphysical source of charge. In the context of the IQHE this source of charge is physical, and is due to the difference of bulk Hall currents $\left\langle J^{y}\right\rangle=\frac{\nu}{2\pi}E_{x}$ on the two sides of the boundary, since $C=\Delta\nu$. This is the anomaly inflow mechanism \cite{callan1985anomalies,naculich1988axionic,harvey2001local}. Running the argument backwards, bulk-boundary correspondence follows from bulk+boundary charge conservation in the presence of an electric field.

The relation between anomaly inflow and topological phases is much more general. It has
been suggested, and to a large extent shown, that the existence of
anomalies in $D-1$ dimensions is equivalent to the existence of corresponding
topological phases in $D$ dimensions, related by the anomaly inflow
mechanism \cite{ryu2012electromagnetic,witten2015fermion}. Moreover,
since anomalies are known to be robust to weak interactions, they
naturally classify topological phases of weakly interacting fermions \cite{witten2015fermion}.
In many instances the anomaly also suggests a topological bulk response.

We can now go back to the $p$-wave SC, and sharpen Questions 1,2 to "what is the topological bulk response, and what is the boundary anomaly corresponding to this response through anomaly inflow?". In the $p$-wave SC, boundaries carry 1+1 dimensional chiral Majorana spinors, which do not carry $U\left(1\right)$
charge. Thus there is no $U\left(1\right)$ anomaly and no corresponding
bulk CS term, or quantized Hall conductivity \cite{read2000paired,stone2004edge}\footnote{Though there is no \textit{quantized} Hall conductivity in a $p$-wave
SC, there is in fact a Hall conductivity, which we discuss in more
detail in our conclusions.}. In fact, the only conserved quantity such a spinor does carry is energy-momentum,
associated with space-time symmetries. The only relevant anomaly is
therefore the \textit{gravitational anomaly}, where energy-momentum conservation is violated. Chiral Majorana spinors in 1+1 dimensions
indeed possess such an anomaly \cite{alvarez1984gravitational,bertlmann2000anomalies,bastianelli2006path}. Just as the $U(1)$ anomaly is manifested in the presence of a background electric field, so does the gravitational anomaly manifests itself in the presence of a background metric with curvature gradients. Like the $U(1)$ anomaly inflow described above, the
gravitational anomaly in 1+1 dimensions can be interpreted as the
inflow of energy-momentum from a 2+1 dimensional bulk with an appropriate
Chern-Simons term, which is the \textit{gravitational Chern-Simons} term (gCS) \cite{stone2012gravitational}, see section \ref{subsec:Effective-action-for} for the definition. Based on these facts it was argued
that a gCS term with coefficient $\alpha=\frac{\nu/2}{96\pi}\in\frac{1}{192\pi}\mathbb{Z}$
should arise when integrating out the bulk fermions in a $p$-wave
SC \cite{read2000paired,ryu2012electromagnetic}. Similar statements were made in \cite{wang2011topological,palumbo2016holographic}. The
gCS term then describes a topological bulk response to the background
metric, from which the Chern number can in principle be measured,
and bulk-boundary correspondence follows from energy-momentum conservation
in the presence of a metric with curvature gradients. One arrives at the appealing
conclusion that a $p$-wave SC is a manifestation of the gravitational
anomaly inflow mechanism, just as the IQHE is a manifestation of the $U(1)$ anomaly inflow mechanism. 

The only problem with the above conclusion is that the actual gravitational field is negligible in condensed matter experiments. The actual metric of space-time is, for all practical purposes, flat. Therefore, in order to find a physically relevant topological bulk response of the $p$-wave SC, one must find some probe that couples to
the fermions as gravity, at least at low energies \footnote{We note that for spin-full $p$-wave SC with an $SU\left(2\right)$
spin rotation symmetry, there is also a spin Hall effect that can be used to probe the Chern number \cite{volovik1989fractional,read2000paired,stone2004edge}.}.


What probe, or background field, could play the role of gravity? One approach is to use real geometry, induced by curving
the 2 dimensional sample in 3 dimensional space. This works well for
the IQHE and has
led to a remarkable body of work on geometric responses of quantum Hall states \cite{ferrari2014fqhe,abanov2014electromagnetic,can2014fractional,gromov2014density,gromov2015geometric,gromov2015framing,klevtsov2015geometric,klevtsov2015quantum,bradlyn2015topological,can2015geometry,gromov2016boundary,schine2016synthetic,wiegmann2017nonlinear,klevtsov2017lowest,klevtsov2017laughlin,schine2018measuring}. For the $p$-wave SC, understanding the effect of real geometry is more complicated, and we will come back to this point in the discussion,
section \ref{sec:Conclusion-and-discussion}. 

A second approach introduces effective gravity  through a space dependent
temperature \cite{read2000paired}. In this approach the corresponding bulk response was suggested to be a quantized bulk \textit{thermal} Hall conductivity 
\cite{wang2011topological,ryu2012electromagnetic}.

The motivation for this suggestion is two fold. First, there is an
argument due to Luttinger that shows that the thermal conductivity
is essentially given by the response of a system to a gravitational
field \cite{luttinger1964theory,cooper1997thermoelectric}. Second,
there is a well known derivation of the thermal Hall \textit{conductance
}(as opposed to conductivity) for 2+1 dimensional topological phases with
chiral boundaries which gives
 $\kappa_{xy}=c_{\text{chiral}}\frac{\pi T}{6}$ \cite{cappelli2002thermal}
where $T$ is the (average) temperature and $c_{\text{chiral}}$
is the chiral central charge of the boundary, given by $C$
for the IQHE and by $C/2$ for the $p$-wave SC.
Using bulk-boundary correspondence one obtains $\kappa_{xy}$ in terms
of the bulk Chern number $\nu$ and the temperature, which is analogous
to $\sigma_{xy}=\frac{\nu}{2\pi}$. This thermal Hall conductance was indeed measured recently in quantum Hall systems \cite{jezouin2013quantum,banerjee2017observed,banerjee2017observation}. One may then hope to obtain the 
same result, now for the bulk thermal conductivity, from the gCS term, by using Luttinger's argument. This, however, cannot be the case, because gCS term is third order in derivatives of the metric, as opposed to a single derivative of the temperature required for
a thermal conductivity \cite{stone2012gravitational,bradlyn2015low}. Some authors argue that there is a quantized bulk thermal Hall conductivity, but relate it to other gravitational terms,
which are first order in derivatives \cite{qin2011energy,shitade2014heat,nakai2016finite},
and to \textit{global} gravitational anomalies \cite{nakai2017laughlin}, which will not
be discussed in this paper. Other authors find that there is no quantized bulk thermal conductivity at all \cite{gromov2015thermal,bradlyn2015low}. In any case, the gCS term and the corresponding
gravitational anomaly have not been interpreted in the context of thermal responses thus far.

We note that on general grounds, the relation between thermal conductivity and conductance is more subtle than the relation between electric conductivity and conductance. First, while there are longitudinal and transverse electric fields, there is no transverse driving force for heat. Second, if one expects a heat current to require the presence of entropy, there cannot be a bulk heat current as long as the temperature is negligible compared with the bulk gap.

In this paper we take a third approach, in which we utilize an additional field which couples to the fermions in a
$p$-wave SC as gravity. This field, which is built into the problem, is the order
parameter itself, as was discovered by Volovik (see e.g \cite{volovik2009universe}), and refined
by Read and Green \cite{read2000paired}. We refer to the gravitational field described by the order parameter
as \textit{emergent gravity}, because the order parameter arises microscopically
from a fermionic two-body interaction. In fact, using this observation,
Volovik suggested early on the existence of a gCS term in a
$p$-wave SC \cite{volovik1990gravitational}.

To gain some intuition into our approach, note that, almost by definition, gravity is a field
that couples to the energy-momentum of matter. The $p$-wave pairing term $\psi^{\dagger}\Delta^{j}\partial_{j}\psi^{\dagger}+h.c$
shows that the order parameter  $\Delta$ couples to derivatives of the fermion field $\psi$, related to
fermionic momentum. More accurately, we will see that the operator
$\psi^{\dagger}\partial_{j}\psi^{\dagger}$ appears in the energy-momentum
tensor of a $p$-wave SC.
The mapping of the order parameter onto gravity is the conceptual starting point of our analysis, which is motivated by the search for edge anomalies and topological bulk responses of the $p$-wave SC.  


\textbf{Outline of this paper:} Our main results along with simple
examples are given in section \ref{sec:Results,-examples,-and}.
In section \ref{sec:Lattice-model} we start our analysis with a simple
lattice model for a $p$-wave SC. We
describe the topological phase diagram of the model and also explain
some ingredients of the emergent geometry which are visible at this
level. In section \ref{sec:Continuum-limits-of} we derive a continuum
description of the lattice model, which is an even number of $p$-wave \textit{superfluids} (SF). In the limit where the order parameter is much larger than the single particle scales, each $p$-wave SF maps to a relativistic Majorana spinor coupled to Riemann-Cartan (RC) geometry, which is
a geometry with both curvature and torsion. We discuss the mapping
of fields, actions, equations of motion, path integrals, symmetries,
conservation laws, and observables in sections \ref{sec:Emergent-Riemann-Cartan-geometry}
and \ref{sec:Symmetries,-currents,-and}, and in appendices \ref{subsec:Equivalent-forms-of}-\ref{subsec:Global-structures-and}.

The rest of the paper is devoted to the application of the above mapping
to the problems described above: finding topological bulk responses
of the $p$-wave SC, and relating them to edge anomalies.  In section
\ref{sec:Bulk-response} we discuss bulk responses.  We verify that the effective action obtained
by integrating over the bulk fermions contains a gCS term, with coefficient
$\alpha=\frac{\nu}{192\pi}\in\frac{1}{192\pi}\mathbb{Z}$, and we
obtain the corresponding topological bulk response of the $p$-wave
SC.  We also find closely related terms, which do not encode topological bulk responses, and are unrelated to edge anomalies. The first, which we refer to a gravitational
\textit{pseudo} Chern-Simons term, is possible due to the spontaneous breaking of $U\left(1\right)$ symmetry, or
in other words, due to the emergent torsion. The second is a difference of two gCS terms, which appears because the different low energy Majorana spinors do not experience the same order parameter, or in other words, the same gravitational background. The calculation of the effective
action within perturbation theory is done in appendix \ref{subsec:Perturbative-calculation-of}.
In section \ref{sec:Boundary-fermions-and} we describe the edge states,
focusing on the physical implication of their gravitational anomaly
in the $p$-wave SC, and the relation to the topological bulk response
from gCS, via the anomaly inflow mechanism. We conclude and discuss
our results in section \ref{sec:Conclusion-and-discussion}. Tables \ref{Table: basic objects}-\ref{Table: Lattice} list our notation, and may be useful for the reader. In particular, Tab.\ref{Table: basic objects} serves as a quick guide for the mapping of the $p$-wave SF to a Majorana spinor in RC geometry.

\begin{table*}[t]

\caption{Notation: basic objects in the $p$-wave superfluid, aligned with the corresponding objects in Riemann-Cartan geometry. All indices are written explicitly, in their natural placement and type. The indices $a,b,\dots\in\left\{ 0,1,2\right\}$  are $SO\left(1,2\right)$ (Lorentz) indices which we refer to as internal indices, while $\mu,\nu,\dots\in\left\{ t,x,y\right\}$  are coordinate indices.
We also use $i,j,\dots\in\left\{ x,y\right\}$  for spatial coordinate indices. Capital letters $A,B,\dots$ take the values 1,2 in bulk objects, and 0,1 in boundary objects. \label{Table: basic objects} }

\begin{ruledtabular}
\renewcommand*{\arraystretch}{1.3}
\begin{tabular}{llll}
\multicolumn{2}{l}{$p$-wave superfluid (SF)} & \multicolumn{2}{l}{Riemann-Cartan (RC) geometry}\tabularnewline
\hline 
$\psi,\Psi$ & Spin-less fermion, Nambu spinor & $\chi$ & Majorana spinor\tabularnewline
\hline 
$\tilde{\xi}$ & boundary chiral Majorana spinor & $\xi$ & Boundary chiral Majorana spinor\tabularnewline
\hline 
$\Delta^{i}$ & $p$-wave order parameter & $e_{a}^{\;\mu}$ & Inverse vielbein\tabularnewline
\hline 
$\Delta^{(i}\Delta^{j)*}$ & Higgs part of $\Delta^{i}$ & $g^{\mu\nu}$ & Inverse metric\tabularnewline
\hline 
$o$ & Orientation of $\Delta^{i}$ & $o$ & Orientation of $e_{a}^{\;\mu}$\tabularnewline
\hline 
$A_{\mu}$ & $U\left(1\right)$ connection & $\omega_{\;b\mu}^{a}$ & Spin connection\tabularnewline
\hline 
$D_{\mu}$ & $U\left(1\right)$ covariant derivative & $D_{\mu}$ & Spin covariant derivative\tabularnewline
\hline 
$F_{\mu\nu}$ & $U\left(1\right)$ curvature, or field strength & $R_{\;b\mu\nu}^{a}$ & Curvature\tabularnewline
\hline 
\multirow{2}{*}{$t_{\text{cov}\;\nu}^{\mu}$} & $U\left(1\right)$-covariant canonical  & \multirow{4}{*}{$\mathsf{J}_{\;a}^{\mu}$} & \multirow{4}{*}{Energy-momentum tensor}\tabularnewline
 & energy-momentum tensor &  & \tabularnewline
\cline{1-2} 
$J_{\varphi}^{\mu}$ & Angular momentum current &  & \tabularnewline
\cline{1-2} 
$J_{E}^{i},\;P_{i}$ & Energy current, Momentum density &  & \tabularnewline
\hline 
$J^{\mu},\;\rho=J^{t}$ & Electric current, charge density & $\mathsf{J}^{ab\mu}$ & Spin current \tabularnewline
\hline 
\multirow{2}{*}{$t_{\text{e}\;\beta}^{\alpha}$} & Boundary, or edge, canonical  & \multirow{2}{*}{$\mathsf{j}_{\;A}^{\alpha}$} & Boundary energy-momentum \tabularnewline
 & Energy-momentum tensor &  & tensor\tabularnewline
\hline 
 & Boundary electric current & $\mathsf{j}^{AB\alpha}$ & Boundary spin current\tabularnewline
\hline 
$-m$ & Chemical potential & $m$ & Relativistic mass\tabularnewline
\hline 
$m^{*}$ & Non-relativistic mass &  & \tabularnewline
\hline 
$S_{\text{SF}}\left[\psi,\Delta,A\right]$ & $p$-wave SF action & \multirow{2}{*}{$S_{\text{RC}}\left[\chi,e,\omega\right]$} & Action for a Majorana spinor \tabularnewline
\cline{1-2} 
$S_{\text{rSF}}\left[\psi,\Delta,A\right]$ & Relativistic limit of $p$-wave SF action &  & in RC geometry\tabularnewline
\hline 
\multirow{2}{*}{$W_{\text{SF}}\left[\Delta,A\right]$} & \multirow{2}{*}{Effective action for the $p$-wave SF } & \multirow{2}{*}{$W_{\text{RC}}\left[e,\omega\right]$} & Effective Action for a Majorana \tabularnewline
 &  &  & spinor in RC geometry\tabularnewline
\hline 
\multirow{2}{*}{$S_{\text{e}}^{\pm}\left[\tilde{\xi},\Delta\right]$ } & Action for a boundary, or edge, & \multirow{2}{*}{$S_{\text{R}}^{\pm}\left[\xi,e\right]$} & Action for a boundary\tabularnewline
 & chiral Majorana spinor &  & chiral Majorana spinor\tabularnewline
\hline 
\multirow{2}{*}{} & Effective action for a  & \multirow{2}{*}{$W_{\text{R}}^{\pm}\left[e\right]$} & Effective action for a \tabularnewline
 & boundary chiral Majorana spinor &  & boundary chiral Majorana spinor\tabularnewline
\end{tabular}
\end{ruledtabular}
\end{table*}

\begin{table}[ht]

\caption{Notation: additional geometric objects.\label{Table: addtional objects}}

\begin{ruledtabular}
\renewcommand*{\arraystretch}{1.3}
\begin{tabular}{ll}
\multicolumn{2}{l}{Additional geometric objects}\tabularnewline
\hline 

$T_{\mu\nu}^{a},\;C_{ab\mu}$ & Torsion, Contorsion \tabularnewline
\hline 
$c$ & Contorsion scalar\tabularnewline
\hline 
$c_{\text{light}}$ & Speed of light\tabularnewline
\hline 
$\Gamma_{\;\nu\rho}^{\mu}$ & Affine connection\tabularnewline
\hline 
$\nabla_{\mu}$ & Coordinate, or total, covariant derivative\tabularnewline
\hline 
$\tilde{\Gamma}_{\;\nu\rho}^{\mu}$ & LC affine connection, or Christoffel symbol\tabularnewline
\hline 
$\tilde{\nabla}_{\mu}$ & LC total covariant derivative\tabularnewline
\hline 
$\tilde{\omega}_{ab\mu}$ & LC spin connection\tabularnewline
\hline 
$\tilde{D}_{\mu}$ & LC spin covariant derivative\tabularnewline
\hline 
$\tilde{R}_{ab\mu\nu}$ & LC curvature, or Riemann tensor\tabularnewline
\hline 
$\tilde{\mathcal{R}}_{\mu\nu},\;\tilde{\mathcal{R}}$ & Ricci tensor, Ricci scalar\tabularnewline
\hline 
$\tilde{C}^{\mu\nu}$ & Cotton tensor\tabularnewline
\hline 
$Diff$ & Diffeomorphism group\tabularnewline
\hline 
$Diff_{+}$ & Orientation preserving subgroup of $Diff$\tabularnewline
\hline 
$Diff_{0}$ & Identity component of $Diff$\tabularnewline
\hline 
$Q_{3}$ & Chern-Simons local three-form\tabularnewline
\end{tabular}
\end{ruledtabular}
\end{table}

\begin{table}[ht]

\caption{Notation: Lattice model and multiple low energy fermions. Geometric
objects are written with indices implicit. The index $n\in\left\{ 1,2,3,4\right\} $
labels the particle-hole invariant point in the Brillouin zone, and the
associated low energy data. As described in section \ref{subsec:Summing-over-lattice},
it is natural to group the $n$s into pairs, which we refer to as
layers, and use the layer index $l=n\;\text{mod}\;2\in\left\{ 1,2\right\} $. \label{Table: Lattice}}

\begin{ruledtabular}
\renewcommand*{\arraystretch}{1.3}

\begin{tabular}{ll}
\multicolumn{2}{l}{Lattice model and multiple low energy fermions }\tabularnewline
\hline 
$\delta=\left(\delta^{x},\delta^{y}\right)$ & $p$-wave order parameter on the lattice\tabularnewline
\hline 
$t$ or $t^{x},t^{y}$ & Hopping amplitudes\tabularnewline
\hline 
$\mu$ & Chemical potential\tabularnewline
\hline 
$a$ & Lattice spacing\tabularnewline
\hline 
$BZ$ & Brillouin zone\tabularnewline
\hline 
$\boldsymbol{K}^{\left(n\right)}$ & Particle-hole invariant point in $BZ$ \tabularnewline
\hline 
$\psi^{\left(n\right)}$ & Low energy fermion around $\boldsymbol{K}^{\left(n\right)}$\tabularnewline
\hline 
$\Lambda_{UV}$ & UV cutoff for the $\psi^{\left(n\right)}$s, of order $a^{-1}$\tabularnewline
\hline 
$\Delta_{\left(n\right)}$ or $e_{\left(n\right)}$ & Order parameter, or vielbein, for $\psi^{\left(n\right)}$\tabularnewline
\hline 
$o_{n}$ & Orientation of $\Delta_{\left(n\right)}$, or of $e_{\left(n\right)}$\tabularnewline
\hline 
$m_{n}$ & Relativistic mass, of $\psi^{\left(n\right)}$\tabularnewline
\hline 
$g_{\left(n\right)},\tilde{\omega}_{\left(n\right)},\tilde{\mathcal{R}}_{\left(n\right)},\dots$ & Geometric objects constructed from $e_{\left(n\right)}$\tabularnewline
\end{tabular}

\end{ruledtabular}
\end{table}

\section{Approach and main results \label{sec:Results,-examples,-and}}



\subsection{Model and  approach}

As a microscopic starting point, we consider a simple model for a
spin-less $p$-wave SC on a square lattice, described in section \ref{sec:Lattice-model}. We analyze the model in the regime
where the order parameter is much larger than the single particle
scales, which we refer to as the relativistic regime. In this regime
the model is essentially a lattice regularization of four, generically
massive, relativistic Majorana spinors, centered at the particle-hole
invariant points $k=-k$ in the Brillouin zone. Around each of these
four points the low energy description is given by a Hamiltonian $H_{\text{SF}}=\psi^{\dagger}\left(-\frac{\delta^{ij}\partial_{i}\partial_{j}}{2m^{*}}+m\right)\psi-\left(\frac{1}{2}\psi^{\dagger}\Delta^{j}\partial_{j}\psi^{\dagger}+h.c\right)$,
which we refer to as a $p$-wave superfluid (SF) Hamiltonian, with
an effective mass $m^{*}$\footnote{The effective mass tensor is actually different for the different
particle-hole invariant points, but this will not be important in
the following.}, chemical potential $-m$, and order parameter $\Delta$, which is
in the $p_{x}\pm ip_{y}$ configuration $\Delta=\left(\Delta^{x},\Delta^{y}\right)=\Delta_{0}e^{i\theta}\left(1,\pm i\right)$,
where $\Delta_{0}>0$ and $\theta$ are constants.  In the relativistic
regime the effective mass $m^{*}$ is large, and in the limit $m^{*}\rightarrow\infty$
one obtains a relativistic Hamiltonian, with mass $m$. This becomes
clear in terms of the Nambu spinor $\Psi^{\dagger}=\left(\psi^{\dagger},\psi\right)$,
which is a Majorana spinor. We refer to the sign $o=\pm$ as the orientation,
and we note that the different Majorana spinors, associated with the
four particle-hole invariant points, have different orientations $o_{n}$
and masses $m_{n}$, where $1\leq n\leq4$. The $n$th Majorana spinor contributes $o_{n}\text{sgn}\left(m_{n}\right)/2$
to the Chern number, and summing over $n$ one obtains the Chern number
of the lattice model $\nu=\sum_{n=1}^{4}\nu_{n}=\sum_{n=1}^{4}o_{n}\text{sgn}\left(m_{n}\right)/2$, 
which gives the topological phase diagram in terms of the low energy data $o_{n},m_{n}$.

In order to probe this topological phase diagram, we perturb the order
parameter out of the $p_{x}\pm ip_{y}$ configuration, and treat $\Delta=\left(\Delta^{x},\Delta^{y}\right)\in\mathbb{C}^{2}$
as a general space-time dependent field, close to the $p_{x}\pm ip_{y}$
configuration. This is analogous to applying an electromagnetic field
in order to probe the topological phase diagram of the IQHE.

Following Volovik \cite{volovik2009universe}, and Read and Green \cite{read2000paired}, we
show that fermionic excitations in each $p$-wave SF experience such
a general order parameter as a non trivial gravitational background.
Some of this gravitational background is described by the (inverse)
metric 
\begin{eqnarray}
 &  & g^{ij}=-\Delta^{(i}\Delta^{j)*},
\end{eqnarray}
where brackets denote the symmetrization, and the sign is a matter
of convention.  We refer to $g^{ij}$ as the Higgs part of the order
parameter. Parameterizing $\Delta=e^{i\theta}\left(\left|\Delta^{x}\right|,e^{i\phi}\left|\Delta^{y}\right|\right)$
with the overall phase $\theta$ and relative phase $\phi\in\left(-\pi,\pi\right]$, the metric is independent of $\theta$
and of the orientation $o=\text{sgn}\phi=\pm$, which splits order parameters into $p_{x}+ip_{y}$-like and $p_{x}-ip_{y}$-like. Note that in the $p_{x}\pm ip_{y}$
configuration the metric is euclidian, $g^{ij}=-\Delta_{0}^{2}\delta^{ij}$.
For our purposes it is important that the metric be perturbed out
of this form, and in particular it is not enough to take the $p_{x}\pm ip_{y}$
configuration with a space-time dependent phase $\theta$.

Before we turn to describe the conclusions that may be drawn from this emergent gravity, we find it instructive to draw analogies to the IQHE. 



\subsection{Electromagnetic response in the IQHE and gravitational response in the $p$-wave SC - analogies and differences}

It is illuminating to examine the gravitational response of the $p$-wave SC that we consider with an eye opened to the electromagnetic response of the IQHE. The defining characteristics of the IQHE, the absence of longitudinal conductivity and the quantization of the Hall conductivity, imply a coupling of charge density to magnetic field. A small local increase of the magnetic field from $B_0$ to $B_0+\delta B$ results in a small local increase of density by $\left\langle \delta\rho \right\rangle=\frac{\nu}{2\pi} \delta B$. This density accumulates as $\delta B$ is turned on, as a consequence of the Hall current that results from the electric field generated when the magnetic field varies. It does not disperse with time, since the bulk is gapped. Since charge is conserved, the density $ \left\langle \delta\rho \right\rangle$ must be supplied by the edges, which forces a correspondence of the bulk and edge responses. As explained in the introduction, this chain of events is encompassed by the bulk $U(1)$ CS term, and the corresponding edge anomaly. 

Roughly speaking, in gravitational response the role of the magnetic field $B$ is played by the curvature, while the role of the vector potential is played by the spin connection, which is first order in derivatives of the inverse metric $g^{ij}$. Thus, the emergent curvature involves two derivatives of the order parameter (see, e.g., \eqref{eq:3-8-8} below). The effect of these derivatives becomes evident when considering the gravitational analog to various electromagnetic vector potentials. For example, the vector potential associated with the Aharonov-Bohm effect decays as $1/r$, with $r$ being the distance from the Aharonov-Bohm flux tube. The analogous spin connection requires the perturbation to the order parameter to scale like $\log r$. 

The observable that responds to the spin connection may be the electronic density and current, but it may also be the density and current of momentum, or energy. A crucial difference between the IQHE and the $p$-wave SC, the absence of fermionic charge conservation in the latter, leads to profound differences between the bulk responses of both systems. In the absence of charge conservation, charge accumulation in the bulk does not necessarily involve the edges, and thus the way is opened to bulk Hall-type responses that do not correlate with the edge, and do not have quantized coefficients. There is a known example for such a response: when a weak magnetic field is introduced into a $p$-wave SC, the fermionic density receives a correction  $ \left\langle \delta\rho \right\rangle \propto \delta B$ \cite{volovik1988quantized,goryo1998abelian,goryo1999observation,stone2004edge,roy2008collective,hoyos2014effective,ariad2015effective}, yet with a proportionality constant that is not quantized. In this paper we find an additional example, where the fermionic density receives a correction proportional to the emergent curvature. These responses originate from bulk terms that carry some similarity to Chern-Simons terms, which we refer to as \textit{pseudo} Chern-Simons terms, see \eqref{eq:15-2} and  \eqref{eq:18-1}.

\subsection{Bulk responses }

\subsubsection{Topological bulk responses from a gravitational Chern-Simons term \label{subsubsec:Topological bulk responses from a gravitational Chern-Simons term}}

We find that  the effective action obtained by integrating over the
bulk fermions in the presence of a general order parameter $\Delta$ contains
a gCS term, with  coefficient $\alpha=\frac{\nu/2}{96\pi}\in\frac{1}{192\pi}\mathbb{Z}$.
Although we obtain this result in the limit $m^{*}\rightarrow\infty$,
we expect it to hold throughout the phase diagram. This is based on
known arguments for the quantization of the coefficient $\alpha$
due to symmetry, and on the relation with the boundary gravitational
anomaly described below. 

The gCS term implies a topological bulk response \eqref{eq:110-1},
where energy-momentum currents and densities appear due to a space-time
dependent order parameter. To gain insight into this result it is
best to examine special cases. Assume that the order parameter is
time independent, and that the relative phase is $\phi=\pm\frac{\pi}{2}$,
as in the $p_{x}\pm ip_{y}$ configuration, so that $\Delta=e^{i\theta}\left(\left|\Delta^{x}\right|,\pm i\left|\Delta^{y}\right|\right)$,
$o=\pm$. Then the metric is time independent, and takes the simple
form 
\begin{eqnarray}
 &  & g^{ij}=-\begin{pmatrix}\left|\Delta^{x}\right|^{2} & 0\\
0 & \left|\Delta^{y}\right|^{2}
\end{pmatrix}.\label{eq:3-3}
\end{eqnarray}
On this background, we find the following contributions to the expectation
values of the  fermionic energy current $J_{E}^{i}$, and momentum
density $P_{i}$ \footnote{$P_{x}$ ($P_{y}$) is the density of the $x$ ($y$) component of
momentum.}, 
\begin{eqnarray}
  \left\langle J_{E}^{i}\right\rangle _{\text{gCS}}&=&-\frac{\nu/2}{96\pi}\frac{1}{\hbar}\varepsilon^{ij}\partial_{j}\tilde{\mathcal{R}},\label{eq:4}\\
  \left\langle P_{i}\right\rangle _{\text{gCS}}&=&-\frac{\nu/2}{96\pi}\hbar g_{ik}\varepsilon^{kj}\partial_{j}\tilde{\mathcal{R}}.\nonumber 
\end{eqnarray}
Here $\tilde{\mathcal{R}}$ is the curvature, or Ricci scalar,
of the metric $g_{ij}$, which is the inverse of $g^{ij}$, and $\varepsilon^{xy}=-\varepsilon^{yx}=1$.
These are written without setting $\hbar$ or an
effective speed of light $c_{\mbox{light}}$ to 1 as we do in the
bulk of the paper. The curvature for the above metric is given explicitly
by 
\begin{eqnarray}
  \tilde{\mathcal{R}}&=&-2\left|\Delta^{x}\right|\left|\Delta^{y}\right|\\\label{eq:3-8-8}
  &\times& \left(\partial_{y}\left(\frac{\left|\Delta^{y}\right|\partial_{y}\left|\ensuremath{\Delta}^{x}\right|}{\left|\ensuremath{\Delta}^{x}\right|^{2}}\right)+\partial_{x}\left(\frac{\left|\Delta^{x}\right|\partial_{x}\left|\ensuremath{\Delta}^{y}\right|}{\left|\ensuremath{\Delta}^{y}\right|^{2}}\right)\right).\nonumber
\end{eqnarray}
It is a nonlinear expression in the order parameter, which is second
order in derivatives. Thus the responses \eqref{eq:4} are third order
in derivatives, and start at linear order but include nonlinear contributions
as well. The first equation in \eqref{eq:4} is analogous to the response $\left\langle J^{i}\right\rangle =\frac{\nu}{2\pi}\varepsilon^{ij}E_{j}$
of the IQHE. The second equation is analogous to the dual response
$\left\langle \rho\right\rangle =\frac{\nu}{2\pi}B$. Unlike the case of charge density, where the role of the magnetic field is played by the curvature (see Eq.\eqref{eq:9-1} below), for the case of the momentum density it is played by curvature gradients. Note that the
dependence on the sign in $\Delta=e^{i\theta}\left(\left|\Delta^{x}\right|,\pm i\left|\Delta^{y}\right|\right)$,
which is the orientation of $\Delta$, hides in the Chern number $\nu$
which is an odd function of the orientation. The above responses are odd under time-reversal, which flips the 
orientation of the order parameter but leaves the metric intact \footnote{The correct notion of time reversal for the $p$-wave SC flips $o$
but not $m$, as opposed to the natural time reversal within the relativistic
description. This is discussed in appendix \ref{subsec:Discrete-symmetries}.}.

Since there is no time dependence, energy is strictly conserved $\partial_{i}J_{E}^{i}=0$,
and it is meaningful to discuss energy transport. Integrating over
any cross section of the sample (a spatial curve $\gamma$ that starts
and ends on the boundary of the sample) we find the net bulk energy current
 through the cross section
\begin{eqnarray}
  \left\langle I_{E}\right\rangle _{\text{gCS}}&=&\int_{\gamma}\left\langle J_{E}^{i}\right\rangle _{\text{gCS}}\text{d}l_{i}\\\label{eq:6}
  &=&\frac{\nu/2}{96\pi}\frac{1}{\hbar}\left[\tilde{\mathcal{R}}\left(\gamma_{1}\right)-\tilde{\mathcal{R}}\left(\gamma_{0}\right)\right],\nonumber
\end{eqnarray}
where $l_{i}$ is a length element perpendicular to the curve, and
$\gamma_{0},\gamma_{1}$ are its end points.  

As an example, consider the order parameter $\Delta=\left(\Delta_{0}+\epsilon\cos\left(y/L\right),i\Delta_{0}\right)$
which is a perturbation to the $p_{x}+ip_{y}$ configuration with
$\epsilon\ll\Delta_{0}$. The scalar curvature for this order parameter is
$\tilde{\mathcal{R}}=\frac{2\epsilon\Delta_{0}}{L^{2}}\cos\left(\frac{y}{L}\right)+O\left(\epsilon^{2}\right)$
so there will be an energy current in the $x$ direction, $\left\langle J_{E}^{x}\right\rangle _{\text{gCS}}=\frac{\nu/2}{96\pi}\frac{1}{\hbar}\frac{2\epsilon\Delta_{0}}{L^{3}}\sin\left(y/L\right)+O\left(\epsilon^{2}\right)$.
If we assume that the system occupies the strip between
$y=0$ to $y=\frac{L\pi}{2}$, as depicted in Fig.\ref{fig:A-comparison-of}(b),
we get the net bulk energy current in the $x$ direction, 
\begin{eqnarray}
 &  & \left\langle I_{E}\right\rangle _{\text{gCS}}
 =\frac{\nu/2}{96\pi}\frac{1}{\hbar}\frac{2\epsilon\Delta_{0}}{L^{2}}+O\left(\epsilon^{2}\right).\label{eq:7}
\end{eqnarray}
The factor $\frac{\nu/2}{96\pi}\frac{1}{\hbar}$ only depends on the
Chern number, and thus on the topological phase, and  $\frac{2\Delta_{0}}{L^{2}}$
is a quantity that only depends on the order parameter. Note that the non-linear nature of the curvature leads to a dependence of the energy current on both the perturbation scale $\epsilon$ and the magnitude
of the unperturbed order parameter $\Delta_{0}$. The topological
invariant $\nu$ can then be measured in a thought experiment
where one tunes the order parameter as in the example and preforms
a measurement of the above contribution to $J_{E}$. In this manner a physical meaning
is assigned to $\nu$ in the bulk.

\subsubsection{Additional bulk responses from a gravitational pseudo  Chern-Simons
term }

Apart from the gCS term, the effective action obtained by integrating over the bulk fermions also contains an additional term of interest, which we refer to as a gravitational \textit{pseudo} Chern-Simons term (gpCS). This term is written explicitly and explained  in section \ref{subsec:Effective-action-for}. To the best of our knowledge, the gpCS term has not appeared previously in the context of the $p$-wave SC. It is possible because $U\left(1\right)$ symmetry is spontaneously broken in the $p$-wave SC. In the geometric point of view, this translates to the emergent geometry in the p-wave SC being not only curved but also torsion-full, see section \ref{sec:Emergent-Riemann-Cartan-geometry}.


The gpCS term  produces bulk responses which are closely
related to those of gCS, despite it being fully gauge invariant. This gauge invariance implies that it is not associated with a boundary anomaly, nor does its coefficient $\beta$ need to be quantized. Hence, gpCS does not encode \textit{topological} bulk responses. Remarkably, we find that $\beta$ is  quantized and identical to the coefficient $\alpha=\frac{\nu/2}{96\pi}$ of the gCS term in the limit of $m^*\rightarrow\infty$, but we do not expect this value to hold outside of this limit. We will put this phenomenon in a broader context in the discussion, section \ref{sec:Conclusion-and-discussion}.


Let us now describe the bulk responses from gpCS, setting
$\beta=\frac{\nu/2}{96\pi}$.
First, we find the following contributions to the fermionic energy
current and momentum density,
\begin{eqnarray}
  \left\langle J_{E}^{i}\right\rangle _{\text{gpCS}}&=&\frac{\nu/2}{96\pi}\varepsilon^{ij}\partial_{j}\tilde{\mathcal{R}},\label{eq:11-1}\\
  \left\langle P_{i}\right\rangle _{\text{gpCS}}&=&-\frac{\nu/2}{96\pi}g_{ik}\varepsilon^{kj}\partial_{j}\tilde{\mathcal{R}}.\nonumber 
\end{eqnarray}
Up to the sign difference in the first equation, these responses are the same as those from gCS \eqref{eq:4}.

As opposed to gCS, the gpCS term also contributes to the fermionic
charge density $\rho=-\psi^{\dagger}\psi$. For the bulk responses
we have written thus far, every Majorana spinor contributed $\nu_{n}=\frac{o_{n}}{2}\text{sgn}\left(m_{n}\right)$,
and summing over $n$ produced the Chern number $\nu$. For the density
response this is not the case. Here, the $n$th Majorana spinor contributes
\begin{eqnarray}
  \left\langle \rho\right\rangle _{\text{gpCS}}=\frac{o_{n}\nu_{n}/2}{24\pi}\sqrt{g}\tilde{\mathcal{R}},\label{eq:9-1}
\end{eqnarray}
where $\sqrt{g}=\sqrt{\text{det}g_{ij}}$ is the emergent volume
element. The orientation $o_{n}$ in Eq. (\ref{eq:9-1}) makes the sum over the four Majorana spinors different from the Chern number, $\sum_{n=1}^{4}o_{n}\nu_{n}=\sum_{n=1}^{4}\frac{1}{2}\text{sgn}\left(m_{n}\right)\neq\nu$.
 The appearance of $o_{n}$ can be understood by considering the
effect of time reversal. Because both the density $\rho$ and the
curvature $\tilde{\mathcal{R}}$ are time reversal even, the coefficient
in \eqref{eq:9-1} must also be even, and cannot be $\nu_{n}$ which
is odd. The response \eqref{eq:9-1} also holds when the order parameter
is time dependent, in which case $\tilde{\mathcal{R}}$ will also contain
time derivatives. One then finds a time dependent density, but there
is no corresponding current response, which is due to the non-conservation
of fermionic charge in a superconductor. It is instructive to compare
\eqref{eq:9-1} to the response $\rho=\frac{\nu}{2\pi}B$ of the IQHE.
Here $B$ is time reversal odd, which is why the coefficient can be
the Chern number $\nu$, and there is also the corresponding current
$\left\langle J^{i}\right\rangle =\frac{\nu}{2\pi}\varepsilon^{ij}E_{j}$
such that $\partial_{\mu}J^{\mu}=0$, as opposed to the $p$-wave
SC. 

To gain some insight into the expressions we have written thus far,
we write the operators $P,J_{E}$ more explicitly. For each Majorana
spinor (suppressing the index $n$), 
\begin{eqnarray}
  P_{j}&=&\frac{i}{2}\psi^{\dagger}\overleftrightarrow{D_{j}}\psi,\label{eq:10-5}\\
  J_{E}^{j}&=&g^{jk}P_{k}+\frac{o}{2}\partial_{k}\left(\frac{1}{\sqrt{g}}\varepsilon^{jk}\rho\right)+O\left(\frac{1}{m^{*}}\right).\nonumber 
 \end{eqnarray}
 
These expressions can be understood from the gravitational
description of the $p$-wave SC, see section \ref{subsec:Currents,-symmetries,-and}.
The momentum density is the familiar expression for free fermions,
but in the energy current we have only written explicitly contributions
that survive the limit $m^{*}\rightarrow\infty$. These contributions
are only possible due to the $p$-wave pairing, and are of order $\Delta^{2}$.

From the relation \eqref{eq:10-5} between $J_{E}$, $P$ and $\rho$ we can understand
that the equality $\left\langle J_{E}^{j}\right\rangle _{\text{gCS}}=g^{jk}\left\langle P_{k}\right\rangle _{\text{gCS}}$
expressed in equation \eqref{eq:4} is a result of the vanishing contribution of gCS to the density $\rho$. We can also understand the sign
difference between the first and second line of \eqref{eq:11-1} as a result
of \eqref{eq:9-1}. The important point is that a measurement of the
charge density $\rho$ can be used to fix the value of the coefficient $\beta$, which is generically unquantized,  and thus separate the contributions of
gpCS to $P,J_{E}$, from those of gCS. In this manner, one can overcome
the obscuring of gCS by gpCS. A more detailed analysis is given in section \ref{subsec:Calculation-of-bulk}.

\subsection{Bulk-boundary correspondence from gravitational anomaly }

Among the two terms in the bulk effective action which we described
above only gCS is related to the boundary gravitational anomaly.
This relation can be fully analyzed in the case where $\Delta=\Delta_{0}e^{i\theta\left(t,x\right)}\left(1+f\left(x,t\right),i\right)$
is a perturbation of the $p_{x}+ip_{y}$ configuration with small
$f$, and there is a domain wall (or boundary) at $y=0$ where the
value of $\nu$ jumps. For simplicity, assume $\nu=1$ for $y<0$
and $\nu=0$ for $y>0$. This situation is illustrated in Fig.\ref{fig:A-comparison-of}(d).
In section \ref{sec:Boundary-fermions-and} we derive the action for the boundary, or edge mode, 
\begin{align}
 &  & S_{\text{e}}=\frac{i}{2}\int\mbox{d}t\text{d}x\tilde{\xi}\left(\partial_{t}-\left|\Delta^{x}\left(t,x\right)\right|\partial_{x}\right)\tilde{\xi},
\end{align}
which describes a chiral $D=1+1$ Majorana fermion $\tilde{\xi}$
localized on the boundary, with a space-time dependent velocity $\left|\Delta^{x}\left(x,t\right)\right|=\Delta_{0}\left|1+f\left(x,t\right)\right|$.
Classically, the edge fermion $\tilde{\xi}$ conserves energy-momentum
in the following sense, 
\begin{eqnarray}
 &  & \partial_{\beta}t_{\mbox{e}\;\alpha}^{\beta}+\partial_{\alpha}\mathcal{L}_{\mbox{e}}=0.\label{eq:12-1}
\end{eqnarray}
Here $t_{\mbox{e}}$ is the canonical energy-momentum tensor for $\tilde{\xi}$,
with indices $\alpha,\beta=t,x$, and $\mathcal{L}_{\mbox{e}}$ is
the edge Lagrangian, $S_{\text{e}}=\int\text{d}t\mathcal{L}_{\mbox{e}}$, see \ref{subsec:Energy-momentum}.
For $\alpha=t$ ($\alpha=x$), equation \eqref{eq:12-1} describes
the sense in which the edge fermion conserves energy (momentum) classically.
The source term $\partial_{\alpha}\mathcal{L}_{\mbox{e}}$ 
follows from the space-time dependence of $\mathcal{L}_{\text{e}}$
through $\Delta^{x}$. Quantum mechanically, the action $S_{\text{e}}$
is known to have a gravitational anomaly, which means that energy-momentum
is not conserved at the quantum level \cite{bertlmann2000anomalies}.
In the context of emergent gravity, this implies that equation \eqref{eq:12-1}
is violated for the expectation values,  
\begin{align}
  \partial_{\beta}\left\langle t_{\mbox{e}\;\alpha}^{\beta}\right\rangle +\partial_{\alpha}\left\langle \mathcal{L}_{\mbox{e}}\right\rangle =-\frac{\nu/2}{96\pi}g_{\alpha\gamma}\varepsilon^{\gamma\beta y}\partial_{\beta}\tilde{\mathcal{R}}.\label{eq:12}
\end{align}
This equation is written with $\hbar=1$ and $c_{\mbox{light}}=\frac{\Delta_{0}}{\hbar}=1$
for simplicity. Since $\Delta^x$ depends on time, $\tilde{\mathcal{R}}$ is not the curvature
of the spatial metric $g_{ij}$, but of a corresponding space-time
metric $g_{\mu\nu}$ \eqref{eq:10-1}, and is given by $\tilde{\mathcal{R}}=\ddot{f}-2\dot{f}^{2}+O\left(f\ddot{f},f\dot{f}^{2}\right)$
in this case. Note that time dependence in this example is crucial.
 From gCS we find for $\Delta=\Delta_{0}e^{i\theta\left(t,x\right)}\left(1+f\left(x,t\right),i\right)$
the bulk energy-momentum tensor
\begin{eqnarray}
 &  & \left\langle t_{\;\alpha}^{y}\right\rangle _{\text{gCS}}=-\frac{\nu/2}{96\pi}g_{\alpha\gamma}\varepsilon^{\gamma\beta y}\partial_{\beta}\tilde{\mathcal{R}},\label{eq:13}
\end{eqnarray}
which explains the anomaly as the inflow of energy-momentum from the
bulk to the boundary,
\begin{eqnarray}
 &  & \partial_{\beta}\left\langle t_{\mbox{e}\;\alpha}^{\beta}\right\rangle +\partial_{\alpha}\left\langle \mathcal{L}_{\mbox{e}}\right\rangle =\left\langle t_{\;\alpha}^{y}\right\rangle _{\text{gCS}}.
\end{eqnarray}
Since $\nu$ jumps from 1 to 0 at $y=0$ the energy-momentum
current \eqref{eq:13}  stops at the boundary and does not extend to the $y>0$ region. The gravitationally anomalous boundary
mode is then essential for the conservation of total energy-momentum to hold. As this example shows, 
bulk-boundary correspondence follows from bulk+boundary conservation
of energy-momentum in the presence of a space-time dependent order
parameter.

\section{Lattice model\label{sec:Lattice-model} }

In this section  we review and slightly generalize a simple lattice
model for a $p$-wave SC \cite{bernevig2013topological}, which will
serve as our microscopic starting point. We describe its band structure
and its symmetry protected topological phases, and also  explain
some of the basics of the emergent geometry which can be seen in this
setting.

 The hamiltonian is given in real space by 
\begin{eqnarray}
  H=&-&\frac{1}{2}\sum_{\boldsymbol{l}}\left[t\psi_{\boldsymbol{l}}^{\dagger}\psi_{\boldsymbol{l}+x}+t\psi_{\boldsymbol{l}}^{\dagger}\psi_{\boldsymbol{l}+y}+\mu\psi_{\boldsymbol{l}}^{\dagger}\psi_{\boldsymbol{l}}\right.\nonumber\\
  &+&\left.\delta^{x}\psi_{\boldsymbol{l}}^{\dagger}\psi_{\boldsymbol{l}+x}^{\dagger} 
  + \delta^{y}\psi_{\boldsymbol{l}}^{\dagger}\psi_{\boldsymbol{l}+y}^{\dagger}+h.c\right].\label{eq:2-1}
\end{eqnarray}
Here the sum is over all lattice sites $\boldsymbol{l}\in L$ of
a 2 dimensional square lattice $L=a\mathbb{Z}\times a\mathbb{Z}$,
with a lattice spacing $a$. $\psi_{\boldsymbol{l}}^{\dagger},\psi_{\boldsymbol{l}}$
are creation and annihilation operators for spin-less fermions on
the lattice, with the canonical anti commutators $\left\{ \psi_{\boldsymbol{l}}^{\dagger},\psi_{\boldsymbol{l}'}\right\} =\delta_{\boldsymbol{l}\boldsymbol{l}'}$.
$\boldsymbol{l}+x$ denotes the nearest neighboring site to $\boldsymbol{l}$
in the $x$ direction. The hopping amplitude $t$ is real and $\mu$
is the chemical potential. Apart from the single particle terms $t\psi_{\boldsymbol{l}}^{\dagger}\psi_{\boldsymbol{l}+x}+t\psi_{\boldsymbol{l}}^{\dagger}\psi_{\boldsymbol{l}+y}+\mu$,
there is also the pairing term $\delta^{x}\psi_{\boldsymbol{l}}^{\dagger}\psi_{\boldsymbol{l}+x}^{\dagger}+\delta^{y}\psi_{\boldsymbol{l}}^{\dagger}\psi_{\boldsymbol{l}+y}^{\dagger}$
, with the order parameter $\delta=\left(\delta^{x},\delta^{y}\right)\in\mathbb{\mathbb{C}}^{2}$.
We think of $\delta$ as resulting from a Hubbard-Stratonovich decoupling
of interactions, in which case we refer to it as intrinsic, or as
being induced by proximity to an $s$-wave SC. In both cases we treat
$\delta$ as a bosonic background field that couples to the fermions.

The generic order parameter is charged under a few  symmetries of
the single particle terms. The order parameter has charge 2 under
the global $U\left(1\right)$ group generated by $Q=-\sum_{\boldsymbol{l}}\psi_{\boldsymbol{l}}^{\dagger}\psi_{\boldsymbol{l}}$,
in the sense that $e^{-i\alpha Q}H\left(e^{2i\alpha}\delta\right)e^{i\alpha Q}=H\left(\delta\right)$,
which physically represents the electromagnetic charge $-2$ of Cooper
pairs\footnote{Since $\delta$ has charge 2, $H$ commutes with the fermion parity
$\left(-1\right)^{Q}$.  The Ground state of $H$ will therefore
be labelled by a fermion parity eigenvalue $\pm1$, in addition to
the topological label which is the Chern number \cite{read2000paired,kitaev2009periodic}.
Fermion parity is a subtle quantity in the thermodynamic limit, and
will not be important in the following.}.  The order parameter is also charged under time reversal $T$,
which is an anti unitary transformation satisfying $T^{2}=1$, that
acts as the complex conjugation of coefficients in the Fock basis
corresponding to $\psi_{\boldsymbol{l}},\psi_{\boldsymbol{l}}^{\dagger}$.
The equation $T^{-1}H\left(\delta^{*}\right)T=H\left(\delta\right)$
shows $\delta\mapsto\delta^{*}$ under time reversal. Finally, $\delta$
is also charged under the point group symmetry of the lattice, which
for the square lattice is the Dihedral group $D_{4}$. The continuum
analog of this is that the order parameter is charged under spatial
rotations and reflections, and more generally, under space-time transformations
(diffeomorphisms), which is due to the orbital angular momentum 1
of Cooper pairs in a $p$-wave SC. This observation will be important
for our analysis, and will be discussed further below. 

 In an intrinsic $p_{x}\pm ip_{y}$ SC, the configuration of $\delta$
which minimizes the ground state energy is given by $\delta=\delta_{0}e^{i\theta}\left(1,\pm i\right)$,
where $\delta_{0}>0$ is determined by the minimization, but the
sign $o=\pm1$ and the phase $\theta$ (which dynamically corresponds
to a goldstone mode) are left undetermined. See \cite{volovik2009universe}
for a pedagogical discussion of a closely related model within mean
field theory. A choice of $\theta$ and $o$ corresponds to a spontaneous
symmetry breaking of the group $U\left(1\right)\rtimes\left\{ 1,T\right\} $
including both the $U\left(1\right)$ and time reversal transformations.
More accurately, in the $p_{x}\pm ip_{y}$ SC, the group $\left(U\left(1\right)\rtimes\left\{ 1,T\right\} \right)\times D_{4}$
is spontaneously broken down to a certain diagonal subgroup. We discuss
the continuum analog of this and its implications in section \ref{subsec:Energy-momentum}. 

Crucially, we do not restrict $\delta$ to the $p_{x}\pm ip_{y}$
configuration, and treat it as a general two component complex vector
$\delta=\left(\delta^{x},\delta^{y}\right)\in\mathbb{\mathbb{C}}^{2}$.
In the following  we will take $\delta$ to be space time dependent,
$\delta\mapsto\delta_{\boldsymbol{l}}\left(t\right)$, and show that
this space time dependence can be thought of as a perturbation to
which there is a topological response, but for now we assume $\delta$
is constant. 

\subsection{\label{subsec:Band-structure-and}Band structure and phase diagram }

Writing the Hamiltonian \eqref{eq:2-1} in Fourier space, and in the
BdG form in terms of the Nambu spinor $\Psi_{\boldsymbol{q}}=\left(\psi_{\boldsymbol{q}},\psi_{-\boldsymbol{q}}^{\dagger}\right)^{T}$
we find
\begin{align}
  H & =\frac{1}{2}\int_{BZ}\frac{\mbox{d}^{2}\boldsymbol{q}}{\left(2\pi\right)^{2}}\Psi_{\boldsymbol{q}}^{\dagger}\begin{pmatrix}h_{\boldsymbol{q}} &  \delta_{\boldsymbol{q}}\\
\delta_{\boldsymbol{q}}^{*}  & -h_{\boldsymbol{q}}
\end{pmatrix}\Psi_{\boldsymbol{q}}+const\nonumber \\
   & =\frac{1}{2}\int_{BZ}\frac{\mbox{d}^{2}\boldsymbol{q}}{\left(2\pi\right)^{2}}\Psi_{\boldsymbol{q}}^{\dagger}\left(\boldsymbol{d}_{\boldsymbol{q}}\cdot\boldsymbol{\sigma}\right)\Psi_{\boldsymbol{q}}+const,\label{eq:3}
\end{align}
with $h_{\boldsymbol{q}}=-t\cos\left(aq_{x}\right)-t\cos\left(aq_{y}\right)-\mu$
real and symmetric, and $\delta_{\boldsymbol{q}}=-i\delta^{x}\sin\left(aq_{x}\right)-i\delta^{y}\sin\left(aq_{y}\right)$
complex and anti-symmetric.  Here $\boldsymbol{\sigma}=\left(\sigma^{x},\sigma^{y},\sigma^{z}\right)$
is the vector of Pauli matrices and $BZ$ is the Brillouin zone $BZ=\left(\mathbb{R}/\frac{2\pi}{a}\mathbb{Z}\right)^{2}$.
By definition, the Nambu spinor obeys the reality condition $\Psi_{\boldsymbol{q}}^{\dagger}=\left(\sigma^{x}\Psi_{-\boldsymbol{q}}\right)^{T}$,
and is therefore a Majorana spinor, see appendix \ref{subsec:Charge-conjugation-(Appendix)}.
Accordingly, the BdG Hamiltonian is particle-hole (or charge conjugation)
symmetric, $\sigma^{x}H\left(\boldsymbol{q}\right)^{*}\sigma^{x}=-H\left(-\boldsymbol{q}\right)$,
and therefore belongs to symmetry class D of the Altland-Zirnbauer
classification of free fermion Hamiltonians \cite{ryu2010topological}.
The constant in \eqref{eq:3} is $\frac{1}{2}\text{tr}h=\frac{V}{2}\int\frac{\text{d}^{2}\boldsymbol{q}}{\left(2\pi\right)^{2}}h_{\boldsymbol{q}}$
where $V$ is the infinite volume. This operator ordering correction
is important as it contributes to physical quantities such as the
energy density and charge density, but we will mostly keep it implicit
in the following. The BdG band structure is given by $E_{\boldsymbol{q},\pm}=\pm\frac{1}{2}E_{\boldsymbol{q}}$
where 
\begin{eqnarray}
 E_{\boldsymbol{q}}=\left|\boldsymbol{d}_{\boldsymbol{q}}\right|=\sqrt{h_{\boldsymbol{q}}^{2}+\left|\delta_{\boldsymbol{q}}\right|^{2}}.
\end{eqnarray}

\begin{figure}[!th]

\subfloat[]{
\includegraphics[width=0.475\linewidth]{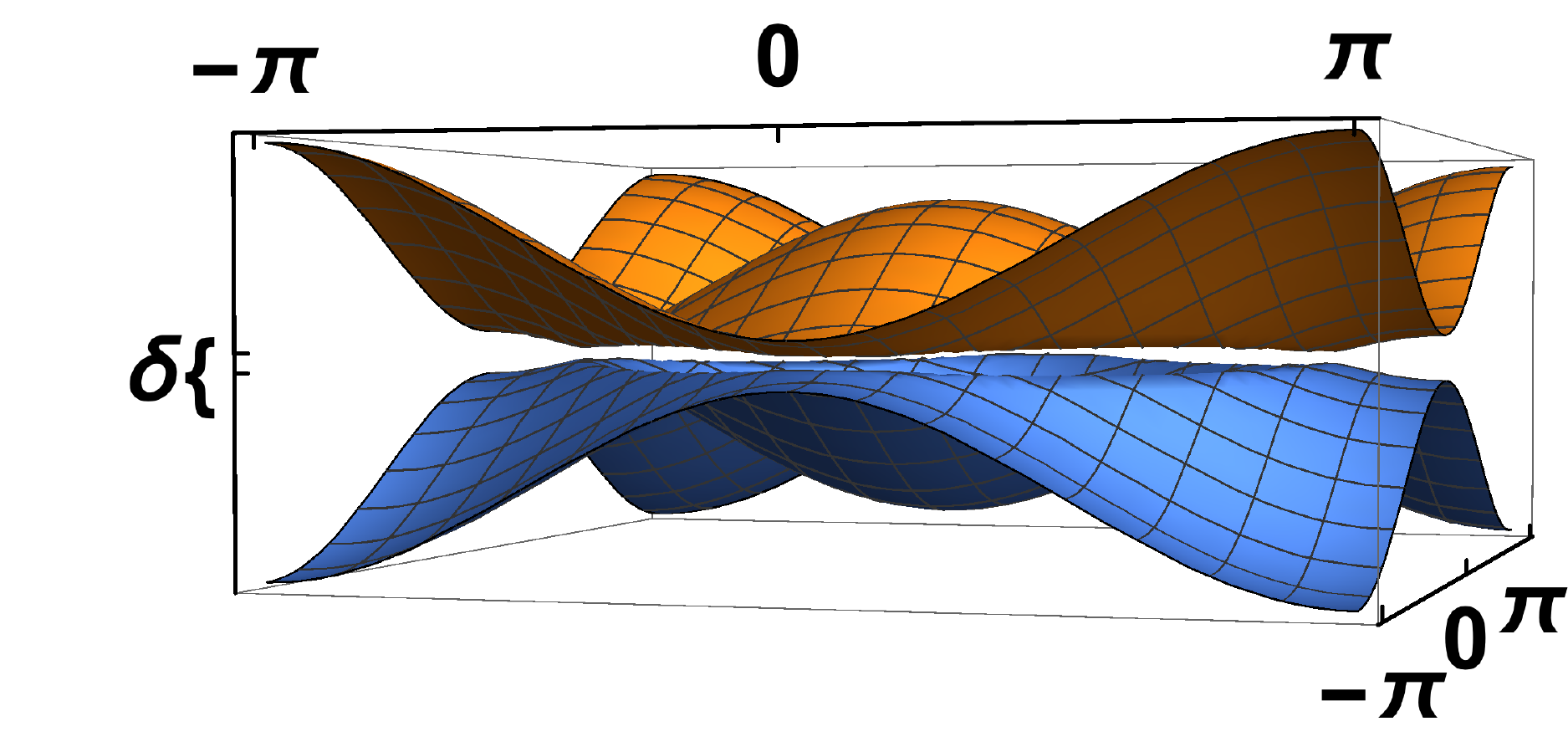}
} \subfloat[]{
\includegraphics[width=0.475\linewidth]{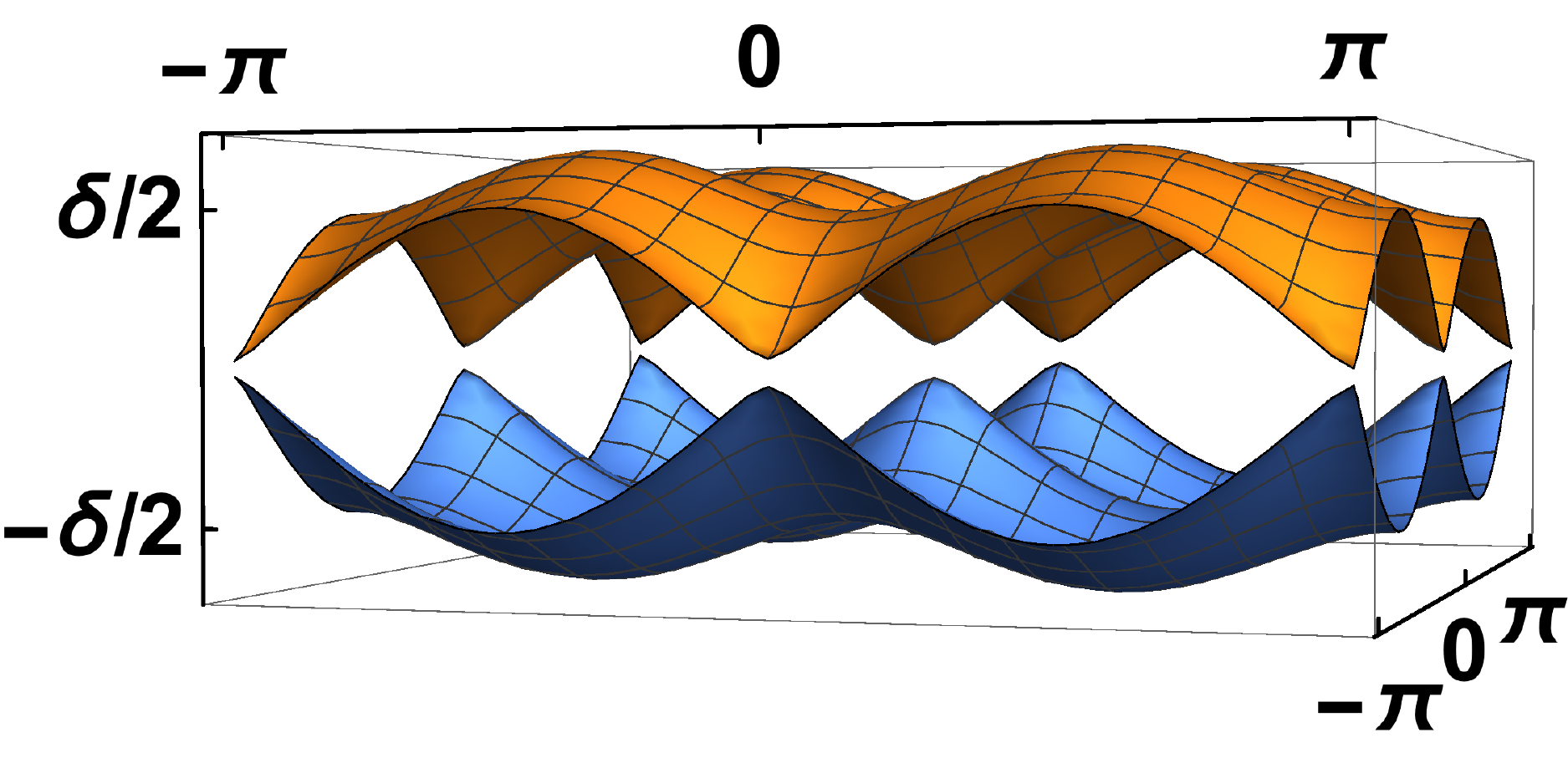}
}
\caption{Generic band structure of the lattice model. (a) When the
order parameter is much smaller than the single particle bandwidth
$\delta\ll t$, the spectrum takes the form of a gapped single particle
Fermi surface with gap $\sim\delta$. This regime describes the onset
of superconductivity, and it is appropriate to refer to $\delta$
as the ``gap function''. (b) When the order parameter is
much larger than the single particle scales $\delta\gg t,\mu$, the
spectrum takes the form of four regulated relativistic fermions centered
at the particle-hole invariant points $\left(0,0\right),\left(0,\pi\right),\left(\pi,0\right),\left(\pi,\pi\right)$,
in units of the inverse lattice spacing $a^{-1}$. We will be working
in this regime. \label{fig:Generic-band-structure}}
\end{figure}

For the $p_{x}\pm ip_{y}$ configuration $\left|\delta_{\boldsymbol{q}}\right|^{2}=\delta_{0}^{2}\left(\sin^{2}aq_{x}+\sin^{2}aq_{y}\right)$,
and therefore $E_{\boldsymbol{q}}$ can only vanish at the particle-hole
invariant points $a\boldsymbol{K}^{\left(1\right)}=\left(0,0\right),a\boldsymbol{K}^{\left(2\right)}=\left(0,\pi\right),a\boldsymbol{K}^{\left(3\right)}=\left(\pi,\pi\right),a\boldsymbol{K}^{\left(4\right)}=\left(\pi,0\right)$,
which happens when $\mu=-2t,0,2t,0$. Representative band structures
are plotted in Fig.\ref{fig:Generic-band-structure}. For $\delta_{0}\ll t$
the spectrum takes the form of a gapped single particle Fermi surface
with gap $\sim\delta_{0}$, while for $\delta_{0}\gg t$ one obtains
Four regulated relativistic fermions centered at the points $\boldsymbol{K}^{\left(n\right)},\;1\leq n\leq4$
with masses $m_{n}=-2t-\mu,-\mu,2t-\mu,-\mu$, speed of light $c_{\text{light}}=a\delta_{0}/\hbar$,
bandwidth $\sim\delta_{0}$ and momentum cutoff $\sim a^{-1}$. 

\begin{figure}[!th]

\begin{centering}

 \subfloat[]{
\includegraphics[width=0.47\linewidth]{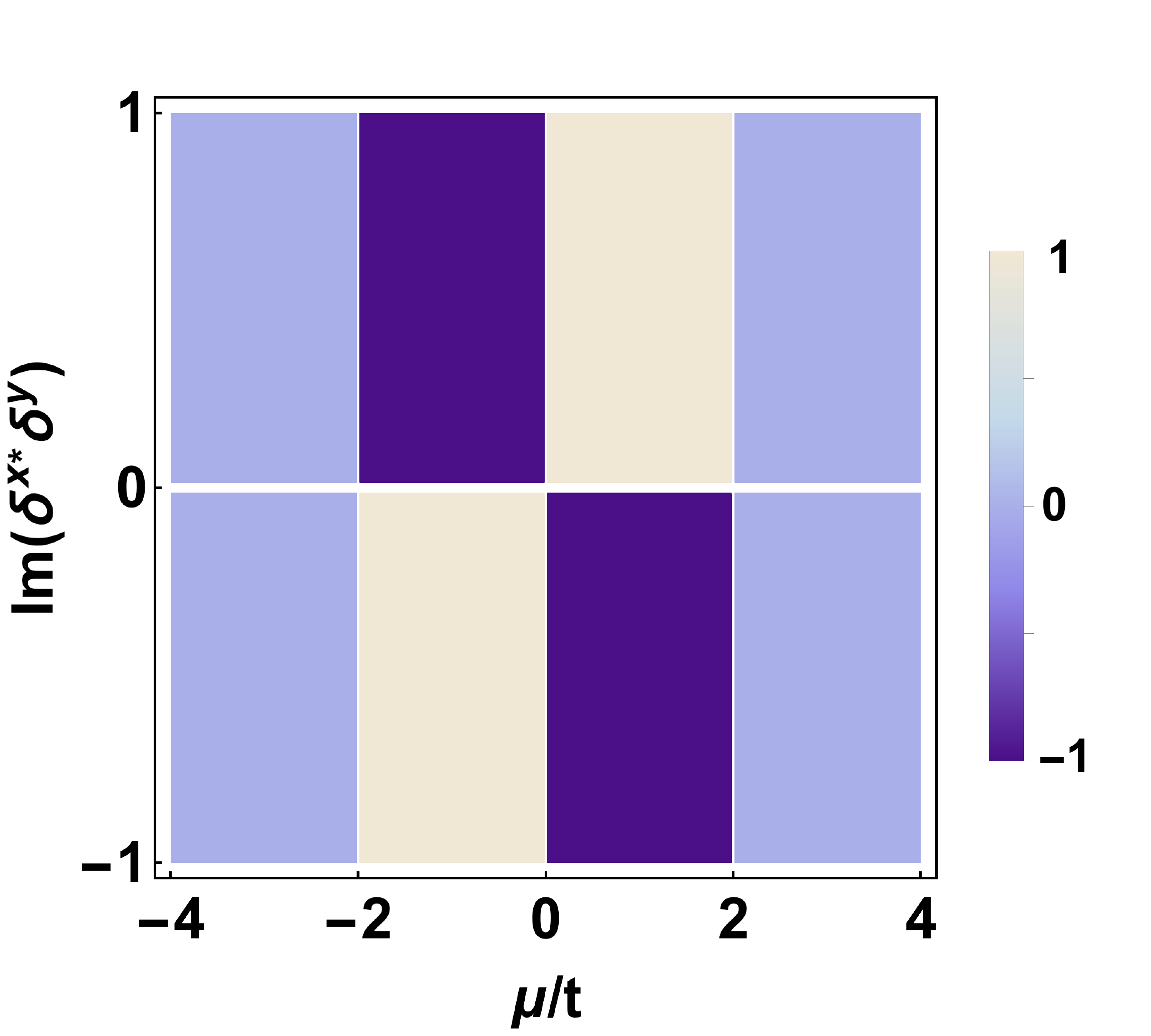}
} \subfloat[]{
\includegraphics[width=0.47\linewidth]{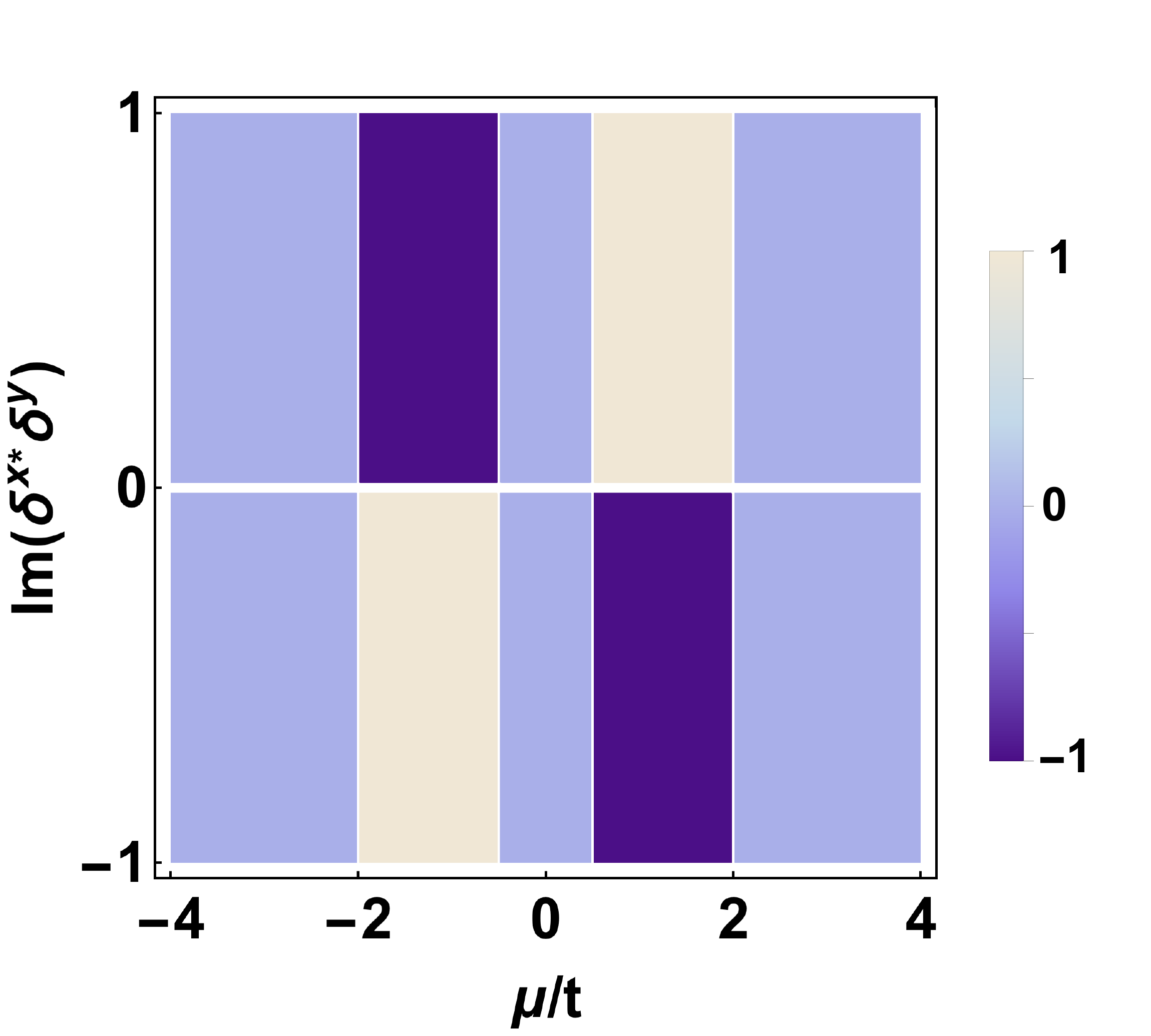}
}
\par\end{centering}

\caption{The topological phase diagram of the lattice model is simplest to
understand from the formula $\nu=\frac{1}{2}\sum_{n=1}^{4}o_{n}\text{sgn}\left(m_{n}\right)$
for the Chern number in terms of the masses and orientations of low energy relativistic
fermions. (a) Topological phase diagram for isotropic hopping
$t$. Units on the vertical axis are arbitrary, the topological phase
diagram only depends  on the orientation $o=\text{sgn}\left(\text{Im}\left(\delta^{x*}\delta^{y}\right)\right)$.
(b) Topological phase diagram for anisotropic hopping $t^{x}\protect\neq t^{y}$,
 additional trivial phases exist around $\mu=0$. Here $t=\frac{t^{x}+t^{y}}{2}$.
\label{fig:Phase-Diagram}}
\end{figure}

With generic $\mu,\delta_{0}$ the spectrum is gapped, and the Chern
number $\nu$ labeling the different topological phases is well defined. It can be calculated by $\nu=\int_{BZ}\frac{\text{d}^{2}k}{2\pi}\text{tr}\left(\mathcal{F}\right)$
where $\mathcal{F}$ is the Berry curvature on the Brillouin zone $BZ$ \cite{ryu2010topological}.
A more general definition is $\nu=\frac{1}{24\pi^{2}}\int_{\mathbb{R}\times BZ}\text{tr}\left(G\text{d}G^{-1}\right)^{3}$\footnote{More explicitly,

$\nu=\frac{1}{24\pi^{2}}\mbox{tr}\int_{\mathbb{R}\times BZ}\mbox{d}^{3}k\varepsilon^{\alpha\beta\gamma}\left(G\partial_{\alpha}G^{-1}\right)\left(G\partial_{\beta}G^{-1}\right)\left(G\partial_{\gamma}G^{-1}\right)$.}, where 
$G\left(k_{0},k_{x},k_{y}\right)$ is the single particle propagator
\cite{volovik2009universe}, which remains valid in the presence of weak interactions, as long as the gap does not close. For two band Hamiltonians such as \eqref{eq:3}, $\nu$ reduces to the homotopy type of the map $\hat{\boldsymbol{d}_{\boldsymbol{q}}}=\boldsymbol{d}_{\boldsymbol{q}}/\left|\boldsymbol{d}_{\boldsymbol{q}}\right|$
from $BZ$ (which is a flat torus) to the sphere, 
\begin{align}
\nu=\frac{a^{2}}{\left(2\pi\right)^{2}}\int_{BZ}\text{d}^{2}\boldsymbol{q}\hat{\boldsymbol{d}_{k}}\cdot\left(\partial_{q_{y}}\hat{\boldsymbol{d}_{\boldsymbol{q}}}\times\partial_{q_{y}}\hat{\boldsymbol{d}_{\boldsymbol{q}}}\right)\in\mathbb{Z}.
\end{align}
One obtains $\nu=0$ for $\left|\mu\right|>2t$, $\nu=\pm1$ for $\mu\in\left(0,2t\right)$
and $\nu=\mp1$ for $\mu\in\left(-2t,0\right)$. The topological phase
diagram is plotted in Fig.\ref{fig:Phase-Diagram}(a). 

Away from the $p_{x}\pm ip_{y}$ configuration, the topological phase
diagram is essentially unchanged. For $\text{Im}\left(\delta^{x*}\delta^{y}\right)\neq0$,
 gap closings happen at the same points $\boldsymbol{K}^{\left(n\right)}$
and the same values of $\mu$ described above. $\nu$ takes the same
values, with the orientation $o=\text{sgn}\left(\text{Im}\left(\delta^{x*}\delta^{y}\right)\right)$,
described below, generalizing the sign $\pm1$ that characterizes
the $p_{x}\pm ip_{y}$ configuration. For $\text{Im}\left(\delta^{x*}\delta^{y}\right)=0$
the spectrum is always gapless. The topological phase diagram is most easily understood from the
formula $\nu=\frac{1}{2}\sum_{n=1}^{4}o_{n}\text{sgn}\left(m_{n}\right)$
where $o_{n}=\pm1$ are orientations associated with the relativistic
fermions which we describe below \cite{sticlet2012edge}. 

It will also be useful consider a slight generalization of the single
particle part of the lattice model, with un-isotropic hopping $t^{x}\psi_{\boldsymbol{l}}^{\dagger}\psi_{\boldsymbol{l}+x}+t^{y}\psi_{\boldsymbol{l}}^{\dagger}\psi_{\boldsymbol{l}+y}$.
This changes the masses to $m_{1}=-\left(t_{1}+t_{2}\right)-\mu,m_{2}=t_{1}-t_{2}-\mu,m_{3}=t_{1}+t_{2}-\mu,m_{4}=-\left(t_{1}-t_{2}\right)-\mu$.
In particular, the degeneracy between the masses $m_{2},m_{4}$ breaks,
and additional trivial phases appear around $\mu=0$. See Fig.\ref{fig:Phase-Diagram}(b).

\subsection{Basics of the emergent geometry \label{subsec:The-order-parameter}}

A key insight which we will extensively use, originally due to Volovik,
is that the order parameter is in fact a \textit{vielbein}. In the
present space-time independent situation, this vielbein is just a
$2\times2$ matrix which generically will be invertible
\begin{eqnarray}
 &  & e_{A}^{\;\;j}=\left(\begin{array}{cc}
\mbox{Re}(\delta^{x}) & \mbox{Re}(\delta^{y})\\
\mbox{Im}(\delta^{x}) & \mbox{Im}(\delta^{y})
\end{array}\right)\in GL\left(2\right),\label{eq:5-1}
\end{eqnarray}
where $A=1,2,\;j=x,y$. More accurately, $e_{A}^{\;\;j}$ is invertible if $\text{det}\left(e_{A}^{\;\;i}\right)=\text{Im}\left(\delta^{x*}\delta^{y}\right)\neq0$.
We refer to an order parameter as singular if $\text{Im}\left(\delta^{x*}\delta^{y}\right)=0$.
From the vielbein one can calculate a metric, which in the present
situation is a general symmetric positive semidefinite matrix 
\begin{align}
 g^{ij}=e_{A}^{\;\;i}\delta^{AB}e_{B}^{\;\;j}&=\delta^{(i}\delta^{j)*}\label{eq:6-2}\\
 &=\left(\begin{array}{cc}
\left|\delta^{x}\right|^{2} & \mbox{Re}\left(\delta^{x}\delta^{y*}\right)\\
\mbox{Re}\left(\delta^{x}\delta^{y*}\right) & \left|\delta^{y}\right|^{2}
\end{array}\right).\nonumber
\end{align}
 Every vielbein determines a metric uniquely, but the converse is
not true. Vielbeins $e,\tilde{e}$ that are related by an internal
reflection and rotation $e_{A}^{\;j}=\tilde{e}{}_{B}^{\;\;j}L_{\;A}^{B}$
with $L\in O\left(2\right)$ give rise to the same metric. By diagonalization,
it is also clear that any metric can be written in terms of a vielbein.
Therefore the set of (constant) metrics can be parameterized by the
coset $GL\left(2\right)/O\left(2\right)$. To see this explicitly
we parameterize $\delta=e^{i\theta}\left(\left|\delta^{x}\right|,e^{i\phi}\left|\delta^{y}\right|\right)$
with the overall phase $\theta$ and relative phase $\phi\in\left(-\pi,\pi\right]$.
Then 
\begin{align}
  g^{ij}=\left(\begin{array}{cc}
\left|\delta^{x}\right|^{2} & \left|\delta^{x}\right|\left|\delta^{y}\right|\cos\phi\\
\left|\delta^{x}\right|\left|\delta^{y}\right|\cos\phi & \left|\delta^{y}\right|^{2}
\end{array}\right)
\end{align}
is independent of $\theta$ which parametrizes $SO\left(2\right)$
and $\text{sgn}\phi$ which parametrizes $O\left(2\right)/SO\left(2\right)$.
Note that the group $O\left(2\right)$ of internal rotations and reflections
is just $U\left(1\right)\rtimes\left\{ 1,T\right\} $ acting on $e_{A}^{\;\;j}$.
In more detail, $\delta\mapsto e^{2i\alpha}\delta$ (or $\delta\mapsto\delta^{*}$)
corresponds to $e_{A}^{\;\;i}\mapsto L_{\;A}^{B}e_{B}^{\;\;i}$ with

\begin{align} 
L=\begin{pmatrix}\cos2\alpha & \sin2\alpha\\
-\sin2\alpha & \cos2\alpha
\end{pmatrix} 
\left(\text{or } 
L=\begin{pmatrix}1 & 0\\
0 & -1 
\end{pmatrix}\right).
\end{align}

The internal reflections, corresponding to a reversal of time,
flip the \textit{orientation} of the vielbein $o=\text{sgn}\left(\text{det}\left(e_{A}^{\;\;i}\right)\right)=\text{sgn}\left(\text{Im}\left(\delta^{x*}\delta^{y}\right)\right)$\textit{,
}and therefore every quantity that depends on $o$ is time reversal
odd\textit{. }We will also refer to $o$ as the orientation of the
order parameter. An order parameter with a positive (negative) orientation
can be thought of as $p_{x}+ip_{y}$-like ($p_{x}-ip_{y}$-like). 

For the $p_{x}\pm ip_{y}$ configuration, $\delta=e^{i\theta}\delta_{0}\left(1,\pm i\right)$,
one obtains a scalar metric $g^{ij}=\delta_{0}\delta^{ij}$, independent
of the phase $\theta$ and the orientation $o=\pm$. We see that $\theta,o$
correspond precisely to the $O\left(2\right)=U\left(1\right)\rtimes\left\{ 1,T\right\} $
degrees of freedom of the vielbein to which the metric is blind to.
Thus the metric $g^{ij}$ corresponds to the Higgs part of the order
parameter, by which we mean the part of the order parameter on which
the ground state energy depends, in the intrinsic case. 

The fact that $U\left(1\right)$ transformations map to internal rotations
also appears naturally in the BdG formalism which we will use in the
following. Consider the Nambu spinor $\Psi=\left(\psi,\psi^{\dagger}\right)^{T}$.
It follows from the $U\left(1\right)$ action $\psi\mapsto e^{i\alpha}\psi$
that $\Psi\mapsto e^{i\alpha\sigma^{z}}\Psi$ where $\sigma^{z}$
is the Pauli matrix. We see that $U\left(1\right)$ acts on $\Psi$
as a spin rotation. Moreover, the fact that $\delta$ has charge $2$
while $\psi$ has charge 1 implies $e$ is an $SO\left(2\right)$
vector while $\Psi$ is a spinor. 

\section{Continuum limits of the lattice model\label{sec:Continuum-limits-of}}

\subsection{The $p$-wave superfluid \label{subsec:Coupling-the-Lattice}}

Consider the lattice model \eqref{eq:2-1}, with a general space time
dependent order parameter $\delta_{\boldsymbol{l}}=\left(\delta_{\boldsymbol{l}}^{x}\left(t\right),\delta_{\boldsymbol{l}}^{y}\left(t\right)\right)$,
and minimally coupled to electromagnetism, 
\begin{eqnarray}
  H=&-&\frac{1}{2}\sum_{\boldsymbol{l}}\left[t\psi_{\boldsymbol{l}}^{\dagger}e^{iA_{\boldsymbol{l},\boldsymbol{l}+x}}\psi_{\boldsymbol{l}+x}+\left(\mu_{\boldsymbol{l}}+A_{t,\boldsymbol{l}}\right)\psi_{\boldsymbol{l}}^{\dagger}\psi_{\boldsymbol{l}}\right.\nonumber\\
 &+&\left.\delta_{\boldsymbol{l}}^{x}\psi_{\boldsymbol{l}}^{\dagger}e^{iA_{\boldsymbol{l},\boldsymbol{l}+x}}\psi_{\boldsymbol{l}+x}^{\dagger}+\left(x\leftrightarrow y\right)+h.c\right].\label{eq:8}
\end{eqnarray}

Here $A_{\boldsymbol{l},\boldsymbol{l}'},A_{t,\boldsymbol{l}}$ are the
components of a $U\left(1\right)$ gauge field describing background
electromagnetism, on the discrete space and continuous time. We will
work in the relativistic regime $\delta_{0}\gg t,\mu$ where $\delta_{0}$
is a characteristic scale for $\delta$. To obtain a continuum description,
we split $BZ$ into four quadrants $BZ=\cup_{n=1}^{4}BZ^{\left(n\right)}$
centered around the four points $\boldsymbol{K}^{\left(n\right)}$,
 and decompose the fermion operator $\psi_{\boldsymbol{l}}$ as a
sum $\psi_{\boldsymbol{l}}=\sum_{n=1}^{4}\psi_{\boldsymbol{l}}^{\left(n\right)}e^{i\boldsymbol{K}^{\left(n\right)}\cdot\boldsymbol{l}}$,
where $\psi_{\boldsymbol{l}}^{\left(n\right)}e^{i\boldsymbol{K}^{\left(n\right)}\cdot\boldsymbol{l}}$
has non zero Fourier modes only in $BZ^{\left(n\right)}$. Thus the
fermions $\psi^{\left(n\right)}$ all have non zero Fourier modes
only in $BZ^{\left(n\right)}-\boldsymbol{K}^{\left(n\right)}=\left[-\frac{\pi}{2a},\frac{\pi}{2a}\right]^{2}$.
This restriction of the quasi momenta provides the fermions $\psi^{\left(n\right)}$
with a \textit{physical} cutoff $\sim a^{-1}$, which will be important
when we compare results from the continuum description to the lattice
model. Assuming $\mu,\delta,A$ have small derivatives relative to
$a^{-1}$, the inter fermion terms in $H$ can be neglected and $H$
splits into a sum $H\approx\sum_{n=1}^{4}H^{\left(n\right)}$, with
$H^{\left(n\right)}$ a Hamiltonian for $\psi^{\left(n\right)}$.
We then expand the Hamiltonians $H^{\left(n\right)}$ in small $\psi^{\left(n\right)}$
derivatives relative to $a^{-1}$. The resulting Hamiltonian, focusing
on the point $\boldsymbol{K}^{\left(1\right)}=\left(0,0\right)$,
is the $p$-wave superfluid (SF) Hamiltonian 
\begin{eqnarray}
  H_{\text{SF}}=\int\text{d}^{2}x\left[\psi^{\dagger}\left( - \frac{D^{2}}{2m^{*}}+m-A_{t}\right)\psi\right.\label{eq:9}\\
  -\left.\left(\frac{1}{2}\psi^{\dagger}\Delta^{j}\partial_{j}\psi^{\dagger}+h.c\right)\right],\nonumber
\end{eqnarray}
where the fermion field has been redefined such that $\left\{ \psi^{\dagger}\left(x\right),\psi\left(x'\right)\right\} =\delta^{\left(2\right)}\left(x-x'\right)$.
Here $D_{\mu}=\partial_{\mu}-iA_{\mu}$ is the $U\left(1\right)$-covariant
derivative, with the connection $A=A_{j}\text{d}x^{j}$ related to
$A_{\boldsymbol{l},\boldsymbol{l}'}$ by $A_{\boldsymbol{l},\boldsymbol{l}'}=\int_{\boldsymbol{l}}^{\boldsymbol{l}'}A$,
and $D^{2}=\delta^{ij}D_{i}D_{j}$ with $i,j=x,y$. Note the appearance
of the flat background spatial metric $\delta^{ij}$. The effective
mass is related to the hopping amplitude $1/m^{*}=a^{2}t$, and the
order parameter is $\Delta=a\delta$, so it is essentially the lattice
order parameter. The chemical potential for the $p$-wave SF is $-m$. The coupling to $A$ in the pairing term is lost,
since $\psi^{\dagger}\psi^{\dagger}=0$. For this reason it is a derivative
and not a covariant derivative that appears in $\psi^{\dagger}\Delta^{j}\partial_{j}\psi^{\dagger}$,
and one can verify that this term is gauge invariant. Moreover, due
to the anti-commutator $\left\{ \psi^{\dagger}\left(x\right),\psi^{\dagger}\left(y\right)\right\} =0$
any operator put between two $\psi^{\dagger}$s is anti-symmetrized,
and in particular $\psi^{\dagger}\Delta^{j}\partial_{j}\psi^{\dagger}=\psi^{\dagger}\frac{1}{2}\left\{ \Delta^{j},\partial_{j}\right\} \psi^{\dagger}$
where $\left\{ \Delta^{j},\partial_{j}\right\} $ is the anti-commutator
of differential operators. This Hamiltonian is essentially the one
considered in \cite{read2000paired} for the $p$-wave SF. The corresponding
action is the $p$-wave SF action 
\begin{eqnarray}
 S_{\text{SF}}\left[\psi,\Delta,A\right]=\int&\text{d}^{2+1}&x\left[\psi^{\dagger}\left(iD_{t}+\frac{D^{2}}{2m^{*}}-m\right)\psi\right.\nonumber\\
 &+&\left.\left(\frac{1}{2}\psi^{\dagger}\Delta^{j}\partial_{j}\psi^{\dagger}+h.c\right)\right],\label{eq:10}
\end{eqnarray}
in which $\psi,\psi^{\dagger}$ are no longer fermion operators, but
independent Grassmann valued fields, $\left\{ \psi\left(x\right),\psi^{\dagger}\left(x'\right)\right\} =0$.
This action comes equipped with a momentum cutoff $\Lambda_{UV}\sim a^{-1}$
inherited from the lattice model. 

For the other points $\boldsymbol{K}^{\left(2\right)},\boldsymbol{K}^{\left(3\right)},\boldsymbol{K}^{\left(4\right)}$
the SF action obtained is slightly different. The chemical potential for the $n$th fermion is $-m_{n}$.The order parameter
for the $n$th fermion is $\Delta_{\left(n\right)}^{x}=a\delta^{x}e^{iK_{x}^{\left(n\right)}},\;\Delta_{\left(n\right)}^{y}=a\delta^{y}e^{iK_{y}^{\left(n\right)}}$,
and we note that $e^{iK_{j}^{\left(n\right)}}=\pm1$. The order parameters
for $\boldsymbol{K}^{\left(1\right)}=\left(0,0\right),\;\boldsymbol{K}^{\left(3\right)}=\left(\pi,\pi\right)$
are related by an overall sign, which is a $U\left(1\right)$ transformation,
and so are the order parameters for $\boldsymbol{K}^{\left(2\right)}=\left(0,\pi\right),\boldsymbol{K}^{\left(4\right)}=\left(\pi,0\right)$.
Thus the order parameters for $n=1,3$ are physically indistinguishable,
and so are order parameters for $n=2,4$. The order parameters for
$n=1$ and $n=2$ are however physically distinct. First, the orientations
$o_{n}=\text{sgn}\left(\text{Im}\left(\Delta_{\left(n\right)}^{x*}\Delta_{\left(n\right)}^{y}\right)\right)$
are different, with $o_{1}=-o_{2}$. Second, the metrics $g_{\left(n\right)}^{ij}=\Delta^{(i}\Delta^{j)*}$
are generically different, with the same diagonal components, but
$g_{\left(1\right)}^{xy}=-g_{\left(2\right)}^{xy}$. We note that
if the relative phase between $\delta^{x}$ and $\delta^{y}$ is $\pm\pi/2$,
as in the $p_{x}\pm ip_{y}$ configuration, then all metrics $g_{\left(n\right)}^{ij}$
are diagonal and therefore equal. These differences between the orientations
and metrics of the different lattice fermions will be important later
on. 

Similarly, the effective mass tensor which in \eqref{eq:9}, for $n=1$,
is $\left(M^{-1}\right)^{ij}=\frac{\delta^{ij}}{m^{*}}$, has different
signatures for different $n$, but this will not be important in this
paper. For now we continue working with the action \eqref{eq:10}
for the $n=1$ fermion, keeping the other lattice fermions implicit
until section \ref{subsec:Summing-over-lattice}.

\subsection{Relativistic limit of the $p$-wave superfluid \label{subsec:Relativistic-limit-of}}

Since we work in the relativistic regime $\delta\gg t,\mu$ we can
treat the term $\psi^{\dagger}\frac{D^{2}}{2m^{*}}\psi$ as a perturbation
and compute quantities to zeroth order in $1/m^{*}$. Then $S_{\text{SF}}$
reduces to what we refer to as the relativistic limit of the $p$-wave
SF action, given in BdG form by 
\begin{eqnarray}
 S_{\text{rSF}}\left[\psi,\Delta,A\right]&\label{eq:14}\\
 =\frac{1}{2}\int\mbox{d}^{2+1}&x\Psi^{\dagger}\begin{pmatrix}i\partial_{t}+A_{t}-m & \frac{1}{2}\left\{ \Delta^{j},\partial_{j}\right\} \\
-\frac{1}{2}\left\{ \Delta^{*j},\partial_{j}\right\}  & i\partial_{t}-A_{t}+m
\end{pmatrix}\Psi.\nonumber
\end{eqnarray}
It is well known that when $\Delta$ takes the $p_{x}\pm ip_{y}$
configuration $\Delta=\Delta_{0}e^{i\theta}\left(1,\pm i\right)$
and $A=0$ this action is that of a relativistic Majorana spinor in
Minkowski space-time, with mass $m$ and speed of light $c_{\text{light}}=\frac{\Delta_{0}}{\hbar}$.
In the following, we will see that for general $\Delta$ and $A$, 
\eqref{eq:14} is the action of a relativistic Majorana spinor in
curved and torsion-full space-time. We wil sometimes refer to the relativistic
limit as $m^{*}\rightarrow\infty$, though this is somewhat loose,
because in the relativistic regime both $m^{*}$ is large and $m$
is small.

Before we go on to analyze the $p$-wave SF in the relativistic limit,
it is worth considering what of the physics of the $p$-wave SF is
captured by the relativistic limit, and what is not. First, the coupling
to $A_{x},A_{y}$ is lost, so the relativistic limit is blind to the
magnetic field. Since superconductors are usually defined by their
interaction with the magnetic field, the relativistic limit is actually
insufficient to describe the properties of the $p$-wave SF as a superconductor.
Of course, a treatment of superconductivity also requires the dynamics
of $\Delta$. Likewise, the term $\frac{1}{2m^{*}}\psi^{\dagger}D^{2}\psi=\frac{1}{2m^{*}}\psi^{\dagger}\delta^{ij}D_{i}D_{j}\psi$
seems to be the only term in $S_{\text{SF}}$ that includes the flat
background metric $\delta^{ij}$, describing the real geometry of
space. It appears that the relativistic limit is insufficient to describe
the response of the system to a change in the real geometry of space\footnote{In fact, some of the response to the real geometry can be obtained,
see our discussion, section \ref{sec:Conclusion-and-discussion}.}. Nevertheless, as is well known, the relativistic limit does suffice
to determine the topological phases of the $p$-wave SC as a free
(and weakly interacting) fermion system. Indeed, the Chern number
labeling the different topological phases can be calculated by the
formula $\nu=\frac{1}{2}\sum_{n=1}^{4}o_{n}\text{sgn}\left(m_{n}\right)$,
which only uses data from the relativistic limit. Here the sum is
over the four particle-hole invariant points of the lattice model,
with orientations $o_{n}$ and masses $m_{n}$. This suggests that
at least some physical properties characterizing the different free
fermion topological phases can be obtained from the relativistic limit.
Indeed, in the following we will see how a topological bulk response
and a corresponding boundary anomaly can be obtained within
the relativistic limit.

\section{Emergent Riemann-Cartan geometry\label{sec:Emergent-Riemann-Cartan-geometry}}

We argue that \eqref{eq:14} is precisely the action which describes
a relativistic massive Majorana spinor in a curved and torsion-full
background known as Riemann-Cartan (RC) geometry, with a particular
form of background fields. We refer the reader to \cite{ortin2004gravity}
parts I.1 and I.4.4, for a review of RC geometry and the coupling
of fermions to it, and provide only the necessary details here, focusing
on the implications for the $p$-wave SF. For simplicity we work locally
and in coordinates, and we differ the treatment of global aspects
to appendix \ref{subsec:Global-structures-and}. 

The action describing the dynamics of a Majorana spinor on RC background
in 2+1 dimensional space-time can be written as 
\begin{eqnarray}
 S_{\text{RC}}\left[\chi,e,\omega\right]&\label{eq:43-1}\\
 =\frac{1}{2}\int&\mbox{d}^{2+1}x\left|e\right|\overline{\chi}\left[\frac{i}{2}e_{a}^{\;\mu}\left(\gamma^{a}D_{\mu}-\overleftarrow{D_{\mu}}\gamma^{a}\right)-m\right]\chi.\nonumber
\end{eqnarray}
 Here $\chi$ is a Majorana spinor with mass $m$  obeying, as a
field operator, the canonical anti-commutation relation $\left\{ \chi\left(x\right),\chi\left(y\right)\right\} =\frac{\delta^{\left(2\right)}\left(x-y\right)}{\left|e\left(x\right)\right|}$,
where we suppressed spinor indices. As a Grassmann field $\left\{ \chi\left(x\right),\chi\left(y\right)\right\} =0$.
The field $e_{a}^{\;\mu}$ is an inverse vielbein which is an invertible
matrix at each point in space-time. The indices $a,b,\dots\in\left\{ 0,1,2\right\} $
are $SO\left(1,2\right)$ (Lorentz) indices which we refer to as
internal indices, while $\mu,\nu,\dots\in\left\{ t,x,y\right\} $
are coordinate indices.

We will also use $A,B,\dots\in\left\{ 1,2\right\} $ for spatial internal
indices and $i,j,\dots\in\left\{ x,y\right\} $ for spatial coordinate
indices

The vielbein $e_{\;\mu}^{a}$, is the inverse of $e_{a}^{\;\mu}$,
such that $e_{\;\mu}^{a}e_{a}^{\;\nu}=\delta_{\mu}^{\nu},\;e_{\;\mu}^{a}e_{b}^{\;\mu}=\delta_{b}^{a}$.
It is often useful to view the vielbein as a set of linearly independent
(local) one-forms $e^{a}=e_{\;\mu}^{a}\text{d}x^{\mu}$. The metric
corresponding to the vielbein is $g_{\mu\nu}=e_{\;\mu}^{a}\eta_{ab}e_{\;\nu}^{b}$
and the inverse metric is $g^{\mu\nu}=e_{a}^{\;\mu}\eta^{ab}e_{b}^{\;\nu}$,
where $\eta_{ab}=\eta^{ab}=\text{diag}\left[1,-1,-1\right]$ is the
flat Minkowski metric. Internal indices are raised and lowered using
$\eta$, while coordinate indices are raised and lowered using $g$
and its inverse. Using $e$ one can replace internal indices with
coordinate indices and vice versa, e.g $v^{a}=e_{\;\mu}^{a}v^{\mu}$.
The volume element is defined by $\left|e\right|=\left|\text{det}e_{\;\mu}^{a}\right|=\sqrt{g}$.
$\left\{ \gamma^{a}\right\} _{a=0}^{2}$ are gamma matrices obeying
$\left\{ \gamma^{a},\gamma^{b}\right\} =2\eta^{ab}$, and we will
work with $\gamma^{0}=\sigma^{z},\;\gamma^{1}=-i\sigma^{x},\;\gamma^{2}=i\sigma^{y}$\footnote{The gamma matrices form a basis for the Clifford algebra  associated
with $\eta$. The above choice of basis is a matter of convention.
}. The covariant derivative $D_{\mu}=\partial_{\mu}+\omega_{\mu}$\footnote{We use the notation $D$ for spin, Lorentz, and $U\left(1\right)$
covariant derivatives in any representation, and the exact meaning
should be clear from the field $D$ acts on. } contains the spin connection $\omega_{\mu}=\frac{1}{2}\omega_{ab\mu}\Sigma^{ab}$,
where $\Sigma^{ab}=\frac{1}{4}\left[\gamma^{a},\gamma^{b}\right]$
generate the spin group $Spin\left(1,2\right)$ which is the double
cover of the Lorentz group $SO\left(1,2\right)$. Note that $\omega_{ab\mu}=-\omega_{ba\mu}$
and therefore $\omega_{\;b\mu}^{a}$ is an $SO\left(1,2\right)$ connection.
It follows that $\omega$ is metric compatible, $D_{\mu}\eta_{ab}=0$.
It is often useful to work (locally) with a connection one-form $\omega=\omega_{\mu}\text{d}x^{\mu}$.
$\overline{\chi}$ is the Dirac conjugate defined as in Minkowski
space-time $\overline{\chi}=\chi^{\dagger}\gamma^{0}$. The derivative
$\overleftarrow{D_{\mu}}$ acts only on $\overline{\chi}$ and is
explicitly given by $\chi\overleftarrow{D_{\mu}}=\partial_{\mu}\overline{\chi}-\overline{\chi}\omega_{\mu}$.

Our statement is that $S_{\text{RC}}\left[\chi,e,\omega\right]$ evaluated
on the fields 
\begin{eqnarray}
 \chi&=&\left|e\right|^{-1/2}\Psi,\label{17}\\
 e_{a}^{\;\mu}&=&\frac{1}{\Delta_{0}}\left(\begin{array}{ccc}
\Delta_{0} & 0 & 0\\
0 & \mbox{Re}(\Delta^{x}) & \mbox{Re}(\Delta^{y})\\
0 & \mbox{Im}(\Delta^{x}) & \text{Im}(\Delta^{y})
\end{array}\right),\hspace{5bp}\omega_{\mu}=-2A_{\mu}\Sigma^{12},\nonumber
\end{eqnarray}
reduces precisely to $S_{\text{rSF}}\left[\psi,\Delta,A\right]$ of
equation \eqref{eq:14}, where one must keep in mind that $S_{\text{RC}}$
is written in relativistic units where $\hbar=1$ and $c_{\text{light}}=\frac{\Delta_{0}}{\hbar}=1$,
which we will use in the following. Moreover, the functional integral
over $\chi$ is equal to the functional integral over $\Psi$. This
is a slight refinement of the original statement by Volovik  and
subsequent work by Read and Green \cite{read2000paired}. We defer
the proof to appendices \ref{subsec:Equivalent-forms-of} and \ref{subsec:Equality-of-path},
where we also address certain subtleties that arise. Here we describe
the particular RC geometry that follows from \eqref{17}, and attempt
to provide some intuition for this geometric description of the $p$-wave
SF.  Starting with the vielbein, note that the only nontrivial part
of $e_{a}^{\;\mu}$ is the spatial part $e_{A}^{\;j}$, which is just
the order parameter $\Delta$, as in \eqref{eq:5-1}. The inverse
metric we obtain from our vielbein is 
\begin{eqnarray}
 g^{\mu\nu}&=&e_{a}^{\;\mu}\eta^{ab}e_{b}^{\;\nu}\label{eq:10-1}\\
 &=&\frac{1}{\Delta_{0}^{2}}\left(\begin{array}{ccc}
\Delta_{0}^{2} & 0 & 0\\
0 & -\left|\Delta^{x}\right|^{2} & -\mbox{Re}\left(\Delta^{x}\Delta^{*y}\right)\\
0 & -\mbox{Re}\left(\Delta^{x}\Delta^{*y}\right) & -\left|\Delta^{y}\right|^{2}
\end{array}\right),\nonumber
\end{eqnarray}
where the spatial part $g^{ij}=-\frac{1}{\Delta_{0}^{2}}\Delta^{(i}\Delta^{j)*}$
is the Higgs part of the order parameter, as in \eqref{eq:6-2}. For
the $p_{x}\pm ip_{y}$ configuration the metric reduces to the Minkowski
metric. If $\Delta$ is time independent $g^{\mu\nu}$ describes a
Riemannian geometry which is trivial in the time direction, but we
allow for a time dependent $\Delta$. A metric of the form \eqref{eq:10-1}
is said to be in gaussian normal coordinates with respect to space
\cite{carroll2004spacetime}. 

The $U\left(1\right)$ connection $A_{\mu}$ maps to a $Spin\left(2\right)$
connection $\omega_{\mu}=-2A_{\mu}\Sigma^{12}=-iA_{\mu}\sigma^{z}$
 which corresponds to spatial spin rotations. This is a special case
of the general $Spin\left(1,2\right)$ connection which appears in
RC geometry. The fact that $U\left(1\right)$ transformations map
to spin rotations when acting on the Nambu spinor $\Psi$ is a general
feature of the BdG formalism as was already discussed in section \ref{subsec:The-order-parameter}.
 From the spin connection $\omega$ it is natural to construct a
curvature, which is a matrix valued two-form defined by $R_{\;b}^{a}=\text{d}\omega_{\;b}^{a}+\omega_{\;c}^{a}\wedge\omega_{\;b}^{c}$.
In local coordinates $x^{\mu}$ it can be written as $R_{\;b}^{a}=\frac{1}{2}R_{\;b\mu\nu}^{a}\text{d}x^{\mu}\wedge\text{d}x^{\nu}$,
where the components are given explicitly by $R_{\;b\mu\nu}^{a}=\partial_{\mu}\omega_{\;b\nu}^{a}-\partial_{\nu}\omega_{\;b\mu}^{a}+\omega_{\;c\mu}^{a}\omega_{\;b\nu}^{c}-\omega_{\;c\nu}^{a}\omega_{\;b\mu}^{c}$.
It follows from \eqref{17} that in our case the only non zero components
are  
\begin{eqnarray}
 &  & R_{12}=-R_{21}=-2F,
\end{eqnarray}
 where the two form $F=\text{d}A$ is the $U\left(1\right)$ field
strength, or curvature, comprised of the electric and magnetic fields.

\subsection{Torsion and additional geometric quantities}

\textcolor{red}{}

Since we treat $A$ and $\Delta$ as independent background fields,
so are the spin connection $\omega$ and vielbein $e$. This situation
is referred to as the first order vielbein formalism for gravity \cite{ortin2004gravity}.
 Apart from the metric $g$ and the curvature $R$ which we already
described, there are a few more geometric quantities which can be
constructed from $e,\omega$, and that will be used in this paper.
These additional quantities revolve around the notion of torsion.

The torsion tensor $T$ is an important geometrical quantity, but
a pragmatic way to view it is as a useful parameterization for the
set of all spin connections $\omega$, for a fixed vielbein $e$.
Thus one can work with the variables $e,T$ instead of $e,\omega$.
We will see later on that the bulk responses in the $p$-wave SC are
easier to describe using $e,T$. This is analogous to, and as we will
see, generalizes, the situation in $s$-wave SC, where the independent
degrees of freedom are $A$ and $\Delta=\left|\Delta\right|e^{i\theta}$,
but it is natural to change variables and work with $\Delta$ and
$D_{\mu}\theta=\partial_{\mu}\theta-2A_{\mu}$ instead. We now provide
the details. 

The torsion tensor, or two-form, is defined in terms of $e,\omega$
as $T^{a}=De^{a}$, or in coordinates $T_{\mu\nu}^{a}=2D_{[\mu}e_{\nu]}^{a}$.
Since our temporal vielbein $e^{0}=\text{d}t$ is trivial and the
connection $\omega$ is only an $SO\left(2\right)$ connection, $T^{0}=0$
for all $A$ and $\Delta$. All other components of the torsion are
in general non trivial, and are given by $T_{ij}^{A}=D_{i}e_{\;j}^{A}-D_{j}e_{\;i}^{A},\;T_{ti}^{A}=-T_{it}^{A}=D_{t}e_{\;i}^{A}$.
This describes the simple change of variables from $\omega$ to $T$. 

Going from $T$ back to $\omega$ is slightly more complicated, and
is done as follows. One starts by finding the $\omega$ that corresponds
to $T=0$. The solution is the unique torsion free spin connection
$\tilde{\omega}=\tilde{\omega}\left(e\right)$ which we refer to as
the Levi Civita (LC) spin connection\footnote{The unique torsion free spin connection $\tilde{\omega}$ is also
referred to as the Cartan connection is the literature.}. This connection is given explicitly by $\tilde{\omega}_{abc}=\frac{1}{2}\left(\xi_{abc}+\xi_{bca}-\xi_{cab}\right)$
where $\xi_{\;bc}^{a}=2e_{b}^{\;\mu}e_{c}^{\;\nu}\partial_{[\mu}e_{\;\nu]}^{a}$.
Now, for a general $\omega$ the difference $C_{\;b\mu}^{a}=\omega_{\;b\mu}^{a}-\tilde{\omega}_{\;b\mu}^{a}$
is referred to as the contorsion tensor, or one-form. It carries the
same information as $T$ and the two are related by $T^{a}=C_{\;b}^{a}\wedge e^{b}$
($T_{\mu\nu}^{a}=2C_{\;b[\mu}^{a}e_{\;\nu]}^{b}$) and $C_{\mu\alpha\nu}=\frac{1}{2}\left(T_{\alpha\mu\nu}+T_{\mu\nu\alpha}-T_{\nu\alpha\mu}\right)$.
One can then reconstruct $\omega$ from $e,T$ as $\omega=\tilde{\omega}\left(e\right)+C\left(e,T\right)$.
Note that $\omega,\tilde{\omega}$ are both connections, but $C,T$
are tensors. 

For the $p_{x}\pm ip_{y}$ configuration $\Delta=\Delta_{0}e^{i\theta}\left(1,\pm i\right)$
one finds $\tilde{\omega}_{12\mu}=-\tilde{\omega}_{21\mu}=-\partial_{\mu}\theta$
(with all other components vanishing), and it follows that $C_{12\mu}=D_{\mu}\theta$.
These are familiar quantities in the theory of superconductivity,
and one can view $\tilde{\omega}$ and $C$ as generalizations of
these. General formulas are given in appendix \ref{subsec:Explicit-formulas-for}.

Using $\tilde{\omega}$ one can define a covariant derivative $\tilde{D}$
and curvature $\tilde{R}$ just as $D$ and $R$ are constructed from
$\omega$. The quantity $\tilde{R}_{\;\nu\rho\sigma}^{\mu}$ is the
usual Riemann tensor of Riemannian geometry and general relativity.
Note that $\tilde{R}_{\;\nu\rho\sigma}^{\mu}$ depends solely on $g$
which is the Higgs part of the order parameter $\Delta$. Since $g$
is flat in the $p_{x}\pm ip_{y}$ configuration, we conclude that
a non vanishing Riemann tensor requires a deviation of the Higgs part
of $\Delta$ from the $p_{x}\pm ip_{y}$ configuration. As in Riemannian
geometry we can define the Ricci tensor $\tilde{\mathcal{R}}_{\nu\sigma}=\tilde{R}_{\;\nu\mu\sigma}^{\mu}$
and Ricci scalar $\tilde{\mathcal{R}}=\tilde{\mathcal{R}}_{\;\nu}^{\nu}$.
Examples for the calculation of $\tilde{\mathcal{R}}$ in terms of
$\Delta$ where given in section \ref{sec:Results,-examples,-and}.

Another important quantity which can be constructed from $e,\omega$
is the affine connection $\Gamma_{\;\beta\mu}^{\alpha}=e_{a}^{\;\alpha}\left(\partial_{\mu}e_{\;\beta}^{a}+\omega_{\;b\mu}^{a}e_{\;\beta}^{b}\right)=e_{a}^{\;\alpha}D_{\mu}e_{\;\beta}^{a}$,
or affine connection (local) one-form $\Gamma_{\;\beta}^{\alpha}=\Gamma_{\;\beta\mu}^{\alpha}\text{d}x^{\mu}$.
It is not difficult to see that $T$ is the anti symmetric part of
$\Gamma$, $T_{\;\mu\nu}^{\rho}=\Gamma_{\mu\nu}^{\rho}-\Gamma_{\nu\mu}^{\rho}$,
and it follows that the LC affine connection $\tilde{\Gamma}_{\;\beta\mu}^{\alpha}=e_{a}^{\;\alpha}\tilde{D}_{\mu}e_{\;\beta}^{a}$,
for which $T=0$, is symmetric in its the two lower indices. This
is the usual metric compatible and torsion free connection of Riemannian
geometry, given by the Christoffel symbol $\tilde{\Gamma}_{\alpha\beta\mu}=\frac{1}{2}\left(\partial_{\mu}g_{\beta\alpha}+\partial_{\beta}g_{\alpha\mu}-\partial_{\alpha}g_{\mu\beta}\right)$.
$\Gamma$ appears in covariant derivatives of tensors with coordinate
indices, for example $\nabla_{\mu}v^{\alpha}=\partial_{\mu}v^{\alpha}+\Gamma_{\;\beta\mu}^{\alpha}v^{\beta}$,
$\nabla_{\mu}v_{\alpha}=\partial_{\mu}v_{\alpha}-v_{\beta}\Gamma_{\;\alpha\mu}^{\beta}$,
and so on. We also denote by $\nabla$ the total covariant derivative
of tensors with both coordinate and internal indices, which includes
both $\omega$ and $\Gamma$. Thus, for example, $\nabla_{\mu}v_{\;\nu}^{a}=\partial_{\mu}v_{\;\nu}^{a}+\omega_{\;b\mu}^{a}v_{\;\nu}^{b}-v_{\;\nu}^{a}\Gamma_{\;\mu\alpha}^{\nu}=D_{\mu}v_{\;\nu}^{a}-v_{\;\nu}^{a}\Gamma_{\;\mu\alpha}^{\nu}$.
The most important occurrence of $\nabla$ is in the identity $\nabla_{\nu}e_{\;\mu}^{a}=0$,
which follows from the definition of $\Gamma$ in this formalism,
and is sometimes called the first vielbein postulate. It means that
the covariant derivative $\nabla$ commutes with index manipulation
preformed using $e,\eta$ and $g$. To obtain more intuition for what
$\Gamma$ is from the $p$-wave SC point of view, we can write it
as $\Gamma_{\;a\mu}^{\alpha}=-D_{\mu}e_{a}^{\;\alpha}$. Then it is
clear that the non vanishing components of $\Gamma_{\;a\mu}^{\alpha}$
are given by $\Gamma_{\;1\mu}^{j}+i\Gamma_{\;2\mu}^{j}=-D_{\mu}\Delta^{j}$
. 

\section{Symmetries, currents, and conservation laws \label{sec:Symmetries,-currents,-and}}

In order to map fermionic observables in the $p$-wave SF to those
of a Majorana fermion in RC space-time, it is usefull is to map the symmetries and the
corresponding conservation laws between the two. We start with $S_{\text{SF}}$,
and then review the analysis of $S_{\text{RC}}$ and show how it maps
to that of $S_{\text{SF}}$, in the relativistic limit. The bottom
line is that there is a sense in which electric charge and energy-momentum
are conserved in a $p$-wave SC, and this maps to the sense in
which spin and energy-momentum are conserved for a Majorana spinor
in RC space-time. 

\subsection{Symmetries, currents, and conservation laws of the $p$-wave superfluid
action\label{subsec:Symmetries,-currents,-and}}

\subsubsection{Electric charge }

$U\left(1\right)$ gauge transformations act on $\psi,\Delta,A$ by
\begin{align}
  \psi\mapsto e^{i\alpha}\psi,\;\Delta\mapsto e^{2i\alpha}\Delta,\;A_{\mu}\mapsto A_{\mu}+\partial_{\mu}\alpha.\label{eq:6.1}
\end{align}
This symmetry of  $S_{\text{SF}}\left[\psi,\Delta,A\right]$
implies a conservation law for electric charge, 
\begin{eqnarray}
  \partial_{\mu}J^{\mu}=-i\psi^{\dagger}\Delta^{j}\partial_{j}\psi^{\dagger}+h.c,\label{eq:15}
\end{eqnarray}
where $J^{\mu}=-\frac{\delta S}{\delta A_{\mu}}$ is the fermion
electric current. Since $A_{\mu}$ does not enter the pairing term,
$J^{\mu}$ is the same as in the normal state where $\Delta=0$, 
\begin{eqnarray}
  J^{t}=-\psi^{\dagger}\psi,\;J^{j}=-\frac{\delta^{jk}}{m^{*}}\frac{i}{2}\psi^{\dagger}\overleftrightarrow{D_{k}}\psi.\label{16}
\end{eqnarray}
Here $\psi^{\dagger}\overleftrightarrow{D_{k}}\psi=\psi^{\dagger}D_{k}\psi-\left(D_{k}\psi^{\dagger}\right)\psi$.
The conservation law \eqref{eq:15} shows that the fermionic charge alone is not conserved
due to the exchange of charge between the fermions $\psi$ and Cooper
pairs $\Delta$. If one adds a ($U\left(1\right)$ gauge invariant)
term $S'\left[\Delta,A\right]$ to the action and  considers $\Delta$
as a dynamical field, then it is possible to use the equation of motion
 $\frac{\delta\left(S'+S\right)}{\delta\Delta}=0$ for $\Delta$ and
the definition $J_{\Delta}^{\mu}=-\frac{\delta S'}{\delta A_{\mu}}$
of the Cooper pair current in order to rewrite \eqref{eq:15} as $\partial_{\mu}\left(J^{\mu}+J_{\Delta}^{\mu}\right)=0$.
This expresses the conservation of total charge in the $p$-wave SC.

\subsubsection{Energy-momentum \label{subsec:Energy-momentum}}

Energy and momentum are at the heart of this paper, and obtaining
the correct expressions for these quantities, as well as interpreting
correctly the conservation laws they satisfy, will be crucial. 

In flat space, one usually starts with the canonical energy-momentum
tensor. For a Lagrangian $\mathcal{L}\left(\phi,\partial\phi,x\right)$,
where $\phi$ is any fermionic of bosonic field, it is given by 
\begin{eqnarray}
 &  & t_{\;\nu}^{\mu}=\frac{\partial\mathcal{L}}{\partial\partial_{\mu}\phi}\partial_{\nu}\phi-\delta_{\nu}^{\mu}\mathcal{L},
\end{eqnarray}
and satisfies, on the equation of motion for $\phi$, 
\begin{eqnarray}
 &  & \partial_{\mu}t_{\;\nu}^{\mu}=-\partial_{\nu}\mathcal{L},\label{eq:18}
\end{eqnarray}
which can be obtained from Noether's first theorem for space-time
translations. Thus $t_{\;\nu}^{\mu}$ is conserved if and only if
the Lagrangian is independent of the coordinate $x^{\nu}$. This motivates
the identification of $t_{\;t}^{\mu}$ as the energy current, and
of $t_{\;j}^{\mu}$ as the current of the $j$th component of momentum
($j$-momentum). $t_{\;t}^{t}$ is just the Hamiltonian density, or
energy density, and $t_{\;j}^{t}$ is the $j$-momentum density. 

It is well known however, that the canonical energy-momentum tensor
may fail to be gauge invariant, symmetric in its indices, or traceless,
in situations where these properties are physically required, and
it is also sensitive to the addition of total derivatives to the Lagrangian.
To obtain the physical energy-momentum tensor one can either ``improve``
$t_{\;\nu}^{\mu}$ or appeal to a geometric (gravitational) definition
which directly provides the physical energy-momentum tensor \cite{ortin2004gravity,forger2004currents}.

We will comment on the coupling of the $p$-wave SF to a real background
geometry our discussion, section \ref{sec:Conclusion-and-discussion},
but here we fix the background geometry to be flat, and instead continue
by introducing the $U\left(1\right)$-covariant canonical energy-momentum
tensor. It can be shown to coincide with the physical energy-momentum
tensor obtained by coupling the $p$-wave SF to a real background
geometry. Since we work with a fixed flat background geometry in this
section, we will only consider space-time transformations which are
symmetries of this background, and it will suffice to consider space-time
translations and spatial rotations. 

The $U\left(1\right)$-covariant canonical energy-momentum tensor
is relevant in the following situation. Assume that the $x$ dependence
in $\mathcal{L}$ is only through a $U\left(1\right)$ gauge field
to which $\phi$ is minimally coupled, $\mathcal{L}\left(\phi,\partial\phi,x\right)=\mathcal{L}\left(\phi,D\phi\right)$.
Then, $t_{\;\nu}^{\mu}$ is not gauge invariant, and therefore physically
ambiguous. This is reflected in the conservation law \eqref{eq:18}
which takes the non covariant form 
\begin{eqnarray}
  \partial_{\mu}t_{\;\nu}^{\mu}=J^{\mu}\partial_{\nu}A_{\mu},\label{19}
\end{eqnarray}
where $J^{\mu}=-\frac{\partial\mathcal{L}}{\partial A_{\mu}}$ is
the $U\left(1\right)$ current. This lack of gauge invariance is to
be expected, as this conservation law follows from translational symmetry,
and translations do not commute with gauge transformations. Instead,
one should use $U\left(1\right)$-covariant space-time translations,
which are translations from $x$ to $x+a$ followed by a $U\left(1\right)$
parallel transport from $x+a$ back to $x$, $\phi\left(x\right)\mapsto e^{iq\int_{x-a}^{x}A}\phi\left(x-a\right)$
where $\phi\mapsto e^{iq\alpha}\phi$ under $U\left(1\right)$ and
the integral is over the straight line from $x-a$ to $a$. This is
still a symmetry because the additional $e^{iq\int_{x-a}^{x}A}$ is
just a gauge transformation. The conservation law that follows from
this modified action of translations is 
\begin{eqnarray}
  \partial_{\mu}t_{\text{cov}\;\nu}^{\mu}=F_{\nu\mu}J^{\mu},\label{eq:21}
\end{eqnarray}
where $F_{\mu\nu}=\partial_{\mu}A_{\nu}-\partial_{\nu}A_{\mu}$ is
the electromagnetic field strength, and 
\begin{eqnarray}
  t_{\text{cov}\;\nu}^{\mu}=\frac{\partial\mathcal{L}}{\partial D_{\mu}\phi}D_{\nu}\phi-\delta_{\nu}^{\mu}\mathcal{L}=t_{\;\nu}^{\mu}-J^{\mu}A_{\nu}\label{eq:32}
\end{eqnarray}
is the $U\left(1\right)$-covariant  version of $t_{\;\nu}^{\mu}$,
which we refer to as the $U\left(1\right)$-covariant canonical energy-momentum
tensor. The right hand side of \eqref{eq:21} is just the usual
Lorentz force, which acts as a source of $U\left(1\right)$-covariant
energy-momentum. We stress that the covariant and non covariant conservation
laws are equivalent, as can be verified by using the fact that $\partial_{\mu}J^{\mu}=0$
in this case. Both hold in any gauge, but in \eqref{eq:21} all quantities
are gauge invariant. 

 For the $p$-wave SF one obtains the $U\left(1\right)$-covariant
energy-momentum tensor 
\begin{eqnarray}
  t_{\text{cov}\;t}^{t}&=&\frac{i}{2}\psi^{\dagger}\overleftrightarrow{D_{t}}\psi-\mathcal{L}\label{25-2}\\
   &=&
  \frac{\delta^{ij}D_{i}\psi^{\dagger}D_{j}\psi}{2m^{*}}+m\psi^{\dagger}\psi-\left(\frac{1}{2}\psi^{\dagger}\Delta^{j}\partial_{j}\psi^{\dagger}+h.c\right),\nonumber\\
  t_{\text{cov}\;j}^{t}&=&\frac{i}{2}\psi^{\dagger}\overleftrightarrow{D_{j}}\psi,\nonumber \\
  t_{\text{cov}\;t}^{i}&=&-\frac{\delta^{ik}\left(D_{k}\psi\right)^{\dagger}D_{t}\psi}{2m^{*}}+\frac{1}{2}\psi^{\dagger}\Delta^{i}\partial_{t}\psi^{\dagger}+h.c,\nonumber \\
  t_{\text{cov}\;j}^{i}&=&-\frac{\delta^{ik}\left(D_{k}\psi\right)^{\dagger}D_{j}\psi}{2m^{*}}+\frac{1}{2}\psi^{\dagger}\Delta^{i}\partial_{j}\psi^{\dagger}+h.c-\delta_{j}^{i}\mathcal{L}.\nonumber 
\end{eqnarray}
 The $U\left(1\right)$-covariant conservation law is slightly
more complicated than \eqref{eq:21} due the additional background
field $\Delta$, 
\begin{align}
  \partial_{\mu}t_{\text{cov}\;\nu}^{\mu}=\frac{1}{2}\psi^{\dagger}\partial_{j}\psi^{\dagger}D_{\nu}\Delta^{j}+h.c+F_{\nu\mu}J^{\mu},\label{32}
\end{align}
where we have used the $U\left(1\right)$ conservation law \eqref{eq:15},
and $D_{\mu}\Delta^{j}=\left(\partial_{\mu}-2iA_{\mu}\right)\Delta^{j}$.
This conservation law shows that ($U\left(1\right)$-covariant) fermionic
energy-momentum is not conserved due to the exchange of energy-momentum
with the background fields $A,\Delta$. Apart from the Lorentz force
there is an additional source term due to the space-time dependence
of $\Delta$. 

 As in the case of the electric charge, if one considers $\Delta$
as a dynamical field and uses its equation of motion, \eqref{32}
can be written as\footnote{$t_{\Delta\;\text{cov}\;\nu}^{\mu}$ is the $U\left(1\right)$-covariant
energy-momentum tensor of Cooper pairs. It is defined by \eqref{eq:32}
with $\phi=\Delta$ and $\mathcal{L}=\mathcal{L}'\left(\Delta,\Delta^{*},D\Delta,D\Delta^{*}\right)$
being the (gauge invariant) term added to the $p$-wave SF Lagrangian.
Here it is important that the coupling of $\Delta$ to $\psi$ in
\eqref{eq:10} can be written without derivatives of $\Delta$. } 
\begin{align}
 &  & \partial_{\mu}\left(t_{\text{cov}\;\nu}^{\mu}+t_{\Delta\;\text{cov}\;\nu}^{\mu}\right)=F_{\nu\mu}\left(J^{\mu}+J_{\Delta}^{\mu}\right),\label{33}
\end{align}
which is of the general form \eqref{eq:21}.

Note that the spatial part $t_{\text{cov}\;j}^{i}$ is not symmetric,
\begin{align}
  t_{\text{cov}\;y}^{x}-t_{\text{cov}\;x}^{y}=\frac{1}{2}\psi^{\dagger}\left(\Delta^{x}\partial_{y}-\Delta^{y}\partial_{x}\right)\psi^{\dagger}+h.c, \label{eq:28}
\end{align}
which physically represents an exchange of angular momentum between
$\Delta$ and $\psi$, possible because of the intrinsic angular momentum
of Cooper pairs in a $p$-wave SC. Explicitly, the ($U\left(1\right)$-covariant)
angular momentum current is given by $J_{\varphi}^{\mu}=t_{\text{cov}\;\varphi}^{\mu}=t_{\text{cov}\;\nu}^{\mu}\zeta^{\nu}$
where $\zeta=x\partial_{y}-y\partial_{x}=\partial_{\varphi}$ is the
generator of spatial rotations around $x=y=0$, and $\varphi$ is
the polar angle. From \eqref{32} and \eqref{eq:28} we find its conservation
law
\begin{align}
  \partial_{\mu}J_{\varphi}^{\mu}=\left(\frac{1}{2}\psi^{\dagger}\partial_{j}\psi^{\dagger}D_{\varphi}\Delta^{j}+h.c+F_{\varphi\mu}J^{\mu}\right)\label{eq:36}\\
  +\frac{1}{2}\psi^{\dagger}\left(\Delta^{x}\partial_{y}-\Delta^{y}\partial_{x}\right)\psi^{\dagger}+h.c,\nonumber 
\end{align}
which shows that even when the Lorentz force in the $\varphi$ direction
vanishes and $\Delta$ is ($U\left(1\right)$-covariantly) constant
in the $\varphi$ direction, $\Delta$ still acts a source for fermionic
angular momentum, due to the last term. 

Even though fermionic angular momentum is never strictly conserved
in a $p$-wave SF, it is well known that a certain combination of
fermionic charge and fermionic angular momentum can be strictly conserved
\cite{shitade2014bulk,tada2015orbital,volovik2015orbital,shitade2015orbital}.
Indeed, using \eqref{eq:36} and \eqref{eq:15},
\begin{align}
  &\partial_{\mu}\left(J_{\varphi}^{\mu}\mp\frac{1}{2}J^{\mu}\right)=\left(\frac{1}{2}\psi^{\dagger}\partial_{j}\psi^{\dagger}D_{\varphi}\Delta^{j}+h.c+F_{\varphi\mu}J^{\mu}\right)\nonumber \\
 & \pm\frac{i}{2}\psi^{\dagger}\left(\Delta^{x}\pm i\Delta^{y}\right)\left(\partial_{x}\mp i\partial_{y}\right)\psi^{\dagger}+h.c.
\end{align}
We see that when $F_{\varphi\mu}=0$, $D_{\varphi}\Delta=0$ and
$\Delta^{y}=\pm i\Delta^{x}$, the above current is strictly conserved
\begin{eqnarray}
 &  & \partial_{\mu}\left(J_{\varphi}^{\mu}\mp\frac{1}{2}J^{\mu}\right)=0,
\end{eqnarray}
which occurs in the generalized $p_{x}\pm ip_{y}$ configuration $\Delta=e^{i\theta\left(r,t\right)}\Delta_{0}\left(r,t\right)\left(1,\pm i\right)$,
written in the gauge $A_{\varphi}=0$, and where $r=\sqrt{x^{2}+y^{2}}$.
This conservation law follows from the symmetry of the generalized
$p_{x}\pm ip_{y}$ configuration under the combination of a spatial
rotation by an angle $\alpha$ and a $U\left(1\right)$ transformation
by a phase $\mp\alpha/2$.

\subsection{\label{subsec:Currents,-symmetries,-and}Symmetries, currents, and
conservation laws in the geometric description}

The symmetries and conservation laws for Dirac fermions have been
described recently in \cite{hughes2013torsional}. Here we review
the essential details (for Majorana fermions) and focus on the mapping
to the symmetries and conservation laws of the $p$-wave SF action
\eqref{eq:14}, which were described in section \ref{subsec:Symmetries,-currents,-and}. 

\subsubsection{Currents in the geometric description \label{currents}}

The natural currents in the geometric description are defined by the
functional derivatives of the action $S_{\text{RC}}$ with respect
to the background fields $e,\omega$,
\begin{eqnarray}
  \mathsf{J}_{\;a}^{\mu}=\frac{1}{\left|e\right|}\frac{\delta S_{\text{RC}}}{\delta e_{\;\mu}^{a}},\;\mathsf{J}^{ab\mu}=\frac{1}{\left|e\right|}\frac{\delta S_{\text{RC}}}{\delta\omega_{ab\mu}}.\label{eq:54}
\end{eqnarray}
$\mathsf{J}_{\;a}^{\mu}$ is the energy momentum (energy-momentum)
tensor and $\mathsf{J}^{ab\mu}$ is the spin current. 
Note that we use $\mathsf{J}$ as opposed to $J$ to distinguish the
geometric currents from the $p$-wave SF currents described in the
previous section, though the two are related as shown below. 

Calculating the geometric currents for the action \eqref{eq:43-1}
one obtains 
\begin{eqnarray}
  2\mathsf{J}_{\;a}^{\mu}&=&\mathcal{L}_{\text{RC}}e_{a}^{\;\mu}-\frac{i}{2}\overline{\chi}\left(\gamma^{\mu}D_{a}-\overleftarrow{D_{a}}\gamma^{\mu}\right)\chi,\label{eq:55}\\
  2\mathsf{J}^{ab\mu}&=&-\frac{1}{4}\overline{\chi}\chi e_{c}^{\;\mu}\varepsilon^{abc},\nonumber 
\end{eqnarray}
where $\mathcal{L}_{\text{RC}}=\overline{\chi}\left[\frac{i}{2}e_{a}^{\;\mu}\left(\gamma^{a}D_{\mu}-\overleftarrow{D_{\mu}}\gamma^{a}\right)-m\right]\chi$
is (twice) the Lagrangian, which vanishes on the $\chi$ equation
of motion. We see that $\mathsf{J}_{\;a}^{\mu}$ is essentially the
$SO\left(1,2\right)$-covariant version of the canonical energy-momentum
tensor of the spinor $\chi$. We also see that the spin current $\mathsf{J}^{ab\mu}$
has a particularly simple form in $D=2+1$, it is just the spin density
$\frac{1}{2}\overline{\chi}\chi$ times a tensor $-\frac{1}{2}e_{c}^{\;\mu}\varepsilon^{abc}$
that only depends on the background field $e$. Using the expressions
\eqref{17} for the geometric  fields we find that $\mathsf{J}_{\;a}^{\mu},\;\mathsf{J}^{ab\mu}$
are related simply to the electric current and the ($U\left(1\right)$-covariant)
canonical energy-momentum tensor described in section \ref{subsec:Symmetries,-currents,-and},
in the limit $m^{*}\rightarrow\infty$,
\begin{eqnarray}
 &  & J^{\mu}=4\left|e\right|\mathsf{J}^{12\mu}=-\psi^{\dagger}\psi\delta_{t}^{\mu},\label{eq:56}\\
 &  & t_{\text{cov}\;\nu}^{\mu}=-\left|e\right|\mathsf{J}_{\;\nu}^{\mu}=\begin{cases}
\frac{i}{2}\psi^{\dagger}\overleftrightarrow{D_{\nu}}\psi & \mu=t\\
\frac{1}{2}\psi^{\dagger}\Delta^{j}\partial_{\nu}\psi^{\dagger}+h.c & \mu=j
\end{cases}.\nonumber 
\end{eqnarray}
Here we have simplified $t_{\text{cov}}$ using the equation of motion
for $\psi$, and one can also use the equation of motion to remove
time derivatives and obtain Schrodinger picture operators. For example,
$t_{\text{cov}\;t}^{t}=\frac{i}{2}\psi^{\dagger}\overleftrightarrow{D_{t}}\psi=m\psi^{\dagger}\psi-\left(\frac{1}{2}\psi^{\dagger}\Delta^{j}\partial_{j}\psi^{\dagger}+h.c\right)$
is just the ($U\left(1\right)$-covariant) Hamiltonian density in
the relativistic limit. The expression for the energy current $t_{\text{cov}\;t}^{i}$
is more complicated, and it is convenient to write it using some of
the geometric quantities introduced above
\begin{align}
  t_{\text{cov }t}^{j}=g^{jk}\frac{i}{2}\psi^{\dagger}\overleftrightarrow{D_{k}}\psi-&\frac{o}{2}\partial_{k}\left(\frac{1}{\left|e\right|}\varepsilon^{jk}\psi^{\dagger}\psi\right)\label{eq:49}\\
  -&\left(\psi^{\dagger}\psi\right)g^{jk}C_{12k}.\nonumber
\end{align}
This is an expression for the energy current in terms of the momentum
and charge densities, and it will be obtained below as a consequence
of Lorentz symmetry in the relativistic limit. We now describe the
symmetries of the action \eqref{eq:43-1} and the conservation laws
they imply for these currents. As expected, these conservation laws
turn out to be essentially the ones derived in section \eqref{subsec:Symmetries,-currents,-and},
in the relativistic limit. 

\subsubsection{Spin \label{spin}}

The Lorentz Lie algebra $so\left(1,2\right)$ is comprised of matrices
$\theta\in\mathbb{R}^{3\times3}$ with entries $\theta_{\;b}^{a}$
such that $\theta_{ab}=-\theta_{ba}$. These can be spanned as $\theta=\frac{1}{2}\theta_{ab}L^{ab}$
where the generators $L^{ab}=-L^{ba}$ are defined such that $\eta L^{ab}$
is the antisymmetric matrix with $1$ ($-1$) at position $a,b$ ($b,a$)
and zero elsewhere. The spinor representation of $\theta$ is 
\begin{eqnarray}
  \hat{\theta}=\frac{1}{2}\theta_{ab}\Sigma^{ab},\;\Sigma^{ab}=\frac{1}{4}\left[\gamma^{a},\gamma^{b}\right].
\end{eqnarray}
Local Lorentz transformations act on $\chi,e,\omega$ by 
\begin{eqnarray}
  \chi&\mapsto& e^{-\hat{\theta}}\chi,\;e_{a}^{\;\mu}\mapsto e_{b}^{\;\mu}\left(e^{\theta}\right)_{\;a}^{b},\nonumber\\
 \omega_{\mu}&\mapsto& e^{-\hat{\theta}}\left(\partial_{\mu}+\omega_{\mu}\right)e^{\hat{\theta}}.\label{eq:20}
\end{eqnarray}
The subgroup of $SO\left(1,2\right)$ that is physical in the $p$-wave
SC is $SO\left(2\right)$ generated by $L^{12}$. Using the relations
\eqref{17} between the $p$-wave SC fields and the geometric fields,
and choosing $\theta=\theta_{12}L^{12}=-2\alpha L^{12}$, the transformation
law \eqref{eq:20} reduces to the $U\left(1\right)$ transformation \eqref{eq:6.1},
\begin{align}
  \psi\mapsto e^{i\alpha}\psi,\;\Delta\mapsto e^{2i\alpha}\Delta,\;A_{\mu}\mapsto A_{\mu}+\partial_{\mu}\alpha.
\end{align} 
The factor of 2 in $\theta_{12}=-2\alpha$ shows that $U(1)$ actually maps to $Spin(2)$, the double cover of $SO(2)$. Moreover, the fact that $\Delta$ has $U\left(1\right)$ charge
2 while $\psi$ has $U\left(1\right)$ charge 1 corresponds to $e_{a}^{\;\mu}$
being an $SO\left(1,2\right)$ vector while $\chi$ is an $SO\left(1,2\right)$
spinor. 
The Lie algebra version of \eqref{eq:20} is 
\begin{align}
  \delta\chi=-\frac{1}{2}\theta_{ab}\Sigma^{ab}\chi,\;\delta e_{\;\mu}^{a}=-\theta_{\;b}^{a}e_{\;\mu}^{b},\;\delta\omega_{\;b\mu}^{a}=D_{\mu}\theta_{\;b}^{a}.
\end{align}
Invariance of $S_{\text{RC}}$ under this variation implies the conservation law 
\begin{eqnarray}
  \nabla_{\mu}\mathsf{J}^{ab\mu}-\mathsf{J}^{ab\rho}T_{\mu\rho}^{\mu}=\mathsf{J}^{[ab]},\label{eq:57}
\end{eqnarray}
valid on the equations of motion for $\chi$ \cite{hughes2013torsional,bradlyn2015low}. This conservation
law relates the anti symmetric part of the energy-momentum tensor
to the divergence of spin current. Essentially, the energy-momentum
tensor isn't symmetric due to the presence of the background field
$\omega$ which transforms under $SO\left(1,2\right)$. From a different
point of view, the vielbein $e$ acts as a source for the fermionic
spin current since it is charged under $SO\left(1,2\right)$. 
Inserting the expressions \eqref{17} into the $\left(a,b\right)=\left(1,2\right)$ component of
\eqref{eq:57} we obtain \eqref{eq:15}, 
\begin{eqnarray}
 &  & \partial_{\mu}J^{\mu}=-i\psi^{\dagger}\Delta^{j}\partial_{j}\psi^{\dagger}+h.c.\label{eq:15-1}
\end{eqnarray}
The other components of \eqref{eq:57} follow from the symmetry under
local boosts, which is only a symmetry of $S_{\text{SF}}$
when $m^{*}\rightarrow\infty$. These can be used to obtain the formula \eqref{eq:49} for the energy current of the $p$-wave
SF, in the limit $m^{*}\rightarrow\infty$, in terms of the momentum
and charge densities.

\subsubsection{Energy-momentum\label{subsec:Diffeomorphism-symmetry}}

A diffeomorphism is a smooth invertible map between manifolds. We
consider only diffeomorphisms from space-time to itself and denote
the group of such maps by $Diff$. Since the flat background metric
$\delta^{ij}$ decouples in the relativistic limit, it makes sense
to consider all diffeomorphisms, and not restrict to symmetries of
$\delta^{ij}$ as we did in section \ref{subsec:Energy-momentum}. 

Locally, diffeomorphisms can be described by coordinate transformations
$x\mapsto x'=f\left(x\right)$. The lie algebra is that of vector
fields $\zeta^{\nu}\left(x\right)$, which means diffeomorphisms in
the connected component of the identity $Diff_{0}$ can be written
as $f\left(x\right)=f_{1}\left(x\right)$ where $f_{\varepsilon}\left(x\right)=\exp_{x}\left(\varepsilon\zeta\right)=x+\varepsilon\zeta\left(x\right)+O\left(\varepsilon^{2}\right)$
is the flow of $\zeta$ \cite{nakahara2003geometry}. $Diff$ acts on
the geometric fields by the pullback
\begin{eqnarray}
  \chi\left(x\right)&\mapsto&\chi\left(f\left(x\right)\right),\;e_{\;\mu}^{a}\left(x\right)\mapsto\partial_{\mu}f^{\nu}e_{\;\nu}^{a}\left(f\left(x\right)\right),\nonumber\\
  \omega_{\mu}\left(x\right)&\mapsto&\partial_{\mu}f^{\nu}\omega_{\nu}\left(f\left(x\right)\right).\label{eq:51}
\end{eqnarray}
The action of $Diff$ on the $p$-wave SF fields is similar, and follows from \eqref{eq:51}
supplemented by the dictionary \eqref{17}. For $f\in Diff_{0}$ generated
by $\zeta$, the Lie algebra version of \eqref{eq:51} is given by the Lie derivative, 
\begin{eqnarray}
 \delta\chi&=&\mathcal{L}_{\zeta}\chi=\zeta^{\mu}\partial_{\mu}\chi,\label{eq:53}\\
  \delta e_{\;\mu}^{a}&=&\mathcal{L}_{\zeta}e_{\;\mu}^{a}=\partial_{\mu}\zeta^{\nu}e_{\;\nu}^{a}+\zeta^{\nu}\partial_{\nu}e_{\;\mu}^{a},\nonumber \\
  \delta\omega_{\mu}&=&\mathcal{L}_{\zeta}\omega_{\mu}=\partial_{\mu}\zeta^{\nu}\omega_{\nu}+\zeta^{\nu}\partial_{\nu}\omega_{\mu}.\nonumber 
\end{eqnarray}
Since these variations are not Lorentz covariant, they will give rise
to a conservation law which is not Lorentz covariant.  This follows
from the fact that the naive $Diff$ action \eqref{eq:51} does not
commute with Lorentz gauge transformations, as was described for the
simpler case of translations and $U\left(1\right)$ gauge transformations
in section \ref{subsec:Energy-momentum}. Instead, one should use
the Lorentz-covariant $Diff$ action, which is the pull back from
$f\left(x\right)$ to $x$ followed by a Lorentz parallel transport
from $f\left(x\right)$ to $x$ along the integral curve $\gamma_{x,\zeta}\left(\varepsilon\right)=\exp_{x}\left(\varepsilon\zeta\right)=f_{\varepsilon}\left(x\right)$,
\begin{eqnarray}
 \chi\left(x\right)&\mapsto& P\chi\left(f\left(x\right)\right),\label{eq:62}\\ e_{\;\mu}^{a}\left(x\right)&\mapsto& P_{\;b}^{a}\partial_{\mu}f^{\nu}e_{\;\nu}^{b}\left(f\left(x\right)\right),\nonumber\\
 \omega_{\mu}\left(x\right)&\mapsto& P\left[\partial_{\mu}f^{\nu}\omega_{\mu}\left(f\left(x\right)\right)+\partial_{\mu}\right]P^{-1},\nonumber
\end{eqnarray}
 where $P=\frac{1}{2}P_{ab}\Sigma^{ab}$ and $P=\mathcal{P}\exp\left(-\int_{\gamma_{x,\zeta}}\omega\right)$
is the spin parallel transport given by the path ordered exponential.
At the Lie algebra level, this modification of \eqref{eq:51} amounts
to an infinitesimal Lorentz gauge transformation generated by $\theta_{ab}=-\zeta^{\rho}\omega_{ab\rho}$,
which modifies \eqref{eq:53} to the covariant expressions 
\begin{eqnarray}
 \delta\chi&=&\zeta^{\mu}\nabla_{\mu}\chi,\label{eq:53-1}\\
  \delta e_{\;\mu}^{a}&=&\nabla_{\mu}\zeta^{a}-T_{\mu\nu}^{a}\zeta^{\nu},\nonumber \\
  \delta\omega_{\mu}&=&\zeta^{\nu}R_{ab\nu\mu}.\nonumber 
\end{eqnarray}
Since the usual $Diff$ and Lorentz actions on the fields are both
symmetries of $S_{\text{RC}}$, so is the Lorenz-covariant $Diff$
action. This leads directly to the conservation law 
\begin{eqnarray}
 &  & \nabla_{\mu}\mathsf{J}_{\;\nu}^{\mu}-\mathsf{J}_{\;\nu}^{\rho}T_{\mu\rho}^{\mu}=T_{\nu\mu}^{b}\mathsf{J}_{\;b}^{\mu}+R_{bc\nu\mu}\mathsf{J}^{bc\mu},\label{eq:71}
\end{eqnarray}
valid on the equations of motion for $\chi$ \cite{bradlyn2015low,hughes2013torsional}. We find it useful
to rewrite \eqref{eq:71} in a way which isolates the effect
of torsion, 
\begin{eqnarray}
 &  & \tilde{\nabla}_{\mu}\mathsf{J}_{\;\nu}^{\mu}=C_{ab\nu}\mathsf{J}^{[ab]}+R_{ab\nu\mu}\mathsf{J}^{ab\mu},\label{eq:60}
\end{eqnarray}
where we note that the curvature also depends on the torsion, $R=\tilde{R}+\tilde{D}C+C\wedge C$.
Equation \eqref{eq:71} can also be massaged to the non-covariant
form 
\begin{align}
  \partial_{\mu}\left(\left|e\right|\mathsf{J}_{\;\nu}^{\mu}\right)=\left(e_{a}^{\;\rho}D_{\nu}e_{\;\mu}^{a}\right)\left|e\right|\mathsf{J}_{\;\rho}^{\mu}+R_{\nu\mu ab}\left|e\right|\mathsf{J}^{ab\mu}.\label{eq:66}
\end{align}
Using the dictionary \eqref{17} and the subsequent paragraph, and
\eqref{eq:56}, this reduces to 
\begin{align}
  \partial_{\mu}t_{\text{cov}\;\nu}^{\mu}=\left(D_{\nu}\Delta^{j}\right)\frac{1}{2}\psi^{\dagger}\partial_{j}\psi^{\dagger}+h.c+F_{\nu\mu}J^{\mu},
\end{align}
which is just the energy-momentum conservation law \eqref{32} for
the $p$-wave SF (with $m^{*}\rightarrow\infty$). 

Writing the conservation law in the form \eqref{eq:66} may not seem
natural from the geometric point of view because it uses the partial
derivative as opposed to a covariant derivative. It is however natural
from the $p$-wave SC point of view, where space-time is actually
flat and $e$ is viewed as a bosonic field with no geometric role,
which is the order parameter $\Delta$. This point will be important
when we discuss the gravitational anomaly in the $p$-wave SC, in
section \ref{subsec:Boundary-gravitational-anomaly}. \textcolor{red}{} 

Similar statements hold for other mechanisms for emergent/analogue
gravity, see section I.6 of \cite{volovik2009universe} and \cite{keser2016analogue},
and were also made in the gravitational context without reference
to emergent phenomena \cite{leclerc2006canonical}.

\section{Bulk response\label{sec:Bulk-response} }

\subsection{Currents from effective action\label{subsec:Bulk-response-from}}

The effective action for the background fields is obtained by integrating
over the spin-less fermion $\psi$,
\begin{eqnarray}
  e^{iW_{\text{SF}}\left[\Delta,A\right]}=\int\text{D}\psi^{\dagger}\text{D}\psi e^{iS_{\text{SF}}\left[\psi,\psi^{\dagger},\Delta,A\right]}.
\end{eqnarray}

The integral is a fermionic coherent state functional integral, over
the Grassmann valued fields $\psi,\psi^{\dagger}$, and the action
$S_{\text{SF}}$ is given in \eqref{eq:10}. 

As described in section \ref{sec:Emergent-Riemann-Cartan-geometry},
in the relativistic limit $W_{\text{SF}}$ is equal to the effective
action obtained by integrating over a Majorana fermion coupled to
RC geometry, 
\begin{eqnarray}
  e^{iW_{\text{SF}}\left[\Delta,A\right]}&=&e^{iW_{\text{RC}}\left[e,\omega\right]}\label{eq:73-1}\\
  &=&\int\text{D}\left(\left|e\right|^{1/2}\chi\right)e^{iS_{\text{RC}}\left[\chi,e,\omega\right]},\nonumber
\end{eqnarray}
where $e,\omega$ are given in terms of $\Delta,A$ by \eqref{17}. 

It follows from the definition \eqref{eq:54} of the spin current
$\mathsf{J}^{ab\mu}$ and the energy-momentum tensor $\mathsf{J}_{\;a}^{\mu}$
as functional derivatives of $S_{\text{RC}}$ that their ground state
expectation values are given by 
\begin{eqnarray}
 &  & \left\langle \mathsf{J}_{\;a}^{\mu}\right\rangle =\frac{1}{\left|e\right|}\frac{\delta W_{\text{RC}}}{\delta e_{\;\mu}^{a}},\;\left\langle \mathsf{J}^{ab\mu}\right\rangle =\frac{1}{\left|e\right|}\frac{\delta W_{\text{RC}}}{\delta\omega_{ab\mu}}.\label{eq:54-1}
\end{eqnarray}
Using the mapping \eqref{eq:56} between $\mathsf{J}_{\;a}^{\mu},\;\mathsf{J}^{ab\mu}$
and $t_{\text{cov}\;\nu}^{\mu},\;J^{\mu}$ 
we see that 
\begin{eqnarray}
 &  & \left\langle J^{\mu}\right\rangle =4\left|e\right|\left\langle \mathsf{J}^{12\mu}\right\rangle =4\frac{\delta W_{\text{RC}}\left[e,\omega\right]}{\delta\omega_{12\mu}},\label{eq:71-1}\\
 &  & \left\langle t_{\text{cov}\;\nu}^{\mu}\right\rangle =-\left|e\right|e_{\;\nu}^{a}\left\langle \mathsf{J}_{\;a}^{\mu}\right\rangle =-e_{\;\nu}^{a}\frac{\delta W_{\text{RC}}\left[e,\omega\right]}{\delta e_{\;\mu}^{a}}.\nonumber 
\end{eqnarray}
This is the recipe we will use to obtain the expectation values $\left\langle J^{\mu}\right\rangle ,\left\langle t_{\text{cov}\;\nu}^{\mu}\right\rangle $
from the effective action $W_{\text{RC}}$
for a Majorana spinor in RC space-time. 

Note that in \eqref{eq:71-1} there are derivatives with respect to
all components of the vielbein, not just the spatial ones which we
can physically obtain from $\Delta$. For this reason, to get all
components of $\left\langle t_{\text{cov}\;\nu}^{\mu}\right\rangle $,
we should obtain $W_{\text{RC}}$ for general $e$, take the functional
derivative in \eqref{eq:71-1}, and only then set $e$ to the configuration
obtained from $\Delta$ according to \eqref{17}. From the $p$-wave
SF point of view, this corresponds to the introduction of a fictitious
background field $e_{0}^{\;\mu}$ which enters $S_{\text{SF}}$ by
generalizing $\psi^{\dagger}iD_{t}\psi$ to $\psi^{\dagger}\frac{i}{2}e_{0}^{\;\mu}\overleftrightarrow{D_{\mu}}\psi$,
and setting $e_{0}^{\;\mu}=\delta_{t}^{\mu}$ at the end of the calculation, as in
\cite{bradlyn2015low}. 

Before we move on, we offer some intuition for the expressions \eqref{eq:71-1}.
The first equation in \eqref{eq:71-1} follows from the definition
$J^{\mu}=-\frac{\delta S_{\text{SF}}}{\delta A_{\mu}}$ of the electric
current and the simple relation $\omega_{12\mu}=-\omega_{21\mu}=-2A_{\mu}$
between the spin connection and the $U\left(1\right)$ connection.
The second equation in \eqref{eq:71-1} is slightly trickier. It implies
that the (relativistic part of the) energy-momentum tensor $t_{\text{cov}\;\nu}^{\mu}$
is given by a functional derivative with respect to the order parameter
$\Delta$, because $\Delta$ is essentially the vielbein $e$. This
may seem strange, and it is certainly not the case in an $s$-wave
SC, where $\frac{\delta H}{\delta\Delta}\sim\psi_{\uparrow}^{\dagger}\psi_{\downarrow}^{\dagger}$
has nothing to do with energy-momentum. In a $p$-wave SC, the operator
$\frac{\delta H}{\delta\Delta^{j}}\sim\psi^{\dagger}\partial_{j}\psi^{\dagger}$
contains a spatial derivative which hints that it is related to fermionic
momentum. More accurately, we see from \eqref{25-2} that the operator
$\psi^{\dagger}\partial_{j}\psi^{\dagger}$ enters the energy-momentum
tensor in a $p$-wave SC. 

\subsection{Effective action from perturbation theory }

\subsubsection{Setup and generalities}

We consider the effective action for a $p$-wave SF on the plane $\mathbb{R}^{2}$,
with the corresponding space-time manifold $M_{3}=\mathbb{R}_{t}\times\mathbb{R}^{2}$,
by using perturbation theory around the $p_{x}\pm ip_{y}$ configuration
$\Delta=\Delta_{0}e^{i\theta}\left(1,\pm i\right)$ with no electromagnetic
fields $\partial_{\mu}\theta-2A_{\mu}=0$. After $U\left(1\right)$
gauge fixing $\theta=0$\footnote{In doing so we are ignoring the possibility of vortices, see \cite{ariad2015effective}.},
we obtain $\Delta=\Delta_{0}\left(1,\pm i\right),\;A=0$. Let us start
with the $p_{x}+ip_{y}$ configuration, which has a positive orientation,
in which case the corresponding (gauge fixed) vielbein and spin connection
are just $e_{a}^{\;\mu}=\delta_{a}^{\mu}$ and $\omega_{ab\mu}=0$.
A perturbation of the $p_{x}+ip_{y}$ configuration corresponds to
$e_{a}^{\;\mu}=\delta_{a}^{\mu}+h_{a}^{\;\mu}$ with a small $h$
and to a small spin connection $\omega_{ab\mu}$. In other words,
a perturbation of the $p_{x}+ip_{y}$ configuration without electromagnetic
fields corresponds to a perturbation of flat and torsion-less space-time. 

The effective action for a Dirac spinor in a background RC geometry
was recently calculated perturbatively around flat and torsionless
space-time, with a positive orientation, in the context of geometric
responses of Chern insulators \cite{hughes2013torsional,parrikar2014torsion}.
This is equal to $2W_{\text{RC}}$ where $W_{\text{RC}}$ is the effective
action for a Majorana spinor in RC geometry. 

At this point is seems that we can apply these results in order to
obtain the effective action for the $p$-wave SC, in the relativistic
limit. There is however, an additional ingredient in the perturbative
calculation of the effective action which we did not yet discuss,
which is the renormalization scheme  used to handle diverging integrals.
We refer to terms in the effective action that involve diverging integrals
as \textit{UV sensitive}. The values one obtains for such terms depend
on the details of the renormalization scheme, or in other words, on
microscopic details that are not included in the continuum action.

For us, the continuum description is simply an approximation to the
lattice model, where space is a lattice but time is continuous. This
implies a physical cutoff $\Lambda_{UV}$ for wave-vectors, but not
for frequencies. In particular, such a scheme is not Lorentz invariant,
even though the action in the relativistic limit is. Lorentz symmetry
is in any case broken down to spatial $SO\left(2\right)$ for finite
$m^{*}$. For these reasons, UV sensitive terms in the effective action
$W_{\text{RC}}$ for the $p$-wave SC will be assigned different values
than those obtained before, using a fully relativistic scheme. 

The perturbative calculation within the renormalization scheme outlined
above is described in appendix \ref{subsec:Perturbative-calculation-of},
where we also demonstrate that it produces physical quantities that
approximate those of the lattice model, and compare to the fully relativistic
schemes used in previous works. In the following we will focus on
the UV \textit{insensitive} part of the effective action, and in doing
so we will obtain results which are essentially\footnote{See the discussion of $O\left(\frac{m}{\Lambda_{UV}}\right)$ corrections
below.} independent of microscopic details that do not appear in the continuum
action. We start by quoting the fully relativistic results of \cite{hughes2013torsional,parrikar2014torsion},
and then restrict our attention to the UV insensitive part of the
effective action, and describe the physics of the $p$-wave SC it
encodes.

\subsubsection{Effective action for a single Majorana spinor \label{subsec:Effective-action-for}}

The results of \cite{hughes2013torsional,parrikar2014torsion} can
be written as 
\begin{eqnarray}
  2W_{\text{RC}}\left[e,\omega\right]&=&\frac{\kappa_{H}}{2}\int_{M_{3}}Q_{3}\left(\tilde{\omega}\right)\label{eq:72}\\
 &+&\frac{\zeta_{H}}{2}\int_{M_{3}}e^{a}De_{a}-\frac{\kappa_{H}}{2}\int_{M_{3}}\tilde{\mathcal{R}}e^{a}De_{a}\nonumber\\
  &+&\frac{1}{2\kappa_{N}}\int_{M_{3}}\left(\tilde{\mathcal{R}}-2\Lambda+\frac{3}{2}c^{2}\right)\left|e\right|\mbox{d}^{3}x+\cdots\nonumber 
\end{eqnarray}
where 
\begin{eqnarray}
Q_{3}\left(\tilde{\omega}\right)=\text{tr}\left(\tilde{\omega}\text{d}\tilde{\omega}+\frac{2}{3}\tilde{\omega}^{3}\right)
\end{eqnarray}
is the Chern-Simons (local) 3-form,  $c=C_{abc}\varepsilon^{abc}$ is the totally antisymmetric
piece of the contorsion tensor, and $\kappa_{H},\zeta_{H},1/\kappa_{N},\Lambda/\kappa_{N}$
are coefficients that will be discussed further below. The first two
lines of \eqref{eq:72} are written in terms of differential forms, and the third line is written in terms of scalars. By scalars we mean $Diff$ invariant
objects.  In the differential forms the wedge product
is implicit, as it will be from now on, so $\tilde{\omega}\wedge\text{d}\tilde{\omega}$
is written as $\tilde{\omega}\text{d}\tilde{\omega}$ and so on. The
integrals over differential forms can be written as integrals over
pseudo-scalars,
\begin{eqnarray}
 &  & e^{a}De_{a}=\left(e_{\;\alpha}^{a}D_{\beta}e_{\;\gamma}^{b}\frac{1}{\left|e\right|}\varepsilon^{\alpha\beta\gamma}\right)\left|e\right|\text{d}^{3}x=-oc\left|e\right|\text{d}^{3}x,\label{eq:76}\\
 &  & Q_{3}\left(\tilde{\omega}\right)=\left(\tilde{\omega}_{\;b\alpha}^{a}\partial_{\beta}\tilde{\omega}_{\;a\gamma}^{b}+\frac{2}{3}\tilde{\omega}_{\;b\gamma}^{a}\tilde{\omega}_{\;c\beta}^{b}\tilde{\omega}_{\;a\gamma}^{c}\right)\frac{1}{\left|e\right|}\varepsilon^{\alpha\beta\gamma}\left|e\right|\text{d}^{3}x,\nonumber 
\end{eqnarray}
which are only invariant under the orientation preserving
subgroup of $Diff$ which we denote $Diff_{+}$. Here $o=\text{sgn}\left(\text{det}\left(e\right)\right)$ is the
orientation of $e$. These expressions are odd under orientation reversing
diffeomorphisms because so are $o$ and the pseudo-tensor
$\frac{1}{\left|e\right|}\varepsilon^{\alpha\beta\gamma}$ \footnote{In this paper $\varepsilon$ always stands for the usual totally anti
symmetric symbol, normalized to 1. Thus $\varepsilon^{123}=\varepsilon^{xyt}=\varepsilon_{xyt}=1$.
Note that $\varepsilon^{abc}$ is an $SO\left(1,2\right)$ tensor,
and an $O\left(1,2\right)$ pseudo-tensor, while $\varepsilon^{\mu\nu\rho}=\text{det}\left(e\right)e_{a}^{\;\mu}e_{b}^{\;\nu}e_{c}^{\;\rho}\varepsilon^{abc}$
is a (coordinate) tensor density, $\frac{1}{\text{det}\left(e\right)}\varepsilon^{\mu\nu\rho}$
is a tensor and $\frac{1}{\left|e\right|}\varepsilon^{\mu\nu\rho}=\frac{1}{\left|\text{det}\left(e\right)\right|}\varepsilon^{\mu\nu\rho}$
is a pseudo-tensor. }.

Equation \eqref{eq:72} can be expanded in the perturbations $h_{a}^{\;\mu}$
and $\omega_{ab\mu}$ to reveal the order in perturbation theory at
which the different terms arise, see appendix \ref{subsec:Perturbative-calculation-of}. Additionally, at every order in the perturbations
the effective action can be expanded in powers of derivatives of the
perturbations over the mass $m$. The terms written explicitly above
show up at first and second order in $h,\omega$ and at up to third order
in their derivatives. They also include higher order corrections that
make them $Diff_{+}$ and Lorentz gauge invariant, or invariant up
to total derivatives. 

All other contributions denoted by $+\cdots$ are at least third order
in the perturbations or fourth order in derivatives. Such
a splitting is not unique \cite{hughes2013torsional}, but the form
\eqref{eq:72} has been chosen because it is well suited for the study
of the bulk responses. 

Let us now describe the different terms in \eqref{eq:72}. The first
term is the gravitational Chern-Simons (gCS) term. It has a similar structure to the more familiar $U\left(1\right)$
CS term $\int A\text{d}A$, and is in fact an $SO\left(1,2\right)$
CS term, but note that the LC spin connection $\tilde{\omega}$ is a functional
of the vielbein $e$. It is important that the spin connection in
gCS is not $\omega$, since through $\omega_{\mu}=-2A_{\mu}\Sigma^{12}$
this would imply a quantized Hall conductivity in a $p$-wave SC,
which does not exist \cite{read2000paired,stone2004edge}. As it is
written in \eqref{eq:72}, gCS is invariant under $Diff_{+}$, but
not under $SO\left(1,2\right)$ if $M_{3}$ has a boundary. This is
the boundary $SO\left(1,2\right)$ anomaly, which is discussed further
in section \ref{subsec:Gauge-symmetry-of}. Using the relation $\tilde{\Gamma}_{\;\beta\mu}^{\alpha}=e_{a}^{\;\alpha}\left(\delta_{b}^{a}\partial_{\mu}+\tilde{\omega}_{\;b\mu}^{a}\right)e_{\;\beta}^{b}$
between $\tilde{\Gamma}$ and $\tilde{\omega}$, one can derive an
important formula,
\begin{align}
 Q_{3}\left(\tilde{\Gamma}\right)-Q_{3}\left(\tilde{\omega}\right)=\text{tr}\left[\frac{1}{3}\left(e\text{d}e^{-1}\right)^{3}+\text{d}\left(\text{d}e^{-1}e\tilde{\Gamma}\right)\right],\nonumber\\\label{eq:70}
\end{align}
where unusually, $e=\left(e_{\;\mu}^{a}\right)$ is treated in this
expression as a matrix valued function \cite{kraus2006holographic,stone2012gravitational}.
The variation with respect to $e$ of the two terms on the right hand
side is a total derivative, which means that they are irrelevant for
the purpose of calculating bulk responses. One can therefore use $Q_{3}\left(\tilde{\Gamma}\right)$,
which only depends on the metric $g_{\mu\nu}$, instead of $Q_{3}\left(\tilde{\omega}\right)$.
The form $\int_{M_{3}}Q_{3}\left(\tilde{\Gamma}\right)$ of gCS is
invariant under $SO\left(1,2\right)$ but not under $Diff_{+}$, as
opposed to $\int_{M_{3}}Q_{3}\left(\tilde{\omega}\right)$. Thus the
right hand side of \eqref{eq:70} has the effect of shifting the boundary
anomaly from $SO\left(1,2\right)$ to $Diff$. 

The second term in \eqref{eq:72} has a structure similar to a CS
term with $e^{a}$ playing the role of a connection, and indeed some
authors refer to it as such \cite{zanelli2012chern}. Nevertheless,
it is $SO\left(1,2\right)$ and $Diff_{+}$ invariant, as can be seen
from \eqref{eq:76}. This term was related to the torsional Hall
viscosity in \cite{hughes2013torsional}, where it was discussed
extensively. The third term in \eqref{eq:72} is also $SO\left(1,2\right)$
and $Diff_{+}$ invariant. We refer to this term as \textit{gravitational
pseudo Chern-Simons} (gpCS), to indicate its similarity to gCS, and
the fact that it is not a Chern-Simons term. The similarity between
gCS and gpCS is demonstrated and put in a broader context in the discussion,
section \ref{sec:Conclusion-and-discussion}. In section \ref{subsec:Calculation-of-bulk} we will see that gCS and gpCS produce similar contributions to bulk responses. For now, we simply note that both terms are second order in $h$ and third order in derivatives
of $h$. 

The third line in \eqref{eq:72} contains the Einstein-Hilbert action
with a cosmological constant $\Lambda$ familiar from general relativity,
and an additional torsional contribution $\propto c^{2}$. The coefficient $1/\kappa_{N}$ of
the Einstein-Hilbert term is usually related to a Newton's constant
$G_{N}=\kappa_{N}/8\pi$. Note that in Riemannian geometry,
where torsion vanishes and $\omega=\tilde{\omega}$, only the gCS
term, the Einstein-Hilbert term, and the cosmological constant survive. 

The coefficients $\kappa_{H},\zeta_{H},1/\kappa_{N},\Lambda/\kappa_{N}$
are given by frequency and wave-vector integrals that arise
within the perturbative calculation, and are described in appendix
\ref{subsec:Perturbative-calculation-of}. In particular $\zeta_{H},1/\kappa_{N},\Lambda/\kappa_{N}$
are dimension-full, with mass dimensions $2,1,3$, and naively diverge. In other words, they are UV sensitive. On the other hand, $\kappa_{H}$ is dimensionless and UV insensitive. With no regularization, one finds
\begin{eqnarray}
 \kappa_{H}=\frac{1}{48\pi}\frac{\text{sgn}\left(m\right)o}{2}.
\end{eqnarray}
Thus, the effective action for a single Majorana spinor can be written
as
\begin{align}
  W_{\text{RC}}\left[e,\omega\right]=\frac{1/2}{96\pi}\frac{\text{sgn}\left(m\right)o}{2}W\left[e,\omega\right]+\cdots\label{eq:77}
\end{align}
where 
\begin{align}
  W\left[e,\omega\right]=\int_{M_{3}}Q_{3}\left(\tilde{\omega}\right)-\int_{M_{3}}\tilde{\mathcal{R}}e^{a}De_{a}
\end{align}
is the sum of gCS and gpCS, and the dots include UV sensitive terms, or terms of a higher order in derivatives or perturbations, as described above. 

Since the lattice model implies a finite physical cutoff $\Lambda_{UV}$
for wave-vectors, \eqref{eq:77} is exact only for $m/\Lambda_{UV}\rightarrow0$.
For non-zero $m$ there are small $O\left(m/\Lambda_{UV}\right)$
corrections\footnote{All expressions here are with $\hbar=c_{\text{light}}=1$. Restoring
units one finds $\frac{m}{\Lambda_{\text{UV}}}\sim\frac{\text{max}\left(t,\mu\right)}{\delta}$
and so $\frac{m}{\Lambda_{\text{UV}}}\ll1$ in the relativistic regime.} to \eqref{eq:77}. We will keep these corrections implicit for now,
and come back to them in section \ref{subsec:Gauge-symmetry-of}.

\subsubsection{Summing over Majorana spinors\label{subsec:Summing-over-lattice}}

As discussed in sections \ref{sec:Lattice-model} and \ref{sec:Continuum-limits-of},
the continuum description of the $p$-wave SC includes four Majorana
spinors labeled by $1\leq n\leq4$, with masses $m_{n}$, which are
coupled to vielbeins $e_{\left(n\right)}$. Let us repeat the necessary details. The vielbein $e_{\left(1\right)}$
is associated with the order parameter $\delta$ of the underlying
lattice model, as in \eqref{17}, up to an unimportant rescaling by
the lattice spacing $a$. For this reason we treat it as a fundamental
vielbein and write $e=e_{\left(1\right)}$ in some expressions. The
other vielbeins $\left(e_{\left(n\right)}\right)_{a}^{\;\mu}$ are
obtained from $e$ by multiplying one of the columns $\mu=x,y$ or
both by $-1$. This implies that $o=o_{1}=o_{3}=-o_{2}=-o_{4}$, and
that the metrics $g_{\left(n\right)}^{\mu\nu}$ are identical apart
from $g^{xy}=g_{\left(1\right)}^{xy}=g_{\left(3\right)}^{xy}=-g_{\left(2\right)}^{xy}=-g_{\left(4\right)}^{xy}$.
With this in mind, we can sum over the four Majorana spinors and obtain
and effective action for the $p$-wave SC, 
\begin{eqnarray}
 W_{\text{SC}}\left[e,\omega\right]&=&\sum_{n=1}^{4}W_{\text{RC}}\left[e_{\left(n\right)},\omega\right]\label{eq:77-1-1}\\
 &=&\frac{1/2}{96\pi}\sum_{n=1}^{4}\frac{\text{sgn}\left(m_{n}\right)o_{n}}{2}W\left[e_{\left(n\right)},\omega\right]+\cdots\nonumber
\end{eqnarray}
Note that the Chern number of the lattice
model is given by $\nu=\sum_{n=1}^{4}\text{sgn}\left(m_{n}\right)o_{n}/2$,
but since $W$ also depends on the different vielbeins $e_{\left(n\right)}$,
\eqref{eq:77-1-1} does not only depend on $\nu$ in the general case. 

Some simplification is possible however. Since $e_{\left(1\right)}=e_{\left(3\right)}$
and $e_{\left(2\right)}=e_{\left(4\right)}$ up to a space-time independent
$SO\left(2\right)$ ($U\left(1\right)$) transformation, 
\begin{eqnarray}
  W_{\text{SC}}\left[e,\omega\right]&=&\sum_{l=1}^{2}\frac{\nu_{l}/2}{96\pi}W\left[e_{\left(l\right)},\omega\right]+\cdots\label{eq:77-3}\\
  &=&\sum_{l=1}^{2}\frac{\nu_{l}/2}{96\pi}\int_{M_{3}}Q_{3}\left(\tilde{\omega}{}_{\left(l\right)}\right)+\cdots\nonumber 
\end{eqnarray}
where in the second line, we have only written explicitly gCS terms.
Here we defined $\nu_{1}=\frac{o_{1}}{2}\left(\text{sgn}\left(m_{1}\right)+\text{sgn}\left(m_{3}\right)\right),\;\nu_{2}=\frac{o_{2}}{2}\left(\text{sgn}\left(m_{2}\right)+\text{sgn}\left(m_{4}\right)\right)$
which are both integers $\nu_{1},\nu_{2}\in\mathbb{Z}.$ The Chern
number of the lattice model is given by the sum $\nu=\nu_{1}+\nu_{2}$.
Thus the lattice model seems to behave like a bi-layer, with layer
index $l=1,2$. In the topological phases of the model $\nu_{1}=0$, $\nu=\nu_{2}=\pm1$, and so 
\begin{eqnarray}
 & W_{\text{SC}}\left[e,\omega\right] & =\frac{\nu/2}{96\pi}W\left[e_{\left(2\right)},\omega\right]+\cdots\label{eq:77-4}\\
 &  & =\frac{\nu/2}{96\pi}\int_{M_{3}}Q_{3}\left(\tilde{\omega}_{\left(2\right)}\right)+\cdots\nonumber 
\end{eqnarray}
where again, in the second line we have only written explicitly the
gCS term. This result is close to what one may have guessed. In the
topological phases with Chern number $\nu\neq0$, the effective action
contains a single gCS term, with coefficient $\frac{\nu/2}{96\pi}$.
A result of this form has been anticipated in \cite{volovik1990gravitational,read2000paired,wang2011topological,ryu2012electromagnetic,palumbo2016holographic},
but there are a few details which are important to note. First, apart
from gCS, $W$ also contains the a gpCS term of the form $\int_{M_{3}}\tilde{\mathcal{R}}e^{a}De_{a}$,
which is possible due to the emergent torsion. Second, the connection
that appears in the CS form $Q_{3}$ is a LC connection, and not the torsion-full connection $\omega$. Moreover, this LC connection is not $\tilde{\omega}$, but
a modification of it $\tilde{\omega}{}_{\left(2\right)}$, where the
subscript $\left(2\right)$ indicates the effect of the multiple Majorana
spinors in the continuum description of the lattice model.  Third, the geometric fields $e,\omega$ are given by $\Delta,A$.

In the trivial phases $\nu_{1}=-\nu_{2}\in\left\{ -1,0,1\right\} $,
$\nu=0$, and we find
\begin{eqnarray}
 & W_{\text{SC}}\left[e,\omega\right] & =\frac{\nu_{1}/2}{96\pi}\left[W\left[e_{\left(1\right)},\omega\right]-W\left[e_{\left(2\right)},\omega\right]\right]+\cdots\label{eq:77-5}\\
 &  & =\frac{\nu_{1}/2}{96\pi}\left[\int_{M_{3}}Q_{3}\left(\tilde{\omega}{}_{\left(1\right)}\right)-\int_{M_{3}}Q_{3}\left(\tilde{\omega}{}_{\left(2\right)}\right)\right]+\cdots\nonumber 
\end{eqnarray}
 This result is quite surprising. Instead of containing no gCS terms,
some trivial phases contain the difference of two such terms, with
slightly different spin connections. One may wonder if these trivial
phases are really trivial after all. This is part of a larger issue
which we now address. 

\subsection{Symmetries of the effective action \label{subsec:Gauge-symmetry-of}}

By considering the gauge symmetry of the effective action we can reconstruct
the topological phase diagram appearing in Fig.\ref{fig:Phase-Diagram}
from \eqref{eq:77-3}. This will also help us understand which of
our results are special to the relativistic limit, and which should
hold throughout the phase diagram. By gauge symmetry we refer in this
section to the $SO\left(2\right)$ subgroup of $SO\left(1,2\right)$,
which corresponds to the physical $U\left(1\right)$ symmetry of the
$p$-wave SC. Equation \eqref{eq:70} shows that we can equivalently
consider $Diff$ symmetry. The physical reason for this equivalence
is that the $p$-wave order parameter is charged under both symmetries,
and therefore maps them to one another. 

The effective action was calculated within perturbation theory on
the space-time manifold $M_{3}=\mathbb{R}_{t}\times\mathbb{R}^{2}$,
but for this discussion, we use its locality to assume it remains
locally valid on more general $M_{3}$, which may be closed (compact
and without a boundary) or have a boundary. A closed space-time is
most simply obtained by working on $M_{3}=\mathbb{R}_{t}\times M_{2}$
with $M_{2}$ closed, and with background fields $\Delta,A$ which
are periodic in time, such that $\mathbb{R}_{t}$ can be  compactified
to a circle. 

As described in appendix \ref{subsec:Global-structures-and}, a non
singular order parameter endows $M_{3}$ with an orientation and a
spin structure, and in particular requires that $M_{2}$ be orientable
\cite{quelle2016edge}, which we assume. Thus, for example, we exclude
the possibility of $M_{2}$ being the Mobius strip. Moreover, a non
singular order parameter on a closed $M_{2}$ requires that $M_{2}$
contain $\left(g-1\right)o$ magnetic monopoles \cite{read2000paired},
where $g$ is the genus of $M_{2}$, and we assume that this condition
is satisfied. For example, if $M_{2}$ is the sphere then it must
contain a single monopole or anti-monopole depending on the orientation
$o$ \cite{kraus2009majorana,moroz2016chiral}. 

\subsubsection{Quantization of coefficients\label{subsec:quantization}}

The first fact about the gCS term that we will need, is that gauge
symmetry of $\alpha\int_{M_{3}}Q_{3}\left(\tilde{\omega}\right)$
for all closed $M_{3}$ requires that $\alpha$ be quantized such
that $\alpha\in\frac{1}{192\pi}\mathbb{Z}$, see equation (2.27) of
\cite{witten2007three}. 
In order to understand how generic is our result \eqref{eq:77-3}, we
will check what quantization condition on $\alpha_{1},\alpha_{2}$
is required for gauge symmetry of $\alpha_{1}\int_{M_{3}}Q_{3}\left(\tilde{\omega}_{\left(1\right)}\right)+\alpha_{2}\int_{M_{3}}Q_{3}\left(\tilde{\omega}_{\left(2\right)}\right)$
on all closed $M_{3}$. Following the arguments of \cite{witten2007three}
we find that $\alpha_{1}+\alpha_{2}\in\frac{1}{192\pi}\mathbb{Z}$,
but $\alpha_{1},\alpha_{2}\in\mathbb{R}$ are not separately restricted, see appendix \ref{subsec:quntization-of-coefficients}.  It is
therefore natural to define $\alpha=\alpha_{1}+\alpha_{2}$ and rewrite
\begin{eqnarray}
 && \alpha_{1}\int_{M_{3}}Q_{3}\left(\tilde{\omega}_{\left(1\right)}\right)+\alpha_{2}\int_{M_{3}}Q_{3}\left(\tilde{\omega}_{\left(2\right)}\right)\\
  &&=\alpha\int_{M_{3}}Q_{3}\left(\tilde{\omega}_{\left(2\right)}\right)+\alpha_{1}\int_{M_{3}}\left[Q_{3}\left(\tilde{\omega}_{\left(1\right)}\right)-Q_{3}\left(\tilde{\omega}_{\left(2\right)}\right)\right],\nonumber 
\end{eqnarray}
where $\alpha\in\frac{1}{192\pi}\mathbb{Z}$ but $\alpha_{1}\in\mathbb{R}$.
Comparing with the result \eqref{eq:77-3}, we identify $\alpha=\frac{\nu/2}{96\pi}$, $\alpha_{1}=\frac{\nu_{1}/2}{96\pi}$,
and we conclude that $\nu$ must be precisely an integer and equal
to the Chern number, while $\nu_{1}$ need not be quantized. We therefore
interpret the $O\left(m/\Lambda_{\text{UV}}\right)$ corrections to
$\alpha=\frac{\nu/2}{96\pi}$ produced in our computation as artifacts
of our approximations\footnote{Specifically, in obtaining the relativistic continuum approximation
we split the Brillouin zone $BZ$ into four quadrants and linearized
the lattice Hamiltonian \eqref{eq:3} in every quadrant. Applying
any integral formula for the Chern number to the approximate Hamiltonian
will give a result $\nu_{\text{apprx}}=\frac{1}{2}\sum_{n=1}^{4}o_{n}\text{sgn}\left(m_{n}\right)+O\left(m/\Lambda_{\text{UV}}\right)$
which is only approximately quantized in the relativistic regime,
simply because the approximate Hamiltonian is discontinuous on $BZ$.
Nevertheless, the known quantization $\nu\in\mathbb{Z}$ and the fact
that $\nu_{\text{apprx}}\approx\nu$ are enough to obtain the exact
result $\nu=\frac{1}{2}\sum_{n=1}^{4}o_{n}\text{sgn}\left(m_{n}\right)$.}, which must vanish due to gauge invariance. On the other hand, we
interpret the quantization $\alpha_{1}=\frac{\nu_{1}/2}{96\pi}$ as
a special property of the relativistic limit with both $m^{*}\rightarrow\infty$ and $m\rightarrow0$,
which should not hold throughout the phase diagram. 

So far we have only considered gCS terms. As already explained, the
gpCS term is gauge invariant on any $M_{3}$, and we therefore see
no reason for the quantization of its coefficient. Explicitly, $-\beta\int_{M_{3}}\tilde{\mathcal{R}}e^{a}De_{a}$
is gauge invariant for all $\beta\in\mathbb{R}$. Thus we interpret
the approximate quantization of the coefficients of gpCS terms as
a special property of the relativistic limit, which should not hold
throughout the phase diagram. We note that even for a relativistic spinor any $\beta\in\mathbb{R}$ can be obtained, by adding a non minimal coupling to torsion \cite{hughes2013torsional}.

In light of the above, it is natural to interpret \eqref{eq:77-3}
as a special case of
\begin{eqnarray}
 && W_{\text{SC}}\left[e,\omega\right]=\frac{\nu/2}{96\pi}\int_{M_{3}}Q_{3}\left(\tilde{\omega}_{\left(2\right)}\right)\label{eq:80-1}\\
  &&+\alpha_{1}\int_{M_{3}}\left[Q_{3}\left(\tilde{\omega}_{\left(1\right)}\right)-Q_{3}\left(\tilde{\omega}_{\left(2\right)}\right)\right]\nonumber\\
  &&-\beta_{1}\int_{M_{3}}\tilde{\mathcal{R}}_{\left(1\right)}e_{\left(1\right)}^{a}De_{\left(1\right)a}-\beta_{2}\int_{M_{3}}\tilde{\mathcal{R}}e_{\left(2\right)}^{a}De_{\left(2\right)a}+\cdots\nonumber 
\end{eqnarray}
where $\nu\in\mathbb{Z}$ is the Chern number and $\alpha_{1},\beta_{1},\beta_{2}$
are additional, non quantized, yet dimensionless, response coefficients. In the relativistic
limit $\alpha_{1},\beta_{1},\beta_{2}$ happen to be quantized, but
this is not generic. Only the first gCS term encodes topological bulk
responses, proportional to the Chern number $\nu$, and below we will
see that only this term is related to an edge anomaly. We can also
write \eqref{eq:80-1} more symmetrically, 
\begin{eqnarray}
 && W_{\text{SC}}\left[e,\omega\right]\label{eq:81-1}\\
 &&=\sum_{l=1}^{2}\left[\alpha_{l}\int_{M_{3}}Q_{3}\left(\tilde{\omega}_{\left(l\right)}\right)-\beta_{l}\int_{M_{3}}\tilde{\mathcal{R}}_{\left(l\right)}e_{\left(l\right)}^{a}De_{\left(l\right)a}\right]+\cdots\nonumber
\end{eqnarray}
but here we must keep in mind the quantization condition $\alpha_{1}+\alpha_{2}=\frac{\nu/2}{96\pi}\in\frac{1}{192\pi}\mathbb{Z}$.

This equation should be compared with the result in the relativistic
limit \eqref{eq:77-3}, where $\alpha_{l},\beta_{l}$ are all quantized,
and $\alpha_{l}=\beta_{l}$. We note that the quantization of $\alpha_{l},\beta_{l}$
in the relativistic limit can be understood on dimensional grounds: in this limit there are simply not enough dimension-full
quantities which can be used to construct dimensionless quantities,
beyond $\text{sgn}\left(m_{n}\right)$ and $o_{n}$. Of course, this
does not explain why $\alpha_{l}=\beta_{l}$ in the relativistic limit.

\subsubsection{Boundary anomalies\label{subsec:Boundary-anomalies}}

We can strengthen the above conclusions by considering space-times
$M_{3}$ with a boundary. The second fact about the gCS term that
we will need is that it is not gauge invariant when $M_{3}$ has a
boundary, even with a properly quantized coefficient. In more detail,
the $SO\left(2\right)$ variation of gCS is given by 
\begin{align}
  \delta_{\theta}\int_{M_{3}}Q_{3}\left(\tilde{\omega}\right)=-\text{tr}\int_{\partial M_{3}}\mbox{d}\theta\tilde{\omega}.
\end{align}
Up to normalization, the boundary term above is called the consistent
Lorentz anomaly, which is one of the forms in which the gravitational
anomaly manifests itself \cite{bertlmann2000anomalies}\footnote{Generally speaking, \textit{consistent} anomalies are given by symmetry
variations of functionals. We will also discuss below the more physical
\textit{covariant} anomalies, which correspond to the actual inflow
of some charge from bulk to boundary}. The anomaly $\text{tr}\int_{\partial M_{3}}\mbox{d}\theta\tilde{\omega}$
is a local functional that can either be written as the gauge variation
of a local bulk functional, as it is written above, or as the gauge
variation of a \textit{nonlocal} boundary functional $F\left[\tilde{\omega}\right]$,
such that $\delta_{\theta}F\left[\tilde{\omega}\right]=\int_{\partial M_{3}}\mbox{d}\theta\tilde{\omega}$,
but cannot be written as the gauge variation of a local boundary functional
\cite{manes1985algebraic}. The difference of two gCS terms is also
not gauge invariant,
\begin{align}
 & \delta_{\theta}\left[\int_{M_{3}}Q_{3}\left(\tilde{\omega}_{\left(1\right)}\right)-\int_{M_{3}}Q_{3}\left(\tilde{\omega}_{\left(2\right)}\right)\right]\\
 &=-\text{tr}\int_{\partial M_{3}}\mbox{d}\theta\left(\tilde{\omega}_{\left(1\right)}-\tilde{\omega}_{\left(2\right)}\right),\nonumber
\end{align}
but here there is a local boundary term that can produce the same
variation, given by $\text{tr}\left(\tilde{\omega}_{(1)}\tilde{\omega}_{(2)}\right)$.

The physical interpretation is as follows. Since $F\left[\tilde{\omega}\right]$
is non local it can be interpreted as the effective action obtained
by integrating over a gapless, or massless, boundary field coupled
to $e$. These are the boundary chiral Majorana fermions of the $p$-wave
SC. The statement that $F$ cannot be local implies that this boundary
field cannot be gapped. In this manner the existence of the gCS term
in the bulk effective action, with a coefficient that is fixed within
a topological phase, implies the existence of gapless degrees of freedom
that cannot be gapped within a topological phase. We will study this
bulk-boundary correspondence in more detail in section \ref{sec:Boundary-fermions-and}.
Naively, the difference of two gCS terms implies the existence of
two boundary fermions with opposite chiralities, one of which is coupled
to $e_{\left(1\right)}$ and the other coupled to $e_{\left(2\right)}$.
The boundary term $\int_{\partial M_{3}}\text{tr}\left(\tilde{\omega}_{\left(1\right)}\tilde{\omega}_{\left(2\right)}\right)$
can only be generated if the two counter propagating fermions are
coupled, and its locality indicates that this coupling can open a
gap. Thus the term $\int_{\partial M_{3}}\text{tr}\left(\tilde{\omega}_{\left(1\right)}\tilde{\omega}_{\left(2\right)}\right)$
represents the effect of a generic interaction between two counter
propagating chiral Majorana fermions.

Again, as opposed to the gCS term, the gpCS term is gauge invariant
on any $M_{3}$, and is therefore unrelated to edge anomalies. Thus,
in the effective action \eqref{eq:80-1}, only the first gCS term
is related to an edge anomaly.

\subsubsection{Time reversal and reflection symmetry of the effective action}

Time reversal $T$ and reflection $R$ are discussed in appendices
\ref{subsec:Spatial-reflections-and} and \ref{subsec:relativisitc Spatial-reflection-and}.
The orientation $o$ of the order parameter is odd under both $T,R$,
and it follows that so are the coefficients $\nu_{l}$. Therefore
$\nu_{l}$ are $T,R$-odd response coefficients. More generally, $\alpha_{l},\beta_{l}$
in \eqref{eq:81-1} are $T,R$-odd response coefficients. As described
in section \ref{subsec:Effective-action-for}, integrals over differential
forms are also odd under the orientation reversing diffeomorphisms
$T,R$, and therefore $W_{\text{SC}}$ is invariant under $T,R$.

\subsection{Calculation of currents\label{subsec:Calculation-of-bulk}}

To derive the currents we start with the expression
\begin{align}
  \alpha_{1}\int_{M_{3}}Q_{3}\left(\tilde{\omega}\right)-\beta_{1}\int_{M_{3}}\tilde{\mathcal{R}}e^{a}De_{a}+\cdots\label{eq:72-1}
\end{align}
which is the effective action for the layer $l=1$. We then sum the results over $l=1,2$, as in
\eqref{eq:81-1}, to get the full low energy response of the lattice
model, keeping in mind that $\alpha_{1}+\alpha_{2}=\frac{\nu/2}{96\pi}\in\frac{1}{192\pi}\mathbb{Z}$. 

\subsubsection{Bulk response from gravitational Chern-Simons terms\label{subsec:Currents-from-the}}

    For the purpose of calculating the contribution of gCS
to the bulk energy-momentum tensor it is easier to use $Q_{3}\left(\tilde{\Gamma}\right)$
instead of $Q_{3}\left(\tilde{\omega}\right)$. The result is \cite{jackiw2003chern,perez2010conserved,stone2012gravitational}
\begin{align}
  \left\langle \mathsf{J}_{\;a}^{\mu}\right\rangle _{\text{gCS}}=\frac{1}{\left|e\right|}\frac{\delta}{\delta e_{\;\mu}^{a}}\left[\alpha_{1}\int_{M_{3}}Q_{3}\left(\tilde{\Gamma}\right)\right]=4\alpha_{1}\tilde{C}_{\;a}^{\mu},\label{eq:74}
\end{align}
where $\tilde{C}$ is the Cotton tensor, which can be written as 
\begin{eqnarray}
  \tilde{C}^{\mu\nu}=-\frac{1}{\sqrt{g}}\varepsilon^{\rho\sigma(\mu}\tilde{\nabla}_{\rho}\tilde{\mathcal{R}}_{\sigma}^{\nu)}.
\end{eqnarray}
Relevant properties of the Cotton tensor are $\tilde{\nabla}_{\mu}\tilde{C}^{\mu\nu}=0$,
$\tilde{C}_{\;\mu}^{\mu}=0$, and $C^{[\mu\nu]}=0$. It follows from \eqref{eq:74} that 
\begin{align}
 \left\langle t_{\text{cov}\;\nu}^{\mu}\right\rangle _{\text{gCS}}=-\left|e\right|\left\langle \mathsf{J}_{\;\nu}^{\mu}\right\rangle _{\text{gCS}}=-4\alpha_{1}\left|e\right|\tilde{C}_{\;\nu}^{\mu}.\label{eq:110}
\end{align}
For order parameters of the form 
\begin{eqnarray}
 &  & \Delta=e^{i\theta}\left(\left|\Delta^{x}\right|,\pm i\left|\Delta^{y}\right|\right)\label{eq:110-10}
\end{eqnarray}
the metrics for both layers $l=1,2$ are identical. Since $\tilde{C}$
only depends on the metric it follows that for such order parameters
the summation over $l=1,2$ gives 

\begin{align}
  \left\langle t_{\text{cov}\;\nu}^{\mu}\right\rangle _{\text{gCS}}=-\left|e\right|\left\langle \mathsf{J}_{\;\nu}^{\mu}\right\rangle _{\text{gCS}}=-\frac{\nu/2}{96\pi}4\left|e\right|\tilde{C}_{\;\nu}^{\mu}.\label{eq:110-1}
\end{align}
Put differently, the difference of gCS terms in \eqref{eq:81-1},
with coefficient $\alpha_{1}$, does not produce a bulk response for
such order parameters. This provides a simple way to separate the
topological invariant $\nu$ from the non quantized $\alpha_{1}$. 

The Cotton tensor takes a simpler form if the geometry is a product
geometry, where the metric is of the form $\text{d}s^{2}=g_{\alpha\beta}\left(x^{\alpha}\right)\text{d}x^{\alpha}\text{d}x^{\beta}+\sigma\text{d}z^{2}$.
Here $\sigma=\pm1$ depends on whether $z$ is a space-like or time-like
coordinate, and we will use both in the following. The two coordinates
$x^{\alpha}$ are space-like if $z$ is time-like and mixed if $z$
is space-like. In this case the curvature is determined by the curvature
scalar, which corresponds to the curvature scalar of the two dimensional
metric $g_{\alpha\beta}$. In particular $\mathcal{R}_{\;\beta}^{\alpha}=\frac{1}{2}\mathcal{R}\delta_{\beta}^{\alpha}$
and the other components of $\mathcal{R}_{\;\nu}^{\mu}$ vanish. Then
\begin{align}
  \left\langle \mathsf{J}^{\alpha z}\right\rangle _{\text{gCS}}=\left\langle \mathsf{J}^{z\alpha}\right\rangle _{\text{gCS}}=\alpha_{1}\frac{1}{\left|e\right|}\varepsilon^{z\alpha\beta}\partial_{\beta}\tilde{\mathcal{R}},\label{eq:111}
\end{align}
and the other components vanish. In terms of $t_{\text{cov}\;\nu}^{\mu}$,
\begin{eqnarray}
 &  & \left\langle t_{\text{cov}\;z}^{\alpha}\right\rangle _{\text{gCS}}=-\alpha_{1}\sigma\varepsilon^{z\alpha\beta}\partial_{\beta}\tilde{\mathcal{R}},\\
 &  & \left\langle t_{\text{cov}\;\alpha}^{z}\right\rangle _{\text{gCS}}=-\alpha_{1}g_{\alpha\beta}\varepsilon^{z\beta\gamma}\partial_{\gamma}\tilde{\mathcal{R}}.\nonumber 
\end{eqnarray}

Taking $z=t$ is natural in the context of the $p$-wave SC, since
the emergent metric \eqref{eq:10-1} is always a product metric if
$\Delta$ is time independent. Then, with a general time independent
order parameter, 
\begin{eqnarray}
 &  & \left\langle J_{E}^{i}\right\rangle _{\text{gCS}}=\left\langle t_{\text{cov}\;t}^{i}\right\rangle _{\text{gCS}}=-\alpha_{1}\varepsilon^{ij}\partial_{j}\tilde{\mathcal{R}}\label{eq:87-2},\\
 &  & \left\langle P_{i}\right\rangle _{\text{gCS}}=\left\langle t_{\text{cov}\;i}^{t}\right\rangle _{\text{gCS}}=-\alpha_{1}g_{ik}\varepsilon^{kj}\partial_{j}\tilde{\mathcal{R}},\nonumber 
\end{eqnarray}
where $\tilde{\mathcal{R}}$ is the curvature associated with the
spatial metric $g^{ij}=-\Delta^{(i}\Delta^{j)*}$. Again, for order
parameters of the form \eqref{eq:110-10} the metrics for both layers $l=1,2$ are identical, and the summation
over $l=1,2$ produces
\begin{eqnarray}
 &  & \left\langle J_{E}^{i}\right\rangle _{\text{gCS}}=-\frac{\nu/2}{96\pi}\varepsilon^{ij}\partial_{j}\tilde{\mathcal{R}}\label{eq:87-2-2},\\
 &  & \left\langle P_{i}\right\rangle _{\text{gCS}}=-\frac{\nu/2}{96\pi}g_{ik}\varepsilon^{kj}\partial_{j}\tilde{\mathcal{R}}.\nonumber 
\end{eqnarray}
These are the topological bulk responses described in section \ref{subsubsec:Topological bulk responses from a gravitational Chern-Simons term}.
It is also usefull to consider order parameters of the form
\begin{eqnarray}
 &  & \Delta=\Delta_{0}e^{i\theta}\left(1,e^{i\phi}\right),\label{eq:87-9}
\end{eqnarray}
where $\phi$ is space dependent. Here the metrics satisfy $g^{xy}=g_{\left(1\right)}^{xy}=-g_{\left(2\right)}^{xy}=\Delta_{0}^{2}\cos\phi$,
with the other components constant, and therefore the Ricci scalars
satisfy $\mathcal{R}=\mathcal{R}_{\left(1\right)}=-\mathcal{R}_{\left(2\right)}$.
The summation over $l=1,2$ for such order parameters then gives
\begin{eqnarray}
 &  & \left\langle J_{E}^{i}\right\rangle _{\text{gCS}}=-\left(\alpha_{1}-\alpha_{2}\right)\varepsilon^{ij}\partial_{j}\tilde{\mathcal{R}}.\label{eq:87-2-1}
\end{eqnarray}
Unlike the sum $\alpha_{1}+\alpha_{2}=\frac{\nu/2}{96\pi}$, the difference
$\alpha_{1}-\alpha_{2}=2\alpha_{1}-\frac{\nu/2}{96\pi}$ is not quantized.
The response \eqref{eq:87-2-1} is therefore not a topological bulk
response. Measuring $\left\langle J_{E}\right\rangle $ for an order
parameter such that $\mathcal{R}=\mathcal{R}_{\left(1\right)}=\mathcal{R}_{\left(2\right)}$,
and then for an order parameter such that $\mathcal{R}=\mathcal{R}_{\left(1\right)}=-\mathcal{R}_{\left(2\right)}$,
allows one to fix both $\alpha_{1},\alpha_{2}$, or both $\nu$ and
$\alpha_{1}$. 

To demonstrate how closely \eqref{eq:87-2-1} can resemble a topological
bulk response, we go back to the lattice model. In the relativistic
limit we found that some trivial phases, where $\nu=0$, have $\alpha_{1}=\frac{\nu_{1}/2}{192\pi}\neq0$.
It follows that these trivial phases have \textit{in the relativistic
limit} a quantized response 
\begin{align}
  \left\langle J_{E}^{i}\right\rangle _{\text{gCS}}=-2\alpha_{1}\varepsilon^{ij}\partial_{j}\tilde{\mathcal{R}}=-\frac{\nu_{1}}{96\pi}\varepsilon^{ij}\partial_{j}\tilde{\mathcal{R}},
\end{align}
for order parameters $\Delta=\Delta_{0}e^{i\theta}\left(1,e^{i\phi}\right)$.

Another case of interest is when $z$ is a spatial coordinate. As
an example, we take $z=y$. This decomposition is less natural in
the $p$-wave SC, as can be seen from \eqref{eq:10-1}. It allows
for time dependence, but restricts the configuration the order parameter
can take at any given time. A simple example for an order parameter
that gives rise to a product metric with respect to $y$ is $\Delta=\Delta_{0}e^{i\theta\left(t,x\right)}\left(1+f\left(t,x\right),\pm i\right)$,
which is a perturbation of the $p_{x}\pm ip_{y}$ configuration with
a small real function $f$. Then 
\begin{eqnarray}
 &  & \left\langle t_{\text{cov}\;\alpha}^{y}\right\rangle _{\text{gCS}}=-\frac{\nu/2}{96\pi}g_{\alpha\beta}\varepsilon^{\beta\gamma y}\partial_{\gamma}\tilde{\mathcal{R}},\label{eq:104}
\end{eqnarray}
where we have summed over $l=1,2$. This an interesting contribution
to the $x$-momentum current and energy current in the $y$ direction.
If we consider, as in Fig.\ref{fig:A-comparison-of}, a boundary or
domain wall at $y=0$, between a topological phase and a trivial phase
where $\nu=0$, we see that there is an inflow of energy and $x$-momentum
into the boundary from the topological phase. This shows that energy
and $x$-momentum are accumulated on the boundary, at least locally,
which corresponds to the boundary gravitational anomaly. We complete
the analysis of this situation from the boundary point of view in
section \ref{subsec:Implication-for-the}. 

\subsubsection{Bulk response from the gravitational pseudo Chern-Simons term \label{subsec:Additional-contributions}}

 The gpCS term $-\beta_{1}\int_{M_{3}}\tilde{\mathcal{R}}e^{a}De_{a}$
contributes to the energy-momentum tensor, and also provides a contribution
to the spin density, 
\begin{align}
  \left\langle \mathsf{J}^{\mu\nu}\right\rangle _{\text{gpCS}}=&\beta_{1}\left\{\frac{1}{\left|e\right|}\varepsilon^{\mu\nu\rho}\partial_{\rho}\tilde{\mathcal{R}}-\frac{1}{\left|e\right|}\varepsilon^{\mu\rho\sigma}\tilde{\mathcal{R}}T_{\rho\sigma}^{\nu}\right.\label{eq:92-1}\\
 &+\left.2o\left[\left(\tilde{\nabla}^{\mu}\tilde{\nabla}^{\nu}-g^{\mu\nu}\tilde{\nabla}^{2}\right)-\tilde{\mathcal{R}}^{\mu\nu}\right]c \vphantom{\frac{1}{\left|e\right|}} \right\}.\nonumber\\
 \left\langle \mathsf{J}^{ab\mu}\right\rangle _{\text{gpCS}}=&\beta_{1}o\tilde{\mathcal{R}}\varepsilon^{abc}e_{c}^{\;\mu}\nonumber 
\end{align}
These are calculated in appendix \ref{subsec:Calculation-of-certain}.

 Using \eqref{eq:71-1}, the above contribution to the spin density
corresponds to a contribution to the charge density, 
\begin{eqnarray}
 &  & \left\langle J^{t}\right\rangle _{\text{gpCS}}=4\beta_{1}o\left|e\right|\tilde{\mathcal{R}}.\label{eq:98}
\end{eqnarray}
The most notable feature of this density is that it is not accompanied
by a current, even for time dependent background fields, where $\partial_{\mu}\left\langle J^{\mu}\right\rangle =\partial_{t}\left\langle J^{t}\right\rangle \neq0$.
This represents the non conservation of fermionic charge in a $p$-wave
SC \eqref{eq:15}. The appearance of $o$ can be understood from \eqref{eq:76}.
One can also understand the appearance of $o$ based on time reversal
symmetry. Since both $J^{t}$ and $\tilde{\mathcal{R}}$ are time
reversal even, the coefficient of the above response cannot be $\beta_{1}$,
which is time reversal odd. 

We now discuss the energy-momentum contributions $\left\langle \mathsf{J}^{\mu\nu}\right\rangle _{\text{gpCS}}$
in \eqref{eq:92-1}, with the purpose of comparing them to the gCS
contributions $\left\langle \mathsf{J}^{\mu\nu}\right\rangle _{\text{gCS}}$.
To do this in the simplest setting, we restrict to a product geometry
with respect to the coordinate $z$ as described in the previous section.
We will also assume for simplicity that torsion vanishes, and generalize
to non-zero torsion in appendix \ref{subsec:Calculation-of-certain}.
For a torsion-less product geometry $\left\langle \mathsf{J}^{\mu\nu}\right\rangle _{\text{gpCS}}$
reduces to 
\begin{align}
  -\left\langle \mathsf{J}^{\alpha z}\right\rangle _{\text{gpCS}}=\left\langle \mathsf{J}^{z\alpha}\right\rangle _{\text{gpCS}}=\beta_{1}\frac{1}{\left|e\right|}\varepsilon^{z\alpha\beta}\partial_{\beta}\tilde{\mathcal{R}}.\label{eq:94-1}
\end{align} 
Note that while the gpCS term vanishes in a torsion-less geometry, the currents it produces, given by its functional derivatives, do not. Comparing with \eqref{eq:111}, we see that $\left\langle \mathsf{J}^{z\alpha}\right\rangle _{\text{gpCS}}\propto\left\langle \mathsf{J}^{z\alpha}\right\rangle _{\text{gCS}}$,
while $\left\langle \mathsf{J}^{\alpha z}\right\rangle _{\text{gpCS}}\propto-\left\langle \mathsf{J}^{\alpha z}\right\rangle _{\text{gCS}}$,
with the proportionality constant $\alpha_{1}/\beta_{1}$, that goes
to 1 in the relativistic limit.  This demonstrates the similarity
between the gpCS and gCS terms. 

In particular, we find in a time independent situation the following
contributions to the energy current and momentum density,
\begin{eqnarray}
 &  & \left\langle J_{E}^{i}\right\rangle _{\text{gpCS}}=\left\langle t_{\text{cov}\;t}^{i}\right\rangle _{\text{gpCS}}=\beta_{1}\varepsilon^{ij}\partial_{j}\tilde{\mathcal{R}},\label{eq:92-2}\\
 &  & \left\langle P_{i}\right\rangle _{\text{gpCS}}=\left\langle t_{\text{cov}\;i}^{t}\right\rangle _{\text{gpCS}}=-\beta_{1}g_{ik}\varepsilon^{kj}\partial_{j}\tilde{\mathcal{R}}.\nonumber 
\end{eqnarray}
Comparing with \eqref{eq:87-2}, we see that $\left\langle P_{i}\right\rangle _{\text{gpCS}}\propto\left\langle P_{i}\right\rangle _{\text{gCS}}$,
while $\left\langle J_{E}^{i}\right\rangle _{\text{gpCS}}\propto-\left\langle J_{E}^{i}\right\rangle _{\text{gCS}}$.
This sign difference can be understood from the density response \eqref{eq:98},
and the relation \eqref{eq:49} between the operators $J_{E}$ and
$P$, in the relativistic limit. With vanishing torsion it reduces
to
\begin{eqnarray}
 &  & J_{E}^{j}-g^{jk}P_{k}=\frac{o}{2}\varepsilon^{jk}\partial_{k}\left(\frac{1}{\left|e\right|}J^{t}\right).
\end{eqnarray}
 Thus the gCS contributions \eqref{eq:87-2} satisfy $\left\langle J_{E}^{j}\right\rangle _{\text{gCS}}-g^{jk}\left\langle P_{k}\right\rangle _{\text{gCS}}=0$
because gCS does not contribute to the density. On the other hand,
the gpCS does contribute to the density, which is why $\left\langle J_{E}^{j}\right\rangle _{\text{gpCS}}-g^{jk}\left\langle P_{k}\right\rangle _{\text{gpCS}}\neq0$.
This conclusion holds regardless of the value of the coefficient $\beta_{1}$
of gpCS. One can therefore fix the value of $\beta_{1}$ by a measurement
of the density, and thus separate the topological bulk responses (gCS)
from the non-topological bulk responses (gpCS). 

More accurately, we have seen that the lattice model behaves as a
bi-layer with layer index $l=1,2$, and there are actually two coefficients
$\beta_{1},\beta_{2}$. As in the previous section, one can extract
both $\beta_{1},\beta_{2}$ by first considering an order parameter \eqref{eq:110-10}
such that $\mathcal{R}=\mathcal{R}_{\left(1\right)}=\mathcal{R}_{\left(2\right)}$,
and then considering an order parameter \eqref{eq:87-9} such that $\mathcal{R}=\mathcal{R}_{\left(1\right)}=-\mathcal{R}_{\left(2\right)}$.

Another case of interest is when $z$ is a spatial coordinate, and
as in the previous section we take $z=y$, $\Delta=\Delta_{0}e^{i\theta\left(t,x\right)}\left(1+f\left(t,x\right),\pm i\right)$.
We then find from \eqref{eq:94-1}, $\left\langle \mathsf{J}^{y\alpha}\right\rangle _{\text{gpCS}}=\beta_{1}\frac{1}{\left|e\right|}\varepsilon^{z\alpha\beta}\partial_{\beta}\tilde{\mathcal{R}}$,
or
\begin{eqnarray}
 &  & \left\langle t_{\text{cov}\;\alpha}^{y}\right\rangle _{\text{gpCS}}=-\beta_{1}g_{\alpha\beta}\varepsilon^{\beta\gamma y}\partial_{\gamma}\tilde{\mathcal{R}}.\label{eq:104-1}
\end{eqnarray}
In the presence of a boundary (or domain wall) at $y=0$, this describes
an inflow of energy and $x$-momentum from the bulk to the boundary,
such that $\left\langle t_{\text{cov}\;\alpha}^{y}\right\rangle _{\text{gpCS}}\propto\left\langle t_{\text{cov}\;\alpha}^{y}\right\rangle _{\text{gCS}}$.
After summing over $l=1,2$ one finds the proportionality constant
$\frac{\alpha_{1}+\alpha_{2}}{\beta_{1}+\beta_{2}}$, that goes to
1 in the relativistic limit. Nevertheless, we argue that $\left\langle t_{\text{cov}\;\alpha}^{y}\right\rangle _{\text{gCS}}$
corresponds to a boundary gravitational anomaly while $\left\langle t_{\text{cov}\;\alpha}^{y}\right\rangle _{\text{gpCS}}$
does not, in accordance with section \ref{subsec:Boundary-anomalies}.
The relation between gCS and the boundary gravitational anomaly is
well known within the gravitational description, and will be described
from the $p$-wave SC point of view in section \ref{subsec:Implication-for-the}.
The fact that $\left\langle t_{\text{cov}\;\alpha}^{y}\right\rangle _{\text{gpCS}}$
is unrelated to any boundary anomaly follows from the fact that it
is $SO\left(1,2\right)$ and $Diff$ invariant. Due to this invariance
the bulk gpCS term produces not only the bulk currents \eqref{eq:92-1},
but also boundary currents, such that bulk+boundary energy-momentum
is conserved. In a product geometry with $z=y$ we find the boundary currents
\begin{align}
  \left\langle \mathsf{j}^{\alpha\beta}\right\rangle _{\text{gpCS}}=&-\beta_{1}\frac{1}{\left|e\right|}\varepsilon^{\alpha\beta y}\tilde{\mathcal{R}},\label{eq:98-3}\\
  \left\langle \mathsf{j}^{ab\mu}\right\rangle _{\text{gpCS}}=&0,\nonumber 
\end{align}
which are calculated in appendix \ref{subsec:Calculation-of-certain}.
We see that 
\begin{align}
  \tilde{\nabla}_{\alpha}\left\langle \mathsf{j}^{\alpha\beta}\right\rangle _{\text{gpCS}}=\left\langle \mathsf{J}^{y\beta}\right\rangle _{\text{gpCS}}.\label{eq:99}
\end{align}
This conservation law is the statement of bulk+boundary conservation
of energy-momentum within the gravitational description. It can be
understood from \eqref{eq:60}, by noting that the source terms in
\eqref{eq:60} vanish because $\left\langle \mathsf{j}^{ab\mu}\right\rangle _{\text{gpCS}}=0$,
and because we assumed torsion vanishes. The additional source term
$\left\langle \mathsf{J}^{y\beta}\right\rangle _{\text{gpCS}}$, absent
in \eqref{eq:60}, represents the inflow from the bulk. In section
\ref{subsec:Implication-for-the} we translate \eqref{eq:99} to the
language of the $p$-wave SC.

\section{\label{sec:Boundary-fermions-and}Boundary fermions and gravitational
anomaly }

It is well known that the $p$-wave SC has localized degrees of freedom
on curves in space where the Chern number $\nu$ jumps, due to boundaries,
or domain walls in $\Delta$ or $\mu$, which at low energies are
$D=1+1$ chiral Majorana spinors \cite{read2000paired}. In this section
we derive the action for the boundary spinor in the presence of a
space-time dependent order parameter, and describe its gravitational
anomaly and corresponding anomaly inflow. 

We start by deriving the boundary action in the geometric description
in section \ref{subsec:Boundary-states-in}, then review the relevant
facts regarding the boundary gravitational anomaly within the gravitational
description in section \ref{subsec:Boundary-gravitational-anomaly},
and finally translate the results back to the $p$-wave SC language
in section \ref{subsec:Implication-for-the}. 

The form of the boundary action in both the geometric description
\eqref{eq:87} and in the $p$-wave SC language \eqref{eq:137} is
not surprising, and within the geometric description the gravitational
anomaly and anomaly inflow are well known. It is the implication of
gravitational anomaly and anomaly inflow for the $p$-wave SC, through
the emergent geometry described in sections \ref{sec:Emergent-Riemann-Cartan-geometry}
and \ref{sec:Symmetries,-currents,-and}, which is the result of this
section.

\subsection{Boundary fermions in a product geometry\label{subsec:Boundary-states-in}}

We take the space time manifold to be $\mathbb{R}\times\mathbb{R}^{2}$,
and assume that the vielbein has a product form with respect to the
spatial coordinate $y$, 
\begin{align}
  e^{A}=e_{\;\alpha}^{A}\left(x^{\alpha}\right)\text{d}x^{\alpha},\;e^{y}=o\text{d}y,\label{eq:81}
\end{align}
 where $\alpha,\beta,\dots\in\left\{ t,x\right\} $ and $A,B,\dots\in\left\{ 0,1\right\} $
(unlike the notation of section \ref{sec:Emergent-Riemann-Cartan-geometry}
where $A,B,\dots\in\left\{ 1,2\right\} $). To account for the orientation
$o=\text{sgn}\left(\text{det}e_{\;\mu}^{a}\right)$ of the vielbein
explicitly, we assumed $e_{\;\alpha}^{A}$ has a positive orientation,
and wrote $e^{y}=o\text{d}y$. To be concrete we take $o=1$ for now.
It follows that the metric also has the product form $\text{d}s^{2}=g_{\alpha\beta}\left(x^{\alpha}\right)\text{d}x^{\alpha}\text{d}x^{\beta}-\text{d}y^{2}$
where $g_{\alpha\beta}=e_{\;\alpha}^{A}\eta_{AB}e_{\beta}^{B}$. The
form of the vielbein implies that the LC spin connection only has
the nonzero components $\tilde{\omega}_{AB\alpha}$, which only depend
on $t,x$. We also assume that the spin connection only has nonzero
components $\omega_{AB\alpha}$ and depends only on $t,x$. Under
these assumptions $c=C_{abc}\varepsilon^{abc}=0$, and therefore torsion
simply drops out from the action, as can be seen from the form \eqref{43-1}.
This is a result of the low dimensionality of the problem. We further
assume that the mass has the form of a flat domain wall in the $y$
direction. By this we mean $m=m\left(y\right)$ with boundary conditions
$m\rightarrow\pm m_{0}$ as $y\rightarrow\pm\infty$, and $m_{0}\neq0$,
which corresponds to an interface between two distinct phases. To
be concrete we take $m_{0}>0$ for now. $S_{\text{RC}}$ then takes
the form 
\begin{align}
  S_{\text{RC}}=\frac{1}{2}\int\mbox{d}^{3}x\left|e\right|\overline{\chi}\left[ie_{A}^{\;\alpha}\gamma^{A}\tilde{D}_{\alpha}+i\gamma^{2}\partial_{y}-m\left(y\right)\right]\chi.
\end{align}
This separable form implies the decomposition described in \cite{chandrasekharan1994anomaly,fosco1999dirac},
which we now apply to the present situation. Defining $a=\partial_{y}-m\left(y\right),\;a^{\dagger}=-\partial_{y}-m\left(y\right),\;P_{\pm}=\frac{1}{2}\left(1\pm i\gamma^{2}\right)$,
the action takes the form 
\begin{align}
  S_{\text{RC}}=\frac{1}{2}\int\mbox{d}^{3}x\left|e\right|\overline{\chi}\left[ie_{A}^{\;\alpha}\gamma^{A}\tilde{D}_{\alpha}+aP_{+}+a^{\dagger}P_{-}\right]\chi.\label{eq:73}
\end{align}
The operators $h_{+}=a^{\dagger}a$ and $h_{-}=aa^{\dagger}$ are
hermitian and non negative. The positive parts of their spectrum coincide.
We denote the positive eigenvalues by $\lambda^{2}>0$, including
both the discrete and continuous parts of the spectrum, with the corresponding
eigenfunctions $\phi_{\lambda,\pm}$ satisfying $h_{\pm}\phi_{\lambda,\pm}=\lambda^{2}\phi_{\lambda,\pm}$.
These eigenfunctions of $h_{\pm}$ are related by $\phi_{\lambda,+}=\frac{1}{\lambda}a^{\dagger}\phi_{\lambda,-},\;\phi_{\lambda,-}=\frac{1}{\lambda}a\phi_{\lambda,+}$,
where the sign chosen for $\lambda$ is arbitrary, and for concreteness
we take $\lambda>0$. Each set of eigenfunctions can be assumed to
be orthonormal $\int_{-\infty}^{\infty}\text{d}y\phi_{\lambda,\pm}^{*}\phi_{\lambda',\pm}=\delta_{\lambda\lambda'}$.
Apart from the positive part of the spectrum, there can also be a
unique eigenfunction with eigenvalue zero, a zero mode, for $h_{+}$
or $h_{-}$ but not both. The only candidates are $\phi_{0,\pm}\left(y\right)\propto e^{\pm\int_{0}^{y}m\left(s\right)\text{d}s}$,
and a zero mode exists when one of these functions is normalizable.
With our choice of boundary conditions for $m$, only $\phi_{0,-}$
is normalizable. In terms of these eigenfunctions, the natural orthogonal
 decomposition of the spinor $\chi$ is 
\begin{align}
  \chi\left(x,y,t\right)=&P_{+}\chi\left(x,y,t\right)+P_{-}\chi\left(x,y,t\right)\nonumber\\
  =&\begin{subarray}{c}
\sum\end{subarray}_{\lambda>0}\left[\chi_{\lambda,+}\left(x,t\right)\phi_{\lambda,+}\left(y\right)+\chi_{\lambda,-}\left(x,t\right)\phi_{\lambda,-}\left(y\right)\right]\nonumber\\
&+\chi_{0,-}\left(x,t\right)\phi_{0,-}\left(y\right),\label{eq:83}
\end{align}
where $\chi_{\lambda,\pm}$ are spinors of definite chirality, $P_{\pm}\chi_{\lambda,\pm}=\chi_{\lambda,\pm}$.
Inserting this decomposition into \eqref{eq:73} we obtain 
\begin{align}
 S_{\text{RC}}=\frac{1}{2}&\int\mbox{d}^{2}x\left|e\right|\overline{\chi_{0,-}}ie_{A}^{\;\alpha}\gamma^{A}\tilde{D}_{\alpha}\chi_{0,-}\label{eq:84}\\
 &+\sum_{\lambda>0}\frac{1}{2}\int\mbox{d}^{2}x\left|e\right|\overline{\chi_{\lambda}}\left[ie_{A}^{\;\alpha}\gamma^{A}\tilde{D}_{\alpha}+\lambda\right]\chi_{\lambda},\nonumber
\end{align}
where $\chi_{\lambda}=\chi_{\lambda,-}+\chi_{\lambda,+}$. Thus the
action splits into an infinite sum of actions for independent $D=1+1$
spinors, coupled to RC geometry, which in the $D=1+1$ case is the
same as the coupling to Riemannian geometry. The spinor corresponding
to the zero mode is chiral, massless, and exponentially localized
on the domain wall as can be seen from the expression $\phi_{0,-}\left(y\right)\propto e^{-\int_{0}^{y}m\left(s\right)\text{d}s}$.
It represents the robust boundary state that exists between two distinct
topological phases. The chiral boundary spinor exhibits a gravitational
anomaly, which we describe in the following.

All other spinors are non chiral and massive with masses $\lambda\neq0$.
It is useful to think of the eigenvalue problems $h_{\pm}\phi_{\lambda,\pm}=\left(-\partial_{y}^{2}+m^{2}\left(y\right)\pm m'\left(y\right)\right)\phi_{\lambda,\pm}$
as one dimensional time independent Schrodinger problems to understand
the eigenvalues $\lambda$ and eigenfunctions $\phi_{\lambda,\pm}$
\cite{chandrasekharan1994anomaly,fosco1999dirac}.  Almost all of
the massive spinors correspond to delocalized bulk degrees of freedom,
with the functions $\phi_{\lambda,\pm}\left(y\right)$ corresponding
to ``scattering states'' of the ``Hamiltonians'' $h_{\pm}$. Additionally,
there can be a finite number of ``bound states'' $\phi_{\lambda,\pm}$,
in which case $\chi_{\lambda}$ corresponds to an additional non-chiral
boundary state, which is not robust, and can always be removed by
making the domain wall narrower, or the bulk masses $\pm m_{0}$ smaller. 

Since the action splits into a sum of $D=1+1$ fermionic actions and
the decomposition \eqref{eq:83} is orthogonal, the effective action
also splits into a sum 
\begin{align}
W_{\text{RC}}\left[e,\omega\right]=W_{\text{R}}^{-}\left[e\right]+\sum_{\lambda>0}W_{\text{R}}\left[e;\lambda\right],
\end{align}
where $W_{\text{R}}\left[e;\lambda\right]$ is the effective action
obtained by integrating over a $D=1+1$ Majorana spinor with mass
$\lambda\neq0$ coupled to Riemannian geometry, and $W_{\text{R}}^{\pm}\left[e\right]$
is the effective action obtained by integrating over a $D=1+1$ massless
chiral Majorana spinor coupled to Riemannian geometry, with chirality
$\pm$. Above we assumed $m\left(\pm\infty\right)=\pm m_{0}$ with
$m_{0}>0$ and $o=1$. Generalizing slightly,  the net chirality of
the boundary spinors is given by 
\begin{align}
  C=\frac{o}{2}\text{sgn}\left(m\left(\infty\right)\right)-\frac{o}{2}\text{sgn}\left(m\left(-\infty\right)\right).\label{eq:101}
\end{align}

The action $S_{\text{R}}^{\pm}=\frac{1}{2}\int\mbox{d}^{2}x\left|e\right|\overline{\chi_{0,\pm}}ie_{A}^{\;\alpha}\gamma^{A}\tilde{D}_{\alpha}\chi_{0,\pm}$
for a single chiral Majorana spinor coupled to Riemannian geometry
can be simplified by using a \textit{Majorana representation} for
the Clifford algebra, as described in appendix
\ref{subsec:Charge-conjugation-(Appendix)}. In the Majorana representation
$\chi_{0,\pm}=\xi v_{\pm}$ where $\xi$ is a single-component \textit{real}
Grassmann field and $v_{\pm}$ are the normalized eigenvectors of
$i\gamma^{2}$, $\left(i\gamma^{2}\right)v_{\pm}=\pm v_{\pm}$. The
action $S_{\text{R}}^{\pm}$ then reduces to 
\begin{align}
 S_{\text{R}}^{\pm}=\frac{i}{2}\int\mbox{d}^{2}x\left|e\right|\xi e_{\mp}^{\;\alpha}\partial_{\alpha}\xi,\label{eq:87}
\end{align}
where $e_{\mp}^{\;\alpha}=e_{0}^{\;\alpha}\mp e_{1}^{\;\alpha}$.

\subsection{Boundary gravitational anomaly and anomaly inflow\label{subsec:Boundary-gravitational-anomaly}}

The chiral boundary spinor does not couple to the spin connection
$\omega$, and therefore does not distinguish the RC background from
a Riemannian background described by the vielbein. This can be seen
by examining the $1+1$ dimensional version of the conservation laws
described in section \ref{subsec:Currents,-symmetries,-and} for the
energy-momentum tensor $\mathsf{j}_{\;A}^{\alpha}=\frac{1}{\left|e\right|}\frac{\delta S_{\text{R}}^{\pm}}{\delta e_{\;\alpha}^{A}}$
and spin current $\mathsf{j}^{AB\alpha}=\frac{1}{\left|e\right|}\frac{\delta S_{\text{R}}^{\pm}}{\delta\omega_{AB\alpha}}$
of the boundary spinor. As in section \ref{subsec:Currents,-symmetries,-and},
these follow from the $Diff$ and Lorentz gauge symmetries of the
``classical'' action $S_{\text{R}}^{\pm}$. Since the boundary
fermion does not couple to $\omega$, its spin current vanishes, $\mathsf{j}^{AB\alpha}=0$.
Therefore \eqref{eq:57} takes the form 
\begin{align}
 \mathsf{j}^{[AB]}=0,
\end{align}
expressing the symmetry of the boundary energy-momentum tensor, as
in Riemannian geometry. The energy-momentum conservation law \eqref{eq:71}
then takes the form $\nabla_{\alpha}\mathsf{j}_{\;\beta}^{\alpha}-\mathsf{j}_{\;\beta}^{\alpha}T_{\gamma\alpha}^{\gamma}=T_{\beta\alpha}^{A}\mathsf{j}_{\;A}^{\alpha}$,
which reduces to 
\begin{align}
  \tilde{\nabla}_{\alpha}\mathsf{j}^{\alpha\beta}=C_{AB}^{\;\;\;\;\beta}\mathsf{j}^{[AB]}=0,
\end{align}
where $\tilde{\nabla}$ is the LC covariant derivative. This is the
energy-momentum conservation law in a background Riemannian geometry.
The energy-momentum tensor is given explicitly by 
\begin{align}
 \mathsf{j}_{\;\beta}^{\alpha}=-\frac{i}{2}e_{\mp}^{\;\alpha}\xi\partial_{\beta}\xi,
\end{align}
up to a term that vanishes on the equation of motion for $\xi$, $ie_{\mp}^{\;\alpha}\partial_{\alpha}\xi+\frac{i}{2}\xi\left|e\right|^{-1}\partial_{\alpha}\left(\left|e\right|e_{\mp}^{\;\alpha}\right)=0$,
which can also be written in a manifestly covariant form. One can
verify that $\mathsf{j}_{\;\beta}^{\alpha}$ is conserved, symmetric,
and traceless on the equation of motion. 

Chiral Majorana fermions in $D=1+1$ coupled to Riemannian geometry
exhibit a gravitational anomaly, which implies that while the ``classical''
action $S_{\text{R}}^{\pm}$ is invariant under both $Diff$ and Lorentz
gauge transformations, the corresponding effective action $W_{\text{R}}^{\pm}$
is not \footnote{$W_{\text{R}}^{\pm}$ is an example for the nonlocal boundary functional
$F$ discussed in section \ref{subsec:Gauge-symmetry-of}.}. A physical manifestation of this phenomena is that the ``classical''
conservation law $\tilde{\nabla}_{\alpha}\mathsf{j}_{\;\beta}^{\alpha}=0$
is violated quantum mechanically, $\tilde{\nabla}_{\alpha}\left\langle \mathsf{j}_{\;\beta}^{\alpha}\right\rangle \neq0$.
The anomaly can be calculated by various techniques \cite{bertlmann2000anomalies,bastianelli2006path},
the simplest of which is the calculation of a single Feynman graph,
as was originally done in \cite{alvarez1984gravitational} for the
two dimensional Weyl spinor, and is reviewed in \cite{bastianelli2006path}
part 5.1.2 for the case of a Majorana-Weyl spinor relevant for this
paper. The gravitational anomaly\footnote{There are a few ambiguities in describing what the gravitational anomaly
is from an intrinsic boundary point of view. First, there is the issue
of covariant versus consistent anomalies which also exists in gauge
anomalies \cite{bertlmann2000anomalies}. See also \cite{stone2012gravitational}
and part 2 of \cite{jensen2013thermodynamics} for a short review.
Then, for the consistent gravitational anomaly, there is the issue
of Lorentz anomalies versus Einstein ($Diff$) anomalies where one
can obtain an effective boundary action that is invariant under local
Lorentz transformations but not under $Diff$, or vice versa \cite{bertlmann2000anomalies}.
It is also useful to discuss linear combinations of the Einstein and
Lorentz anomalies, related to the symmetry of the effective action
under the Lorentz-covariant $Diff$ action \eqref{eq:62}, see part
6.3 of \cite{bastianelli2006path}. All of these ambiguities are resolved
when calculating the boundary energy-momentum tensor within the anomaly
inflow mechanism: the bulk gCS term contributes to the boundary energy-momentum
tensor, assuring it is symmetric and covariant, so that the physically
relevant gravitational anomaly is the covariant Einstein anomaly \cite{stone2012gravitational},
which is what we refer to here as ``the gravitational anomaly''.} is given by \cite{bertlmann2000anomalies}
\begin{align}
 \tilde{\nabla}_{\alpha}\left\langle \mathsf{j}^{\alpha\beta}\right\rangle =\frac{\nu/2}{96\pi}\frac{1}{\left|e\right|}\varepsilon^{y\beta\alpha}\partial_{\alpha}\tilde{\mathcal{R}}.\label{eq:88}
\end{align}

The physical interpretation of the anomaly, within the gravitational
theory, is obtained by identifying the right hand side with the energy-momentum
inflow from the bulk, \eqref{eq:111}. Then \eqref{eq:88} can be
written as
\begin{align}
  \tilde{\nabla}_{\alpha}\left\langle \mathsf{j}^{\alpha\beta}\right\rangle =\left\langle \mathsf{J}^{y\beta}\right\rangle ,
\end{align}
which, together with the bulk conservation equation \eqref{eq:71},
is just the statement of energy-momentum conservation for a system
with a boundary. This is the anomaly inflow mechanism, recasting what
appears to be energy-momentum non-conservation in a $D=1+1$ system,
as energy-momentum conservation in a $D=2+1$ system with a boundary. 

\subsection{Implication for the $p$-wave SC\label{subsec:Implication-for-the}}

Let us now apply the above to the $p$-wave SC with a flat domain
wall in the chemical potential, $\mu\left(y\right)$, which physically
represents a fixed chemical potential and an additional $y$-dependent
electric potential. To obtain an emergent geometry which is a product
geometry, we take the order parameter to be of the form $\Delta=\left(\Delta^{x},\Delta^{y}\right)=\Delta_{0}e^{i\theta\left(t,x\right)}\left(1+f\left(t,x\right),\pm i\right)$
\footnote{Assuming that $\mu$ depends on $y$ but $\Delta$ is independent
of $y$ may not be self consistent. Nevertheless, it is a simple ansatz
that allows for a description of the boundary fermion and its anomaly,
which is fixed within a topological phase \cite{read2000paired}.} with $\Delta_{0}>0$ and small $f$. We also assume that $A_{y}=0$
and $A_{t},A_{x}$ are functions of $t,x$. This corresponds to a
perturbation of the $p_{x}\pm ip_{y}$ configuration. Note that assuming
$A_{y}=0$ involves a partial $U\left(1\right)$ gauge fixing, leaving
only $y$ independent gauge transformations $\alpha\left(t,x\right)$.
These are the $U\left(1\right)$ gauge transformations that will be
considered in this subsection. After further $U\left(1\right)$ gauge
fixing such that $\theta\mapsto0$, using a gauge transformation $\alpha\left(t,x\right)=-\theta\left(t,x\right)/2$
\footnote{Here we are explicitly assuming that there are no vortices, such that
$\alpha=-\theta/2$ is a gauge transformation.}, the inverse vielbein will be of the form 
\begin{eqnarray}
 &  & e_{a}^{\;\mu}=\left(\begin{array}{ccc}
1 & 0 & 0\\
0 & 1+f\left(t,x\right) & 0\\
0 & 0 & \pm1
\end{array}\right),\label{eq:88-1}
\end{eqnarray}
so the vielbein is of the product form \eqref{eq:81} with $o=\pm1$.
The corresponding inverse metric is given by 
\begin{eqnarray}
 &  & g^{\mu\nu}=\left(\begin{array}{ccc}
1 & 0 & 0\\
0 & -\left(1+f\left(t,x\right)\right)^{2} & 0\\
0 & 0 & -1
\end{array}\right).
\end{eqnarray}
We will also need the Ricci scalar for this metric, 
\begin{eqnarray}
 &  & \tilde{\mathcal{R}}=\frac{2\left(\left(1+f\right)\partial_{t}^{2}f-2\left(\partial_{t}f\right)^{2}\right)}{\left(1+f\right)^{2}}.
\end{eqnarray}
Recalling that $\mu$ determines the bulk masses $m_{n}$, we can
use the formula $\nu=\frac{1}{2}\sum_{n=1}^{4}o_{n}\text{sgn}\left(m_{n}\right)$
for the Chern number in terms of the low energy data, and \eqref{eq:101},
to express the net chirality of the boundary spinors as $C=\sum_{n=1}^{4}C_{n}=\Delta\nu$,
where $C_{n}=\frac{o_{n}}{2}\text{sgn}\left(m_{n}\left(y=\infty\right)\right)-\frac{o_{n}}{2}\text{sgn}\left(m\left(y=-\infty\right)\right)$
and $\Delta\nu=\nu\left(y=\infty\right)-\nu\left(y=-\infty\right)$.
This relation between the boundary net chirality $C$ and the Chern
number difference $\Delta\nu$ is the well known bulk-boundary correspondence.
It can be derived from index theorems as described in \cite{volovik2009universe},
but in the following we will place it on a more physical footing by
describing it as a consequence of energy-momentum conservation. 
Let us now rewrite the action \eqref{eq:87} in terms of the $p$-wave
SC quantities and in physical units (without setting the emergent
speed of light $\Delta_{0}$ to 1, but with $\hbar=1$), 
\begin{align}
 S_{\text{e}}^{\pm}=\frac{i}{2}\int\mbox{d}t\text{d}x\tilde{\xi}\left(\partial_{t}\mp\left|\Delta^{x}\left(t,x\right)\right|\partial_{x}\right)\tilde{\xi}.\label{eq:137}
\end{align}
Here $\left|\Delta^{x}\right|=\Delta_{0}\left(1+f\right)$ and $\tilde{\xi}=\left|e\right|^{1/2}\xi$
is a chiral Majorana spinor \textit{density} from the geometric point
of view, but a chiral Majorana spinor from the physical flat space
point of view. As an operator $\tilde{\xi}$ satisfies $\left\{ \tilde{\xi}\left(x_{1}\right),\tilde{\xi}\left(x_{2}\right)\right\} =\delta\left(x_{1}-x_{2}\right)$.
We see that $\left|\Delta^{x}\right|$ acts as a space-time dependent
velocity for the boundary fermions, which reduces to a constant $\Delta_{0}$
in the $p_{x}\pm ip_{y}$ configuration. Note that both fields $\left|\Delta^{x}\right|,\tilde{\xi}$
are uncharged under $U\left(1\right)$. This is clear for $\left|\Delta^{x}\right|$,
and to see this explicitly for $\tilde{\xi}$ we relate it to the
original spin-less fermion $\psi$ and the (phase of the) order parameter
$\Delta$, 
\begin{eqnarray}
 &  & \psi\left(t,x,y\right)\propto\tilde{\xi}\left(t,x\right)e^{i\theta\left(t,x\right)/2}\phi_{0,\pm}\left(y\right)+\cdots
\end{eqnarray}
where $\phi_{0,\pm}\left(y\right)\propto e^{\pm\int_{0}^{y}m\left(s\right)\text{d}s}$
was defined in section \ref{subsec:Boundary-states-in} and the dots
represent the massive bulk modes and additional non robust massive
boundary modes. From this expression it is clear that $\tilde{\xi}$
is uncharged even though $\psi$ is.

Let us now consider the energy-momentum conservation law for the boundary.
The expression $\tilde{\nabla}_{\alpha}\left\langle \mathsf{j}_{\;\beta}^{\alpha}\right\rangle =0$
involves the covariant derivative, and is therefore inappropriate
from the $p$-wave SC point of view, where space-time is flat and
$e$ is just the order parameter and has no geometric role. We already
described how to interpret covariant energy-momentum conservation
laws from the flat space-time point of view in section \ref{subsec:Energy-momentum},
where we studied the bulk conservation laws. Here we simply repeat
the procedure. We first relate the energy-momentum tensor $\mathsf{j}_{\;\beta}^{\alpha}$
to the canonical boundary (or edge) energy-momentum tensor $t_{\text{e}\;\beta}^{\alpha}$,
and write it in terms of $\tilde{\xi}$
\begin{align}
  t_{\text{e}\;\beta}^{\alpha}=-\left|e\right|\mathsf{j}_{\;\beta}^{\alpha}=&\frac{i}{2}e_{\mp}^{\;\alpha}\tilde{\xi}\partial_{\beta}\tilde{\xi}\nonumber\\
  =&\begin{cases}
\frac{i}{2}\tilde{\xi}\partial_{\beta}\tilde{\xi} & \alpha=t\\
\mp\frac{i}{2}\left|\Delta^{x}\left(t,x\right)\right|\tilde{\xi}\partial_{\beta}\tilde{\xi} & \alpha=x
\end{cases}.
\end{align}
This is the correct notion of energy and momentum from the physical
flat space-time point of view. Note that the relation between $t_{\text{e}\;\beta}^{\alpha}$
and $\mathsf{j}_{\;\beta}^{\alpha}$ is the same as for the bulk quantities
\eqref{eq:56}, and that since $\tilde{\xi}$ is uncharged the canonical
energy-momentum tensor $t_{\text{e}\;\beta}^{\alpha}$ is automatically
$U\left(1\right)$-covariant. We then write the conservation law $\tilde{\nabla}_{\alpha}\left\langle \mathsf{j}_{\;\beta}^{\alpha}\right\rangle =0$
in terms of $t_{\text{e}\;\beta}^{\alpha}$ and using partial derivatives
as $\partial_{\alpha}t_{\text{e}\;\beta}^{\alpha}+\frac{i}{2}\tilde{\xi}\partial_{\alpha}\tilde{\xi}\partial_{\beta}e_{\mp}^{\;\alpha}=0$,
or more explicitly, 
\begin{eqnarray}
 &  & \partial_{\alpha}t_{\text{e}\;\beta}^{\alpha}\mp\frac{i}{2}\tilde{\xi}\partial_{x}\tilde{\xi}\partial_{\beta}\left|\Delta^{x}\right|=0.
\end{eqnarray}
This is just a special case of the usual conservation law \eqref{eq:18}
for the canonical energy-momentum tensor. As usual, it describes the
space-time dependence of the background field $\left|\Delta^{x}\right|$
as a source of energy-momentum for the boundary fermion $\tilde{\xi}$.
This is the ``classical'' analysis of energy momentum-conservation
for the boundary fermion. Quantum mechanically, this equation acquires
a correction due to the anomaly and the presence of the bulk. Translating
the anomaly equation \eqref{eq:88} to the flat space-time point of
view, we obtain 
\begin{align}
 \partial_{\alpha}\left\langle t_{\text{e}\;\beta}^{\alpha}\right\rangle \mp\frac{i}{2}\left\langle \tilde{\xi}\partial_{x}\tilde{\xi}\right\rangle \partial_{\beta}\left|\Delta^{x}\right|=-\frac{\nu}{192\pi}g_{\beta\gamma}\varepsilon^{y\gamma\alpha}\partial_{\alpha}\tilde{\mathcal{R}}.
\end{align}

As in the gravitational point of view, the right hand side is actually
the inflow of energy-momentum from the bulk \eqref{eq:104}, 
\begin{align}
 \partial_{\alpha}\left\langle t_{\text{e}\;\beta}^{\alpha}\right\rangle \mp\frac{i}{2}\left\langle \tilde{\xi}\partial_{x}\tilde{\xi}\right\rangle \partial_{\beta}\left|\Delta^{x}\right|=\left\langle t_{\text{cov }\beta}^{y}\right\rangle. \label{eq:89-1-1}
\end{align}

This equation expresses the conservation of energy ($\beta=t$) and
$x$-momentum ($\beta=x$) on the domain wall. Along with the bulk
conservation equation \eqref{32}, $\partial_{\mu}\left\langle t_{\text{cov}\;\nu}^{\mu}\right\rangle =\frac{1}{2}\left\langle \psi^{\dagger}\partial_{j}\psi^{\dagger}\right\rangle D_{\nu}\Delta^{j}+h.c+F_{\nu\mu}\left\langle J^{\mu}\right\rangle $
\footnote{We note that the domain wall acts as a source for $y$-momentum, which
is included in the term $F_{\nu\mu}\left\langle J^{\mu}\right\rangle $
since $\mu\left(y\right)$ is part of the electric potential $A_{t}$.}, it expresses the sense in which energy-momentum is conserved in
a $p$-wave SC in the presence of a boundary, or domain wall. 

We thus obtain the equation $\Delta\nu=C$, usually referred to as
bulk boundary correspondence, as a direct consequence of bulk+boundary
energy-momentum conservation in the presence of a space-time dependent
order parameter. 


\section{Conclusion and discussion \label{sec:Conclusion-and-discussion}}

\subsection{Chern-Simons terms and pseudo Chern-Simons terms}

In this paper we have shown that there is a topological bulk response
of the $p$-wave SC to a perturbation of its order parameter, which
follows from a gCS term, and we have described a corresponding gravitational
anomaly of the edge states. The coefficient of gCS was found to be
$\alpha=\frac{\nu/2}{96\pi}$ where $\nu$ is the Chern number, as
anticipated in previous work. These results are based on a mapping
of the $p$-wave SC, in the regime where the order parameter is very
large, to a relativistic Majorana spinor in a curved and torsion-full
space-time. We provided arguments for the validity of these results
beyond the relativistic limit in which they were computed, but it
is of interest to preform explicit computations beyond the relativistic
limit. 

The appearance of torsion in the emergent geometry brought about a
surprise: we found an additional term, closely related but distinct
from gCS, which we referred to as gravitational pseudo Chern-Simons
(gpCS), with a dimensionless coefficient $\beta=\frac{\nu/2}{96\pi}=\alpha$.
The gpCS term is fully invariant under the symmetries we 
considered, and is therefore unrelated to edge anomalies and does
not have to have a quantized coefficient. Therefore, the quantization
of $\beta$ seems to be a property of the relativistic limit, which
will not hold throughout the phase diagram (this can be understood
on dimensional grounds, as explained below \eqref{eq:81-1}). Computations
beyond the relativistic limit are required to test this expectation. 

To put the gpCS term in a broader context, we would like to draw an
analogy to the behavior of the Hall conductivity of the $p$-wave
SC. There is theoretical work that predicts a Hall conductivity in
the $p$-wave SC, which, as opposed to the IQHE, is not quantized
\cite{volovik1988quantized,goryo1998abelian,goryo1999observation,furusaki2001spontaneous,stone2004edge,roy2008collective,lutchyn2008gauge,hoyos2014effective,ariad2015effective}.
This Hall conductivity can be traced back to the following term in
the effective action for a $p$-wave SC, obtained by integrating over
the bulk fermions in the presence of $\Delta=\Delta_{0}e^{i\theta}\left(1,\pm i\right)$
and $A$,
\begin{align}
  -\beta'\int\text{d}^{3}xD_{t}\theta\varepsilon^{ij}\partial_{i}A_{j}=2\beta'\int\text{d}^{3}xA_{t}B-\beta'\int\text{d}^{3}x\partial_{t}\theta B,\label{eq:15-2}
\end{align}
where $\beta'$ is a coefficient to be discussed below. We will refer
to this term as a $U\left(1\right)$ pseudo Chern-Simons (pCS) term,
though it has been referred to as Chern-Simons-like, Chern-Simons-type,
and also partial Chern-Simons in previous work. This terminology
reflects the similarity to the $U\left(1\right)$ CS term, which occurs
in the IQHE but \textit{not} in a $p$-wave SC, 
\begin{align}
  \alpha'\int A\text{d}A=2\alpha'\int\text{d}^{3}xA_{t}B-\alpha'\int\text{d}^{3}x\varepsilon^{ij}A_{i}\partial_{t}A_{j}.\label{eq:16}
\end{align}
The $U\left(1\right)$ pCS term was not obtained in this paper because,
as explained in section \ref{subsec:Relativistic-limit-of},  in the
relativistic limit the coupling to the magnetic field is lost. This term is fully gauge invariant, owing to the presence of the phase of the charged order parameter in Eq. \eqref{eq:16}. Thus, it is
unrelated to edge anomalies, and $\beta'$ need not
be quantized. Explicit computation yields an unquantized  $\beta'$ that reduces to $\frac{\nu/2}{4\pi}$ in the limit
$\Delta_{0}\rightarrow0$ \cite{ariad2015effective, goryo1999observation}, which may be partially understood by dimensional analysis, as explained above for $\beta$. In contrast, the behavior of $U\left(1\right)$
CS under gauge transformations implies that it is related to a boundary
$U\left(1\right)$ anomaly, and that $\alpha'\in\frac{1}{4\pi}\mathbb{Z}$. The integer is the Chern number $\nu$. 

Let us now see how this is related to our results. We found two terms
in the effective action for a $p$-wave SC that have dimensionless
and UV insensitive coefficients. The first is gpCS, which for vielbeins
of the form \eqref{17}, and to first order in time derivatives (see
appendix \ref{subsec:Explicit-formulas-for}), can be written as
\begin{align}
 2o&\beta\int\text{d}^{3}x\left|e\right|\left(-\tilde{\omega}_{12t}-2A_{t}\right)\tilde{\mathcal{R}}^{\left(2\right)}\label{eq:18-1}\\
=&-2o\beta\int\text{d}^{3}x\left|e\right|\tilde{\omega}_{12t}\tilde{\mathcal{R}}^{\left(2\right)}-4o\beta\int\text{d}^{3}x\left|e\right|A_{t}\tilde{\mathcal{R}}^{\left(2\right)}, \nonumber
\end{align}
where $\tilde{\mathcal{R}}^{\left(2\right)}=\frac{2o}{\left|e\right|}\varepsilon^{ij}\partial_{i}\tilde{\omega}_{12j}$, the curvature of a spatial slice, is the geometric analog
of $B=\varepsilon^{ij}\partial_{i}A_{j}$. The second term, gCS, can be written as
\begin{align}
 -2&\alpha\int\tilde{\omega}_{12}\text{d}\tilde{\omega}_{12}\label{eq:17} \\
 =&-2o\alpha\int\text{d}^{3}x\left|e\right|\tilde{\omega}_{12t}\tilde{\mathcal{R}}^{\left(2\right)}
 +2\alpha\int\text{d}^{3}x\varepsilon^{ij}\tilde{\omega}_{12i}\partial_{t}\tilde{\omega}_{12j},\nonumber
\end{align}
under the same assumptions. The similarity between the two gravitational terms \eqref{eq:18-1}-\eqref{eq:17},
as well as the analogy with the two $U\left(1\right)$ terms \eqref{eq:15-2}-\eqref{eq:16}
is now manifest. The gCS coefficient $\alpha$ must be quantized such
that $\alpha\in\frac{1}{192\pi}\mathbb{Z}$, and is given by the Chern
number $\alpha=\frac{\nu/2}{96\pi}$, while $\beta$ need not be quantized,
but takes the value $\beta=\frac{\nu/2}{96\pi}=\alpha$ in the relativistic
limit. With a properly quantized $\alpha$ and away from boundaries,
gCS is gauge invariant. In this sense both gravitational terms are
gauge invariant. For the purpose of computing the bulk response, gCS
only depends on the metric (see equation \eqref{eq:70}), which is
the \textit{uncharged} Higgs part of the order parameter. On the other
hand, gpCS depends also on the charged phase $\theta$. 

The main point is that both $U\left(1\right)$ and gravitational pseudo
Chern-Simons terms are possible due to the spontaneous breaking of
$U\left(1\right)$ symmetry in the $p$-wave SC. They encode interesting
bulk responses, which are closely related but distinct from \textit{topological}
bulk responses. We expect similar phenomena to occur in other topological
phases of matter with a spontaneously broken symmetry, and a more
general study awaits future work. 

\subsection{Real background geometry and manipulation of the order parameter}

In this paper we have considered the $p$-wave SC in flat space, and
focused on the emergent geometry described by a general $p$-wave
order parameter. It is also natural to consider the effect of a real
background geometry, obtained by deforming the 2-dimensional sample
in 3-dimensional space, possibly in a time dependent manner. Treating
this at the level of the lattice model is beyond the scope of this
paper, but we can take the $p$-wave SF as a starting point.   On
a deformed sample the $p$-wave SF action \eqref{eq:10} generalizes
to 
\begin{align}
  S&\left[\tilde{\psi};\Delta,A,G\right]=\int\mbox{d}^{2+1}x\sqrt{G}\left[\tilde{\psi}^{\dagger}\frac{i}{2}\overleftrightarrow{D_{t}}\tilde{\psi}\right.\label{eq:120}\\
  -&\left.\frac{1}{2m^{*}}G^{ij}D_{i}\tilde{\psi}^{\dagger}D_{j}\tilde{\psi}-m\tilde{\psi}^{\dagger}\tilde{\psi}
  +\left(\frac{1}{2}\Delta^{j}\tilde{\psi}^{\dagger}\partial_{j}\tilde{\psi}^{\dagger}+h.c\right)\right], \nonumber 
\end{align}
which now depends on three background fields: the order parameter
$\Delta$, the $U\left(1\right)$ connection $A$, and the real background
metric $G$, coming from the embedding of the 2-dimensional sample
in 3-dimensional space. This action is written for the fermion $\tilde{\psi}$,
which satisfies $\left\{ \tilde{\psi}^{\dagger}\left(x\right),\tilde{\psi}\left(y\right)\right\} =\delta^{\left(2\right)}\left(x-y\right)/\sqrt{G\left(x\right)}$
as an operator. In this problem there are two (inverse) metrics, the
real $G^{ij}$ and emergent $g^{ij}=\Delta^{(i}\Delta^{j)*}$, and
it is interesting to study their interplay. In our analysis we have
focused on the relativistic limit, where $m^{*}\rightarrow\infty$.
In this limit the metric $G$ completely decouples from the action,
when written in terms of the fundamental fermion density $\psi=G^{1/4}\tilde{\psi}$,
see appendix \ref{subsec:Equality-of-path}. Thus, results obtained
within the relativistic limit, are essentially unaffected by the background
metric $G$. This conclusion is appropriate as long as the order
parameter is treated as an independent background field, which is
always suitable for the purpose of integrating out the gapped fermion
density $\psi$. One then obtains the bulk currents and densities
that we have described, which depend on the configuration of $\Delta$,
and the question that remains is what this configuration physically
is. Two scenarios are of importance.

The first is when the order parameter is induced by proximity to a
3-dimensional $s$-wave SC. In this case both $G$ and $g$ are background metrics, a scenario similar to the bi-metric description of anisotropic quantum Hall states \cite{gromov2017investigating}. In this case the magnitude of the order
parameter depends on the distance between the sample and the $s$-wave
SC, so if the position of the $s$-wave SC is fixed but the sample
is deformed, a space-time dependent order parameter is obtained. Of
course, one can also obtain the same effect by considering a flat
sample and an $s$-wave SC with a non flat surface. This provides
one route to a manipulation of the order parameter that will result
in the bulk effects we have described. 
Since vanishing torsion, in the setting of this paper, is a compatibility condition on $A$ and $\Delta$, and in this setup $\Delta$ and $A$ are independent, the emergent geometry will in general be torsion-full. For example, one may set $A=0$ and manipulate $\Delta$ as described above to obtain any emergent torsion tensor. This provides a rather flexible setup in which torsion-full geometries can be realized, compared to the more standard approach in which the torsion describes lattice dislocations \cite{hughes2013torsional}. 

The second important scenario is that of an intrinsic order parameter,
in which case it is a dynamical field. The order parameter naturally
splits into a massive Higgs part, which is precisely the emergent
metric $g^{ij}$, and a massless Goldstone part which is the overall
phase $\theta$. The quantum theory of the emergent metric $g^{ij}$
is on its own an interesting problem, which should be similar to theories
of \textit{massive gravity} \cite{hassan2012ghost}, and to a recent
bi-metric theory of quantum Hall states with a gapped collective excitation
\cite{gromov2017bimetric}. Nevertheless, as long as the probes $A,G$
are slow compared to the Higgs gap, $g^{ij}$ can be treated as fixed
to its instantaneous ground state configuration, and it remains to
find this configuration, which in general will depend on the details
of the microscopic fermionic interaction. A common assumption in the
literature is that, for an interaction that depends only on the geodesic
distance, the long wavelength ground state configuration will be the
curved space $p_{x}\pm ip_{y}$ configuration, where the pairing term
is $\frac{1}{2}\Delta_{0}e^{i\theta}\psi^{\dagger}\left(E_{1}^{\;j}\partial_{j}\pm iE_{2}^{\;j}\partial_{j}\right)\psi^{\dagger}$
\cite{read2000paired,hoyos2014effective,moroz2015effective,quelle2016edge,moroz2016chiral,kvorning2017geo}.
Here $\Delta_{0}$ is a constant, $\theta$ is the Goldstone phase,
and $E$ is a vielbein for the real metric $G$, such that $G^{ij}=E_{A}^{\;i}\delta^{AB}E_{B}^{\;j}$,
which is a fixed background field\footnote{The $SO\left(2\right)$ ambiguity in choosing $E$ is incorporated
into $\theta$, which has $SO\left(2\right)$ charge 1 and $U\left(1\right)$
charge 2.}. What this means, in the language of this paper, is that the emergent
metric is proportional to the real metric, $g^{ij}=\Delta_{0}^{2}G^{ij}$.
It follows that the responses to the emergent metric $g$ that we
have described, are in this case, and under the above assumption,
responses to the real metric $G$. This suggests a second route to
a manipulation of the order parameter that will result in the bulk
effects we have described. 

Of course, in the intrinsic case one cannot ignore the dynamics of
the Goldstone phase $\theta$, which will be gapless as long as $A$
is treated as a background field. When $\theta$ is a dynamical field, charge conservation is restored. One may then inquire what is the fate of the $U(1)$ pCS and gpCS responses, which we explained as originating from non-conserved quantities. The answer to this question is known for the $U(1)$ pCS contribution to the Hall conductivity. The gaplessness of $\theta$ makes the Hall conductivity sensitive to the order of limits between the wave vector $q$ and the frequency $\omega$ \cite{goryo1998abelian,roy2008collective,lutchyn2008gauge,hoyos2014effective}. While the Hall conductivity goes to the constant $-2\beta'$ as $\omega\rightarrow 0$ before $q$, it vanishes in the opposite limit, insuring that the total charge is magnetic field independent. Thus, anomalous edge states are not required for charge conservation. We expect a similar state of affairs to occur also for the gpCS responses. This will be discussed elsewhere. 
 Would the emergent torsion vanish in this case? To answer
this question we use our expressions \eqref{eq:146-1} and \eqref{eq:147-1}
for the emergent LC spin connection and contorsion tensors, and insert
$g^{ij}=\Delta_{0}^{2}G^{ij}$. We find $C_{12\mu}=\partial_{\mu}\theta-2A_{\mu}-o\tilde{\omega}_{12\mu}^{\left(E\right)}$,
where $\tilde{\omega}_{12\mu}^{\left(E\right)}$ is a LC spin connection
constructed from $E$. Taking the exterior derivative we find $\frac{1}{\sqrt{G}}\varepsilon^{ij}\partial_{i}C_{j}=v-2B-\frac{o}{2}\mathcal{R}^{\left(G\right)}$
where $v=\frac{1}{\sqrt{G}}\varepsilon^{ij}\partial_{i}\partial_{j}\theta$
is the vorticity and $\mathcal{R}^{\left(G\right)}$ is the background
Ricci scalar. Comparing with the Goldstone action of \cite{hoyos2014effective}
we conclude that torsion should dynamically vanish due
to the formation of vortices such that $v=2B+\frac{o}{2}\mathcal{R}^{\left(G\right)}$.
If $A$ is also treated as dynamical, we expect the torsion to vanish due
to the formation of vortices or magnetic flux such that $v-2B=\frac{o}{2}\mathcal{R}^{\left(G\right)}$
\cite{kvorning2017geo}.

\subsection{Towards experimental observation }

There are a few basic questions that arise when trying to make contact
between the phenomena described in this paper and a possible experimental
observation. Here we take as granted that one has at one's disposal
either a $p$-wave SC, or a candidate material. The first question
is how to manipulate the Higgs part of the order parameter, which
is the emergent metric, and was discussed above. 

The second natural question is how to measure energy currents and
momentum densities. Also relevant, though not accentuated in this
paper, is a measurement of the stress tensor, comprised of the spatial
components of the energy-momentum tensor. One possible approach, which
provides both a means to manipulate the order parameter, and a measurement
of energy-momentum-stress is a measurement of the phonon spectrum
a la \cite{barkeshli2012dissipationless,schmeltzer2014propagation,schmeltzer2017detecting}.
 For the gpCS term, apart from energy-momentum-stress, there is also
the density response \eqref{eq:9-1}, which is a simpler quantity
for measurement, though not a \textit{topological} bulk response.
A possible way to avoid the need to measure energy-momentum-stress
is possible in a Galilean invariant system, where electric current
and momentum density are closely related. The simplest scenario is
that of the $p$-wave SF on a curved sample \eqref{eq:120}, where
one assumes that the emergent metric follows the real metric, $g^{ij}=\Delta_{0}^{2}G^{ij}$.
Here the electric current is related to the momentum density by 
\begin{align}
  J^{i}=-\frac{G^{ij}}{m^{*}}P_{j}.
\end{align}
Our result \eqref{eq:4} then implies that the expectation value $\left\langle J^{i}\right\rangle $
has a contribution related to the gCS term, 
\begin{align}
 &  & \left\langle J^{i}\right\rangle _{\text{gCS}}=-\frac{G^{ij}}{m^{*}}\left\langle P_{j}\right\rangle _{\text{gCS}}=\frac{1}{m^{*}}\frac{\nu/2}{96\pi}\hbar\varepsilon^{ij}\partial_{j}\tilde{\mathcal{R}}.\label{eq:18-6} 
\end{align}
This is not a topological bulk response due to the appearance of
$m^{*}$, but if $m^{*}$ is known, then the Chern number $\nu$ can
be extracted from a measurement of the electric current, which may
be simpler to measure than energy-momentum-stress. It should be noted that there will be additional contributions, similar to \eqref{eq:18-6}, from the gpCS term, which can be distinguished from \eqref{eq:18-6} by the corresponding density response to curvature. There will also be  contributions similar to \eqref{eq:18-6} that originate from integrating out the Goldstone phase, which depend on the combination  $B+\frac{o}{2}\tilde{\mathcal{R}}$ \cite{hoyos2014effective}, and can therefore be separated from \eqref{eq:18-6} by a measurement of the current in response to a magnetic field.

\subsection{Implications for related phases of matter}

The integer quantum Hall effect is the basis for our understanding
of the closely related time reversal invariant topological insulators
in 2 and 3 dimensions, and the fractional quantum Hall effect. In
the same manner, one may hope to utilize the understanding of the
$p$-wave SC gained in this paper in order to better understand the
physics of time reversal invariant topological superconductors in
2 and 3 dimensions, and of recently proposed fractional topological
superconductors \cite{vaezi2013fractional,sagi2017fractional}. It
is also of interest to study the implications for the $\nu=5/2$ fractional
quantum Hall state. We hope to address these issues in future work.

\begin{acknowledgments}
The authors would like to thank Carlos Hoyos and Sergej Moroz for collaboration in a closely related project. The authors benefited from discussions with Yuval Baum, Zohar Komargodski, Paul Wiegmann, Semyon Klevtsov, Michael Stone, Andrey Gromov, Barry Bradlyn, Thors Hans Hansson, Thomas Kvorning, Luca V. Delacretaz, and Ryan Thorngren. OG acknowledges support during early stages of this project by the Simons Center for Geometry and Physics, Stony Brook University, and the 2016 Boulder Summer School for Condensed Matter and Materials Physics through NSF grant DMR-13001648. This work was supported by the Israel Science Foundation, the European Research Council under the Project MUNATOP, and the DFG (CRC/Transregio 183, EI 519/7-1).
\end{acknowledgments}

\appendix

\section{\label{subsec:Equivalent-forms-of}Equivalent forms of $S_{\text{RC}}$
and equality to $S_{\text{rSF}}$ }

It is useful to write the action $S_{\text{RC}}$ in a few equivalent
forms \cite{bertlmann2000anomalies,hughes2013torsional}. To pass
between these equivalent forms one only needs the identity 
\begin{eqnarray}
 &  & \partial_{\nu}\left(\left|e\right|e_{a}^{\;\nu}\right)=\left|e\right|\tilde{\omega}_{\;ab}^{b}\label{40}
\end{eqnarray}
 relating $e$ to the LC spin connection, and the following identity,
which holds for any spin connection $\omega$ but relies on the property
$\gamma^{a}\gamma^{b}\gamma^{c}=i\varepsilon^{abc}$ of $\gamma$
matrices in 2+1 dimensions, 
\begin{align}
  ie_{a}^{\;\mu}\gamma^{a}\omega_{\mu}=&\frac{1}{4}ie_{a}^{\;\mu}\omega_{bc\mu}\left\{ \gamma^{a},\Sigma^{bc}\right\} +\frac{1}{4}ie_{a}^{\;\mu}\omega_{bc\mu}\left[\gamma^{a},\Sigma^{bc}\right]\nonumber\\
  =&-\frac{1}{4}\omega_{abc}\varepsilon^{abc}+\frac{1}{2}i\omega_{\;ab}^{b}\gamma^{a}.\label{eq:135}
\end{align}
The most explicit form of the action is
\begin{align}
 S_{\text{RC}}=\frac{1}{2}\int\mbox{d}^{2+1}x\left|e\right|\overline{\chi}\left[\frac{1}{2}ie_{a}^{\;\mu}\gamma^{a}\overleftrightarrow{\partial_{\mu}}-\frac{1}{4}\omega_{abc}\varepsilon^{abc}-m\right]\chi,\label{eq:40}
\end{align}
where the derivatives act only on the spinors. Here we see that in
2+1 dimensions the spin connection only enters through the scalar
$\omega_{abc}\varepsilon^{abc}$ as a correction to the mass. It also
makes it rather simple to see why $S_{\text{RC}}$ is equal to $S_{\text{rSF}}$
from \eqref{eq:14}, 
\begin{align}
  &S_{\text{RC}}  =\frac{1}{2}\int\mbox{d}^{2+1}x\left|e\right|\overline{\chi}\left[\frac{1}{2}ie_{a}^{\;\mu}\gamma^{a}\overleftrightarrow{\partial_{\mu}}-\frac{1}{4}\omega_{abc}\varepsilon^{abc}-m\right]\chi\nonumber\\
   & =\frac{1}{2}\int\mbox{d}^{2+1}x\Psi^{\dagger}\gamma^{0}\left[\frac{1}{2}ie_{a}^{\;\mu}\gamma^{a}\overleftrightarrow{\partial_{\mu}}-\frac{1}{4}\omega_{abc}\varepsilon^{abc}-m\right]\Psi\nonumber \\
   & =\frac{1}{2}\int\mbox{d}^{2+1}x\Psi^{\dagger}\gamma^{0}\left[\frac{i}{2}\gamma^{0}\overleftrightarrow{\partial_{t}}+\frac{1}{2}ie_{A}^{\;j}\gamma^{A}\overleftrightarrow{\partial_{j}}+A_{t}-m\right]\Psi\nonumber \\
   & =\frac{1}{2}\int\mbox{d}^{2+1}x\Psi^{\dagger}\begin{pmatrix}\frac{i}{2}\overleftrightarrow{\partial_{t}}+A_{t}-m & \frac{1}{2}\Delta^{j}\overleftrightarrow{\partial_{j}}\\
-\frac{1}{2}\Delta^{j*}\overleftrightarrow{\partial_{j}} & \frac{i}{2}\overleftrightarrow{\partial_{t}}-A_{t}+m
\end{pmatrix}\Psi\nonumber \\
 & =\frac{1}{2}\int\mbox{d}^{2+1}x\Psi^{\dagger}\begin{pmatrix}i\partial_{t}-m+A_{t} & \frac{1}{2}\left\{ \Delta^{j},\partial_{j}\right\} \\
-\frac{1}{2}\left\{ \Delta^{j*},\partial_{j}\right\}  & i\partial_{t}+m-A_{t}
\end{pmatrix}\Psi\nonumber\\&=S_{\text{rSF}},\label{eq:43}
\end{align}
where we have used the dictionary \eqref{17}, and also integrated by parts. In going from the third to the fourth line we
have reinstated the emergent speed of light $c_{\text{light}}=\frac{\Delta_{0}}{\hbar}$,
but kept $\hbar=1$. This completes that proof of the equality $S_{\text{rSF}}=S_{\text{RC}}$,
which was stated and explained in section \ref{sec:Emergent-Riemann-Cartan-geometry}. 

Before we move on, an important comment is in order. Since $A_{j}$ does not appear in $S_{\text{rSF}}$, it is clear that for the above equality of
 actions only the identification $\omega_{t}=-2A_{t}\Sigma^{12}$ is required, rather than the full $\omega_{\mu}=-2A_{\mu}\Sigma^{12}$ of \eqref{eq:17}. Accordingly,  $\omega_{j}$ does not appear in $S_{\text{RC}}$ when $\omega, e$ are both spatial ($\omega_{0A\mu}=0, e_{0}^{\;\mu}=\delta_{t}^{\mu}$), because then 
\begin{align}
\omega_{abc}\varepsilon^{abc}=2e_{0}^{\mu}\omega_{12\mu}=2\omega_{12t}.\label{eq:400}
\end{align}
Thus, for the equality of actions $S_{\text{rSF}}=S_{\text{RC}}$  it is not required that $\omega_{j}=-2A_{j}\Sigma^{12}$. Nevertheless, in this work we are actually identifying two QFTs as equal, and there is more to a QFT than its classical action. One must also compare symmetries, observables, and path integral measures (that latter is discussed in appendix \ref{subsec:Equality-of-path}). The mapping of symmetries 
and observables is the subject of section \ref{sec:Symmetries,-currents,-and}, and only holds if the full identification $\omega_{\mu}=-2A_{\mu}\Sigma^{12}$ is made:

In section \ref{spin}, we identify the physical $U\left(1\right)$ symmetry group with the $Spin\left(2\right)$ subgroup of $Spin\left(1,2\right)$ in Riemann-Cartan geometry. For this reason $A_{\mu}$, which is $U\left(1\right)$ connection, really maps to a $Spin\left(2\right)$ connection in the geometric point of view, even if certain components of it do not appear in the action $S_{\text{rSF}}$. 

In section \ref{currents} we discuss the mapping of observables. In particular, even though $A_{j}$ disappears from the action in the relativistic limit, it does not disappear from the energy-momentum tensor (see \eqref{eq:56}, \eqref{eq:49}, where the derivative $D_{\mu}$ contains $A_{\mu}$). Moreover, as explained below \eqref{eq:71-1}, even though the order parameter $\Delta$ corresponds to the spatial vielbein in \eqref{17}, in order to obtain the expectation value of full energy-momentum tensor we must take derivatives of the effective action with respect to all components of the vielbein, not only the spatial ones obtained from $\Delta$. This corresponds to adding to $S_{\text{rSF}}$ a fictitious background field $e_{0}^{\;\mu}$ which is set to zero after the expectation value is computed. In the presence of $e_{0}^{\;\mu}$ the potential $A_{t}$ generalizes to $e_{0}^{\;\mu}A_{\mu}$, and so $A_{j}$ does appear in $S_{\text{rSF}}$. Accordingly, with a general $e_{0}^{\;\mu}$ we see from \eqref{eq:400} that $\omega_{12j}$ appears in $S_{\text{RC}}$. The equality $S_{\text{RC}}=S_{\text{rSF}}$ in the presence of $e_{0}^{\;\mu}$ is then obtained only if $\omega_{j}=-2A_{j}\Sigma^{12}$. 

To close this discussion, we note that the identification of $\Delta^{j}$ as a spatial vielbein and $A_{\mu}$ as a $Spin\left(2\right)$ connection actually holds beyond the relativistic limit, though this is not discussed in this paper. Beyond the relativistic limit  $A_{j}$ will appear in both the action and observables, and identifying the full $\omega_{12\mu}$ with $A_{\mu}$ will be crucial also at the level of the fermionic action.  

Going back to equivalent forms of $S_{\text{RC}}$,  if we wish to isolate the effect of torsion, we can also write 
\begin{align}
  S_{\text{RC}} =\frac{1}{2}\int\mbox{d}^{2+1}x\left|e\right|\overline{\chi}\left[\frac{1}{2}\right.ie_{a}^{\;\mu}&\gamma^{a}\overleftrightarrow{\partial_{\mu}}-\frac{1}{4}\tilde{\omega}_{abc}\varepsilon^{abc}\\
  &\left.-\frac{1}{4}C_{abc}\varepsilon^{abc}-m\right]\chi, \nonumber
\end{align}
or 
\begin{align}
    S_{\text{RC}} =\frac{1}{2}\int\mbox{d}^{2+1}x\left|e\right|\overline{\chi}\left[\frac{1}{2}\right.ie_{a}^{\;\mu}&\left(\gamma^{a}\tilde{D}_{\mu}-\overleftarrow{\tilde{D}_{\mu}}\gamma^{a}\right)\\
    &\left.-\left(m+\frac{1}{4}c\right)\right]\chi,\nonumber
\end{align}
where we see that in 2+1 dimensions torsion enters only trough the
scalar $c=C_{abc}\varepsilon^{abc}$ as a correction to the mass.
One can also integrate by parts in order
to obtain a form from which it is simple to derive the equation of
motion, 
\begin{align}
  S_{\text{RC}} & =\frac{1}{2}\int\mbox{d}^{2+1}x\left|e\right|\overline{\chi}\left[ie_{a}^{\;\mu}\gamma^{a}\tilde{D}_{\mu}-\frac{1}{4}C_{abc}\varepsilon^{abc}-m\right]\chi\nonumber\\
   & =\frac{1}{2}\int\mbox{d}^{2+1}x\left|e\right|\overline{\chi}\left[ie_{a}^{\;\mu}\gamma^{a}D_{\mu}-\frac{1}{2}iC_{\;ab}^{b}\gamma^{a}-m\right]\chi.\label{43-1}
\end{align}
The form in the first equation is special to 2+1 dimensions, but
the form in the second equation holds in any dimension.

\section{Dirac and BdG equations\label{subsec:Dirac-and-BdG}}

Since the $p$-wave SF action is equal to $S_{\text{RC}}$ in the
relativistic limit, and the fermions $\chi$ and $\Psi$ are related
simply, the equation of motion for $\chi$, which is the Dirac equation
in RC background, maps to the equation of motion for $\Psi$, which
is the BdG equation (in the relativistic limit). 

The equation of motion for the Majorana spinor $\chi$ needs to be
derived carefully, because $\chi$ is Grassmann valued and $\overline{\chi}=\chi^{T}\gamma^{0}$ cannot
be treated as independent of $\chi$. Nevertheless, if the operator between $\chi^{T}$ and $\chi$
is particle-hole symmetric, the equations of motion are the same as those of a Dirac
spinor, which are easy to read from \eqref{43-1}, 
\begin{align}
 0=\left[ie_{a}^{\;\mu}\gamma^{a}D_{\mu}-\frac{1}{2}iC_{\;ab}^{b}\gamma^{a}-m\right]\chi.
\end{align}
This is the Dirac equation in RC background. When inserting $\chi=\left|e\right|^{-1/2}\Psi$
and using the identity 
\begin{align}
  \partial_{\mu}\left|e\right|=\left|e\right|\Gamma_{\mu\rho}^{\rho}=\left|e\right|\tilde{\Gamma}_{\mu\rho}^{\rho},\label{47}
\end{align}
we obtain 
\begin{align}
 0=\left[i\gamma^{\mu}\left(D_{\mu}-\frac{1}{2}\Gamma_{\mu\rho}^{\rho}\right)-\frac{1}{2}iC_{\;ab}^{b}\gamma^{a}-m\right]\Psi.
\end{align}
The expression in brackets is the appropriate covariant derivative
for a spinor density of weight 1/2 \cite{ortin2004gravity}, which
is what $\Psi=\left|e\right|^{1/2}\chi$ is from the geometric point
if view. Simplifying this equation using \eqref{40} and \eqref{47},
we arrive at 
\begin{align}
 0=\left[\frac{1}{2}i\gamma^{a}\left\{ e_{a}^{\;\mu},\partial_{\mu}\right\} -\frac{1}{4}\omega_{abc}\varepsilon^{abc}-m\right]\Psi.
\end{align}
By using the dictionary \eqref{17} and multiplying by $\gamma^{0}=\sigma^{z}$
 this reduces to 
\begin{align}
 0=\begin{pmatrix}i\partial_{t}+A_{t}-m & \frac{1}{2}\left\{ \Delta^{j},\partial_{j}\right\} \\
-\frac{1}{2}\left\{ \Delta^{j*},\partial_{j}\right\}  & i\partial_{t}-A_{t}+m
\end{pmatrix}\Psi,
\end{align}
which is the BdG equation in the relativistic limit.  Thus the BdG equation in the relativistic limit is not quite the Dirac equation,
because $\Psi$ is a spinor density, though it is the Dirac equation for the spinor $\chi$.

\section{Equality of path integrals \label{subsec:Equality-of-path}}

In appendix \ref{subsec:Equivalent-forms-of} we showed that the action
for the $p$-wave SF in the relativistic limit, is equal to the action
for a Majorana fermion coupled to RC geometry. To conclude that the
corresponding fermionic path integrals are equal, we also need to
verify that the path integral measure for the $p$-wave SF is equal
to that of the Majorana fermion in RC background. For the $p$-wave
SF \eqref{eq:10}, the path integral measure is written formally as
$\text{D}\psi^{\dagger}\text{D}\psi=\prod_{x}\text{d}\psi^{\dagger}\left(x\right)\text{d}\psi\left(x\right)$
where $x$ runs over all points in space time. In the BdG formalism
we work with the Nambu (or Majorana) spinor $\Psi=\left(\psi,\psi^{\dagger}\right)^{T}$,
in terms of which the measure takes the form $\text{D}\psi^{\dagger}\text{D}\psi=\text{D}\Psi$.
As described in section \ref{sec:Emergent-Riemann-Cartan-geometry}
and appendix \ref{subsec:Dirac-and-BdG}, from the geometric point
of view $\Psi$ is a Majorana spinor density of weight 1/2, and $\chi=\left|e\right|^{-1/2}\Psi$
is a Majorana spinor. In terms of the spinor $\chi$, the measure
takes the form $\text{D}\Psi=\text{D}\left(\left|e\right|^{1/2}\chi\right)$,
which is the correct measure for a matter field in curved background
\cite{hawking1977zeta,fujikawa1980comment,abanov2014electromagnetic}.
With this measure, the path integral over the Majorana spinor $\chi$
formally computes functional pfaffians as in flat space, $e^{iW_{M}\left[A\right]}=\int\text{D}\left(\left|e\right|^{1/2}\chi\right)e^{\frac{i}{2}\int\text{d}^{d}x\left|e\right|\chi^{T}A\chi}=\text{Pf}\left(iA\right)=\sqrt{\text{Det}iA}$,
where $A$ is an antisymmetric hermitian operator with respect to
the inner product $\left\langle f,g\right\rangle =\int\text{d}^{d}x\left|e\right|f^{\dagger}Ag$,
and the determinant $\text{Det}$ is defined by the product of eigenvalues.
For a Dirac spinor $\chi$ the fermionic path integral formally computes
functional determinants, $e^{iW_{D}\left[D\right]}=\int\text{D}\left(\left|e\right|^{1/2}\chi^{\dagger}\right)\text{D}\left(\left|e\right|^{1/2}\chi\right)e^{i\int\text{d}^{d}x\left|e\right|\chi^{\dagger}D\chi}=\text{Det}\left(iD\right)$,
where $D$ is hermitian. In particular, the effective action for a
Majorana spinor is half that of a Dirac spinor with the same operator,
$W_{M}\left[A\right]=\frac{1}{2}W_{D}\left[A\right]$.

\section{Explicit formulas for certain geometric quantities\label{subsec:Explicit-formulas-for}}

Using  $\tilde{\omega}_{\;b\mu}^{a}=e_{\;\alpha}^{a}\left(\partial_{\mu}e_{b}^{\;\alpha}+\tilde{\Gamma}_{\;\beta\mu}^{\alpha}e_{b}^{\;\beta}\right)$
we can calculate the LC spin connection for a vielbein of the form
\begin{align}
 e_{a}^{\;\mu}=\frac{1}{\Delta_{0}}\left(\begin{array}{ccc}
\Delta_{0} & 0 & 0\\
0 & \mbox{Re}(\Delta^{x}) & \mbox{Re}(\Delta^{y})\\
0 & \mbox{Im}(\Delta^{x}) & \text{Im}(\Delta^{y})
\end{array}\right)=\left(\begin{array}{cc}
1\\
 & e_{A}^{\;j}
\end{array}\right)
\end{align}
that occurs in the $p$-wave SC,
\begin{widetext}
\begin{align}
  \tilde{\omega}_{A0t}=&0,\\
  \tilde{\omega}_{A0j}=&e_{A}^{\;i}\frac{1}{2}\partial_{t}g_{ij},\nonumber \\
  \tilde{\omega}_{12t}=&\frac{1}{2}\varepsilon^{AB}e_{Ai}\partial_{t}e_{B}^{\;i}=-\frac{1}{2}\frac{1}{\text{det}\left(e\right)}\varepsilon^{ij}e_{Ai}\partial_{t}e_{\;j}^{A},\nonumber \\
 \tilde{\omega}_{12j}=&\frac{1}{2}\left(\varepsilon^{AB}e_{Ai}\partial_{j}e_{B}^{\;i}-\frac{1}{\text{det}\left(e\right)}\varepsilon^{kl}\partial_{k}g_{lj}\right)=-\frac{1}{2}\frac{1}{\text{det}\left(e\right)}\varepsilon^{kl}\left(e_{Ak}\partial_{j}e_{\;l}^{A}+\partial_{k}g_{lj}\right).\nonumber 
\end{align}
In terms of the parameterization $\Delta=e^{i\theta}\left(\left|\Delta^{x}\right|,e^{i\phi}\left|\Delta^{y}\right|\right)$,
as in section \ref{subsec:The-order-parameter}, the $SO\left(2\right)$
part can be written as

\begin{eqnarray}
 &  & \tilde{\omega}_{12t}=o\left[\frac{1}{2}\cot\left|\phi\right|\partial_{t}\log\frac{\left|\Delta^{y}\right|}{\left|\Delta^{x}\right|}-\frac{1}{2}\partial_{t}\left|\phi\right|\right]-\partial_{t}\theta, \label{eq:146-1}\\
 &  & \tilde{\omega}_{12x}=o\left[\frac{\left|\Delta^{y}\right|}{\left|\Delta^{x}\right|}\frac{\cot\left|\phi\right|}{\sin\left|\phi\right|}\partial_{y}\left|\phi\right|+\left(\frac{1}{\sin^{2}\left|\phi\right|}-1\right)\partial_{x}\left|\phi\right|+\cot\left|\phi\right|\partial_{x}\log\left|\Delta^{y}\right|+\frac{1}{\sin\left|\phi\right|}\frac{\left|\Delta^{y}\right|}{\left|\Delta^{x}\right|}\partial_{y}\log\left|\Delta^{x}\right|\right]-\partial_{x}\theta,\nonumber \\
 &  & \tilde{\omega}_{12y}=o\left[-\frac{\left|\Delta^{x}\right|}{\left|\Delta^{y}\right|}\frac{\cot\left|\phi\right|}{\sin\left|\phi\right|}\partial_{x}\left|\phi\right|-\left(\frac{1}{\sin^{2}\left|\phi\right|}\right)\partial_{y}\left|\phi\right|-\cot\left|\phi\right|\partial_{y}\log\left|\Delta^{x}\right|-\frac{1}{\sin\left|\phi\right|}\frac{\left|\Delta^{x}\right|}{\left|\Delta^{y}\right|}\partial_{x}\log\left|\Delta^{y}\right|\right]-\partial_{y}\theta,\nonumber 
\end{eqnarray}
\end{widetext}
where $o=\text{sgn}\phi$ is the orientation. Note that the terms
in square brackets only depend on the metric degrees of freedom $\left|\phi\right|,\left|\Delta^{x}\right|,\left|\Delta^{y}\right|$,
and that this reduces to $\tilde{\omega}_{12\mu}=-\partial_{\mu}\theta$
in the $p_{x}\pm ip_{y}$ configuration. We can then obtain explicit
formulas for the contorsion using $\omega_{ab\mu}=-2A_{\mu}\left(\delta_{a}^{1}\delta_{b}^{2}-\delta_{b}^{1}\delta_{a}^{2}\right)$
and $C_{ab\mu}=\omega_{ab\mu}-\tilde{\omega}_{ab\mu}$, 
\begin{eqnarray}
 &  & C_{12\mu}=-2A_{\mu}-\tilde{\omega}_{12\mu}, \label{eq:147-1}\\
 &  & C_{A0j}=-e_{A}^{\;i}\frac{1}{2}\partial_{t}g_{ij}.\nonumber 
\end{eqnarray}

We also consider the quantity $c=\varepsilon^{abc}C_{abc}$ which
appears in certain forms of the action $S_{\text{RC}}$ \eqref{eq:43-1},
and of the effective action \eqref{eq:76}. Evaluated in terms of
$\Delta$ and $A$ we find 
\begin{align}
  \frac{1}{2}c=&C_{12t}\\
  =&\partial_{t}\theta-2A_{t}-o\left[\frac{1}{2}\cot\left|\phi\right|\partial_{t}\log\frac{\left|\Delta^{y}\right|}{\left|\Delta^{x}\right|}-\frac{1}{2}\partial_{t}\left|\phi\right|\right],\nonumber
\end{align}
which reduces to $\frac{1}{2}c=D_{t}\theta=\partial_{t}\theta-2A_{t}$
in the $p_{x}\pm ip_{y}$ configuration.

\section{\label{subsec:Discrete-symmetries}Discrete symmetries}

\subsection{Charge conjugation and particle-hole \label{subsec:Charge-conjugation-(Appendix)}}

Our conventions for gamma matrices and spinors follow appendix B
of \cite{ortin2004gravity}. In three dimensions, if the matrices
$\gamma^{a}$ define a representation of the Clifford algebra then
$-\left(\gamma^{a}\right)^{T}$ define an equivalent representation.
The matrix $\mathcal{C}$ relating the two representations by $-\left(\gamma^{a}\right)^{T}=\mathcal{C}_{\;b}^{a}\gamma^{b}=\mathcal{C}\gamma^{a}\mathcal{C}^{-1}$
is called charge conjugation. In our representation $\gamma^{0}=\sigma^{z},\;\gamma^{1}=-i\sigma^{x},\;\gamma^{2}=i\sigma^{y}$,
one finds that $\mathcal{C}=\sigma^{y}$ up to a phase and $\mathcal{C}_{\;b}^{a}=\text{diag}\left[-1,-1,1\right]$,
so we see that $\mathcal{C}$ is unitary and $\mathcal{C}^{2}=1$.
Likewise, the matrices $\left(\gamma^{a}\right)^{\dagger}$ also define
an equivalent representation, and are therefore related by $\left(\gamma^{a}\right)^{\dagger}=\mathcal{D}_{\;b}^{a}\gamma^{b}=\mathcal{D}\gamma^{a}\mathcal{D}^{-1}$
where $\mathcal{D}$ is the Dirac conjugation. In any unitary representation
$\mathcal{D}=i\gamma^{0}$ up to a phase and $\mathcal{D}_{\;b}^{a}=\text{diag}\left[1,-1,-1\right]$.
Using $\mathcal{D}$ we define the conjugate spinor $i\overline{\Psi}=\Psi^{\dagger}\mathcal{D}$.
We also note that $-\left(\gamma^{a}\right)^{*}=\mathcal{B}_{\;b}^{a}\gamma^{b}=\mathcal{B}\gamma^{a}\mathcal{B}^{-1}$
with $\mathcal{B}=\mathcal{D}\mathcal{C}$, which will also show up
in our discussion of time reversal. In our representation, $\mathcal{B}=\sigma^{x}$
and $\mathcal{B}_{\;b}^{a}=\text{diag}\left[-1,1-1\right]$. 

A spinor $\Psi$ is called a Majorana spinor if it satisfies the reality
condition $i\overline{\Psi}=\Psi^{T}\mathcal{C}$, which can also
be written as $\Psi^{\dagger}\mathcal{B}=\Psi^{T}$. In our representation
this condition reads $\Psi^{\dagger}=\Psi^{T}\sigma^{x}$, which is
the reality condition satisfied by the Nambu spinor $\Psi=\left(\psi,\psi^{\dagger}\right)^{T}$.
We see that the Nambu spinor is a Majorana spinor. The reality condition
can also be written as $\Psi=P\Psi$ where $P=\sigma^{x}K$ and $K$
is the complex conjugation. $P$ is usually referred to as a particle-hole
symmetry \cite{ryu2010topological}, and it is anti-unitary and $P^{2}=1$.
Eventually, the particle-hole symmetry of the $p$-wave SC maps to
the charge conjugation symmetry of the relativistic Majorana fermion,
with the differences between the two being a matter of convention. 

For any Hamiltonian $H=\frac{1}{2}\int\text{d}^{2}x\Psi^{\dagger}\left(x\right)H_{\text{BdG}}\Psi\left(x\right)$,
the BdG Hamiltonian $H_{\text{BdG}}$ can be assumed to satisfy a
reality condition, $\left\{ H_{\text{BdG}},P\right\} =0$. An example
is given by \eqref{eq:14}. To make a similar statement for actions,
where $\psi,\psi^{\dagger}$ are Grassmann valued, we need to clarify
how the conjugation $K$ acts on the Grassmann algebra generated $\psi,\psi^{\dagger}$.
This is defined by $K\psi=\psi^{\dagger},\;K\psi^{\dagger}=\psi$,
anti-linearity, and a reversal of the ordering of Grassmann numbers.
For example, $K\left(\psi\psi^{\dagger}\right)=K\psi^{\dagger}K\psi=\psi\psi^{\dagger}$,
$K\left(i\psi\right)=-i\psi^{\dagger}$. It is under this complex
conjugation that a fermionic action, such as \eqref{eq:10}, is ``real'',
$K\left(S_{\text{SF}}\left[\psi,\psi^{\dagger},\Delta,A\right]\right)=S_{\text{SF}}\left[\psi,\psi^{\dagger},\Delta,A\right]$,
and it is due to this reality of $S_{\text{SF}}$ that we expect to
obtain a real effective action after integrating out the fermions
\cite{wetterich2011spinors}. Then, for any action $S=\frac{1}{2}\int\text{d}^{2+1}x\Psi^{\dagger}\left(x\right)S_{\text{BdG}}\Psi\left(x\right)$,
the operator $S_{\text{BdG}}$ can then be assumed to satisfy $\left\{ S_{\text{BdG}},P\right\} =0$,
and an example is given by the Dirac operator in \eqref{43-1}.

When working with Majorana fermions it is useful to use gamma matrices
$\gamma^{a}$ that form a \textit{Majorana representation} \cite{ortin2004gravity},
which means that $\gamma^{a}$ are all imaginary . In a Majorana
representation $\Psi^{\dagger}\mathcal{B}=\Psi^{T}$ simplifies to
$\Psi^{\dagger}=\Psi^{T}$, so a Majorana spinor in a Majorana representation
has real components.  To obtain a Majorana representation from our
representation we change basis in the space of spinors using the unitary
matrix $U=\frac{1}{\sqrt{2}}\begin{pmatrix}1 & 1\\
-i & i
\end{pmatrix}$. Then $\gamma^{a}\mapsto\tilde{\gamma}^{a}=U\gamma^{a}U^{\dagger}$
and $\Psi\mapsto\tilde{\Psi}=U\Psi$. Explicitly, $\tilde{\gamma}^{0}=-\sigma^{y},\;\tilde{\gamma}^{1}=-i\sigma^{z},\;\tilde{\gamma}^{2}=-i\sigma^{x}$,
and the Nambu spinor $\Psi=\left(\psi,\psi^{\dagger}\right)^{T}$
maps to $\tilde{\Psi}=\begin{pmatrix}\tilde{\Psi}_{1}\\
\tilde{\Psi}_{2}
\end{pmatrix}=\frac{1}{\sqrt{2}}\begin{pmatrix}\psi+\psi^{\dagger}\\
\frac{1}{i}\left(\psi-\psi^{\dagger}\right)
\end{pmatrix}$, where $\tilde{\Psi}_{1},\tilde{\Psi}_{2}$ are both real as Grassmann
valued fields. As operators $\tilde{\Psi}_{1},\tilde{\Psi}_{2}$ are
hermitian and $\left\{ \tilde{\Psi}_{i},\tilde{\Psi}_{j}\right\} =\delta_{ij}$,
so they are \textit{Majorana operators} in the sense of \cite{kitaev2009periodic}.
In the Majorana representation $H_{\text{BdG}}$ is imaginary and
antisymmetric, and so is $S_{\text{BdG}}$.

\subsection{Spatial reflection and time reversal in the $p$-wave superfluid\label{subsec:Spatial-reflections-and}}

In section \ref{subsec:Symmetries,-currents,-and} we discussed the
sense in which energy, momentum, and angular momentum are conserved
in a $p$-wave SF, which followed from the symmetry of the $p$-wave
SF action under space-time translations and spatial rotations. There
are also discrete (or large) space-time transformations which are
of interest. Spatial reflections reverse the orientation of space,
and are generated by a single arbitrary reflection, which we take
to be $R:y\mapsto-y$, followed by the spatial rotations and translations
described previously. $R$ acts naturally on the fields $\psi,\Delta,A$:
\begin{align}
  \psi\left(y\right)\mapsto&\psi\left(-y\right),\label{eq:35}\\
  \left(\Delta^{x},\Delta^{y}\right)\left(y\right)\mapsto&\left(\Delta^{x},-\Delta^{y}\right)\left(-y\right),\nonumber \\
  \left(A_{t},A_{x},A_{y}\right)\left(y\right)\mapsto&\left(A_{t},A_{x},-A_{y}\right)\left(-y\right),\nonumber 
\end{align}
where we suppressed the dependence on the coordinated $t,x$ which do not transform. One can verify that $R$ is a symmetry of the $p$-wave SF action
\eqref{eq:10}. The best way to understand these transformations is
to identify the fields as space-time tensors: $\psi$ is a scalar,
$\Delta^{j}\partial_{j}$ is a vector field, and $A_{\mu}\text{d}x^{\mu}$
is a differential 1-form. The above transformation laws are then a
special case of how space-time transformations act on space-time tensors,
by the pullback/push forward. 

Time reversal transformations reverse the orientation of time, and
are generated by a single arbitrary time reversal, which we take to
be $T:t\mapsto-t$, followed by the time translations described previously.
The action of $T$ on the fields includes the transformation laws
analogous to \eqref{eq:35}, but additionally involves a complex conjugation,
as follows from the Schrodinger equation in the Fock space $i\partial_{t}\ket{\Omega\left(t\right)}=H\left(t;A,\Delta\right)\ket{\Omega\left(t\right)}$.
In our case $H\left(t;A,\Delta\right)$ is the $p$-wave SF Hamiltonian
\eqref{eq:9}, in a notation that stresses the time dependence through
the background fields. On the Fock space the complex conjugation is
the usual complex conjugation of coefficients in the position basis,
defined by $K\psi\left(x,y\right)K^{-1}=\psi\left(x,y\right),\;K\psi^{\dagger}\left(x,y\right)K^{-1}=\psi^{\dagger}\left(x,y\right),\;K\ket 0=\ket 0$
and anti-linearity. Acting with it on the $p$-wave SF Hamiltonian
\eqref{eq:9} we find that the action of $T$ on the background fields
$\Delta,A$ is 
\begin{align}
  \left(\Delta^{x},\Delta^{y}\right)\left(t\right)\mapsto&\Delta^{T}\left(t\right)=\left(\Delta^{x},\Delta^{y}\right)^{*}\left(-t\right),\label{eq:29}\\
 \left(A_{t},A_{x},A_{y}\right)\left(t\right)\mapsto& A^{T}\left(t\right)=-\left(-A_{t},A_{x},A_{y}\right)\left(-t\right),\nonumber 
\end{align}
where we suppressed the dependence on the coordinates $x,y$ which do not transform. If $\ket{\Omega\left(t\right)}$ satisfies the Schrodinger equation
with Hamiltonian $H\left(t;A,\Delta\right)$ and initial condition
$\ket{\Omega}$ then $K\ket{\Omega\left(-t\right)}$ satisfies the
Schrodinger equation with time reversed Hamiltonian $KH\left(-t;A,\Delta\right)K^{-1}=H\left(t;A^{T},\Delta^{T}\right)$
and time reversed initial state $K\ket{\Omega}$.  As a result one
obtains the following relation between expectation values of operators,
\begin{align}
 \bra{\Omega}O_{A,\Delta}\left(-t\right)\ket{\Omega}=\bra{K\Omega}\left(KOK\right)_{A^{T},\Delta^{T}}\left(t\right)\ket{K\Omega}.\label{eq:30-1}
\end{align}
Here $O$ is a Schrodinger operator considered as an operator at
time $t=0$, and $O_{A,\Delta}\left(t\right)$ is its time evolution
using $H\left(t;A,\Delta\right)$. $\ket{K\Omega}=K\ket{\Omega}$
is the time reversed state, and $KOK$ is the time reversed Schrodinger
operator.   

 To describe how time reversal acts on the action, we need to use
the complex conjugation $K$ on the Grassmann algebra, described in
\ref{subsec:Charge-conjugation-(Appendix)}. We then define the action
of time reversal on the Grassmann fields $\psi,\psi^{\dagger}$ as
the analog of \eqref{eq:35}, but with an additional conjugation by
$K$,
\begin{align}
  \psi\left(t,x,y\right)\mapsto&\psi^{T}\left(t,x,y\right)=\psi^{\dagger}\left(-t,x,y\right),\label{eq:30}\\
  \psi^{\dagger}\left(t,x,y\right)\mapsto&\left(\psi^{\dagger}\right)^{T}\left(t,x,y\right)=\psi\left(-t,x,y\right).\nonumber 
\end{align}
Using the transformations \eqref{eq:29},\eqref{eq:30} and the ``reality''
of the action \eqref{eq:10} one finds
\begin{align}
  S_{\text{SF}}\left[\psi^{T},\left(\psi^{\dagger}\right)^{T},\Delta^{T},A^{T}\right]=&-K\left(S_{\text{SF}}\left[\psi,\psi^{\dagger},\Delta,A\right]\right)\nonumber\\
  =&-S_{\text{SF}}\left[\psi,\psi^{\dagger},\Delta,A\right],
\end{align}
so that up to a sign, time reversal is a symmetry of the action. 
It was shown in \cite{wetterich2011spinors} that, at least formally,
this sign does not effect the value of the fermionic functional integral,
and can therefore be ignored. Then time reversal symmetry defined
by \eqref{eq:29}, \eqref{eq:30} can be regraded as a symmetry of
the action in the usual sense, and one can use this fact to derive
\eqref{eq:30-1} using functional integrals.  

\subsection{Spatial reflection and time reversal in the geometric description\label{subsec:relativisitc Spatial-reflection-and}}

In this section we map and slightly generalize $R,T$, as defined
in appendix \ref{subsec:Spatial-reflections-and}, to the geometric
description of the $p$-wave SC in terms of a Majorana spinor in RC
space, given in section \ref{sec:Emergent-Riemann-Cartan-geometry}.
We will see that there is a difference between the standard notion
of $R,T$ for a spinor in $2+1$ dimensions \cite{witten2015fermion}
and the notion of $R,T$ for the $p$-wave SC, described in appendix
\ref{subsec:Spatial-reflections-and}. The reason is that our mapping
of the $p$-wave SC to a Majorana spinor maps charge to spin, and
charge is $R,T$-even, while spin is $R,T$-odd. This is a general
property of the BdG formalism. The main point is that the physical
$R,T$, coming from the $p$-wave SC, leave the mass $m$ invariant
and flip the orientation $o$, as opposed to the standard $R,T$ for
a spinor in $2+1$ dimensions, which map $m\mapsto-m$ and leave $o$
invariant. Thus, the contribution $\frac{1}{2}o\cdot\text{sgn}\left(m\right)$
of a single Majorana spinor to the Chern number is $R,T$-odd under
both notions of $R,T$, but for different reasons. 

First, by spatial reflection we mean an element of the Diffeomorphism
group that reverses the orientation of space but not of time, and
not to an internal Lorentz transformation. Since the composition
of any spatial reflection with $Diff_{0}$ is again a spatial reflection,
it suffices to consider a single spatial reflection $R$. Since spatial
reflections are just diffeomorphisms, their action on the fields is
already defined by \eqref{eq:51}, which is just the pullback 
\begin{eqnarray}
 &  & \chi\mapsto R^{*}\chi,\;e^{a}\mapsto R^{*}e^{a},\;\omega\mapsto R^{*}\omega,\label{eq:67}
\end{eqnarray}
and is a symmetry of the action $S_{\text{RC}}$. If space-time is
$\mathbb{R}_{t}\times\mathbb{R}^{2}$ it suffices to consider $R:\;y\mapsto-y$,
as was done in appendix \ref{subsec:Spatial-reflections-and}. Then
\eqref{eq:67} takes the explicit form 
\begin{align}
  \chi\left(y\right)\mapsto&\chi\left(-y\right),\label{eq:51-1}\\
  \left(e_{\;t}^{a},e_{\;x}^{a},e_{\;y}^{a}\right)\left(y\right)\mapsto&\left(e_{\;t}^{a},e_{\;x}^{a},-e_{\;y}^{a}\right)\left(-y\right),\nonumber \\
 \left(\omega_{t},\omega_{x},\omega_{y}\right)\left(y\right)\mapsto&\left(\omega_{t},\omega_{x},-\omega_{y}\right)\left(-y\right),\nonumber 
\end{align}
which maps to the transformation laws \eqref{eq:35} of the $p$-wave
SF. The orientation of space-time $o=\text{sgn}\left(\text{det}e\right)$
is odd under spatial reflections, like the orientation of space.
Note that even the flat vielbein $e_{\;\mu}^{a}=\delta_{\mu}^{a}$
transforms under $R$, which corresponds to the mapping of a $p_{x}+ip_{y}$
order parameter to a $p_{x}-ip_{y}$ order parameter by $R$.

A time reversal is any diffeomorphism that reverses the orientation
of time but not of space. It suffices to consider a single representative,
and since we work with space-times of the form $\mathbb{R}_{t}\times M_{2}$
we may take $\tau:t\mapsto-t$. Apart from the pullback by $\tau$
analogous to \eqref{eq:51-1}, $T$ also includes additional ``external''
transformations of the fields, which all trace back to the complex
conjugation included in the time reversal operation in quantum mechanics,
as in appendix \ref{subsec:Spatial-reflections-and}. As reviewed
in appendix \ref{subsec:Charge-conjugation-(Appendix)}, a complex
conjugation of the gamma matrices is implemented by $-\left(\gamma^{a}\right)^{*}=\mathcal{B}\gamma^{a}\mathcal{B}^{-1}=\mathcal{B}_{\;b}^{a}\gamma^{b}$
where $\mathcal{B}=\sigma^{x}$, $\mathcal{B}_{\;b}^{a}=\text{diag}\left[-1,1-1\right]$
in our representation. We then define the action of $T$ on the fields
by 
\begin{align}
  \chi\mapsto K\left(\tau^{*}\chi\right),\;e^{a}\mapsto\mathcal{B}_{\;b}^{a}\tau^{*}e^{b},\;\omega\mapsto\mathcal{B}\left(\tau^{*}\omega\right)\mathcal{B}^{-1},
\end{align}
where $\tau^{*}$ is the pullback by $\tau$, and $K$ is the complex
conjugation of the Grassmann algebra defined in appendix \ref{subsec:Charge-conjugation-(Appendix)}.
One can check that this is a symmetry of the action $S_{\text{RC}}$
up to an irrelevant sign already explained in appendix \ref{subsec:Spatial-reflections-and},
and that this action of $T$ reduces to the transformation laws \eqref{eq:29}
and \eqref{eq:30} of the $p$-wave SF fields.

The standard time reversal for spinors in 2+1 dimensions is given
by $T_{\text{s}}=i\sigma^{y}K$, where the phase $i$ is a matter
of convention. It is anti-unitary and $T_{\text{s}}^{2}=-1$. This
is related to $T$ through the charge conjugation matrix defined in
\ref{subsec:Charge-conjugation-(Appendix)}, 
\begin{eqnarray}
 &  & T_{\text{s}}=i\mathcal{C}T\text{ or }T=-i\mathcal{C}T_{\text{s}}.
\end{eqnarray}
This relates the time reversal $T$ that is natural in this paper,
to the standard time reversal $T_{\text{s}}$ and standard charge
conjugation $\mathcal{C}$. 

\textcolor{red}{}

\section{Global structures and obstructions\label{subsec:Global-structures-and}}

We already described the emergent geometry in a $p$-wave SC  locally
in section \ref{sec:Emergent-Riemann-Cartan-geometry}. Here we complete
the description by considering global aspects. We use some elements
from the theory of fiber bundles and characteristic classes, which
are reviewed in \cite{friedrich2000dirac,nakahara2003geometry} for
example. 

We work with space-time manifolds of the form $M_{3}=\mathbb{R}_{t}\times M_{2}$,
which represent the world volume of the $p$-wave SF. $M_{2}$ is
the sample, the two dimensional spatial surface occupied by the $p$-wave
SF, and $\mathbb{R}_{t}$ is the real line parameterizing time. Because
the order parameter is locally a vector $\Delta^{j}$ with $U\left(1\right)$
charge 2, at any time $t\in\mathbb{R}_{t}$ it is globally a map between
vector bundles $\Delta:T^{*}M_{2}\rightarrow E^{2}$, that acts by
$v_{j}\mapsto\Delta^{j}v_{j}$ \footnote{Equivalently, $\Delta$ is globally a section of $TM_{2}\otimes E^{2}$.}.
Here $T^{*}M_{2}$ is the co-tangent bundle of the sample $M_{2}$
and $E^{2}$ is the square of the electromagnetic $U\left(1\right)$
vector bundle. $E$ has fibers $\mathbb{C}$ and $U\left(1\right)$-valued
transition functions, and its topology is labeled by the monopole
number (first Chern number)  $\Phi=\frac{1}{2\pi}\int_{M_{2}}F\in\mathbb{Z}$
if $M_{2}$ has no boundary, and it is otherwise trivial. $E^{2}$
is obtained from $E$ by replacing every transition function by its
square, and therefore the topology of $E^{2}$ is labeled by $2\Phi\in2\mathbb{Z}$.
If $M_{2}$ has no boundary, the topology of the tangent bundle $TM_{2}$
(and that of $T^{*}M_{2}$) is labeled by the Euler characteristic
$\chi=2\left(1-g\right)\in2\mathbb{Z}$ where $g$ is the genus of
$M_{2}$. 

As a map $\Delta:T^{*}M_{2}\rightarrow E^{2}$, if $\Delta$ is non
singular in the sense of section \ref{subsec:The-order-parameter}
($\text{det}e\neq0$), it defines three geometric structures on $M_{2}$:
a metric, which is $g^{ij}$, an orientation, $o=\text{sgn}\left(\text{det}e\right)$,
and a spin structure, which follows from the fact that $\Delta$ has
charge 2.

To see this, we can think of $E^{2}$ as an $SO\left(2\right)$ vector
bundle, with fibers $\mathbb{R}^{2}$ and $SO\left(2\right)$ valued
transition functions. The map $\Delta$ then gives a reduction of
the structure group of $T^{*}M_{2}$ from $GL\left(2\right)$ to $SO\left(2\right)$,
thus defining a metric and an orientation. Since the transition functions
of $E^{2}$ are obtained by squaring the transition functions of $E$,
it is natural to think of $E$ as a $Spin\left(2\right)$ vector bundle\footnote{Both $Spin\left(2\right)$ and $SO\left(2\right)$ are isomorphic
as Lie groups to $U\left(1\right)$, but are related by the double
cover $Spin\left(2\right)\ni e^{i\alpha}\mapsto e^{2i\alpha}\in SO\left(2\right)$.}. $E^{2}$ therefore naturally carries a spin structure \cite{nakahara2003geometry},
and the mapping $\Delta:T^{*}M_{2}\rightarrow E^{2}$ then endows
$M_{2}$ with a spin structure. 

 The different possible spin structures correspond to an assignment
of signs $\pm1$ to non contractible loops in $M_{2}$, or more precisely
to elements of $H^{1}\left(M_{2},\mathbb{Z}_{2}\right)$. Generally,
this identification of spin structures with $H^{1}\left(M_{2},\mathbb{Z}_{2}\right)$
is not canonical, which means that there is no natural way to declare
one of the spin structures as ``trivial''.

In the simple case where $TM_{2}$ is trivial as in the case of the
torus $M_{2}=\mathbb{R}^{2}/\mathbb{Z}^{2}$, spin structures correspond
canonically to elements of $H^{1}\left(M_{2},\mathbb{Z}_{2}\right)$,
which in turn correspond to a choice of periodic or anti-periodic
boundary conditions for spinors around the non contractible loops.
The boundary condition for the BdG spinor $\Psi=\left(\psi,\psi^{\dagger}\right)^{T}$
follows from that of the microscopic spin-less fermion $\psi$, for
which it is natural to take fully periodic boundary conditions, which
is the ``trivial'' spin structure. Other boundary conditions have
been discussed in \cite{read2000paired,read2009non}.

A non singular $\Delta$ is not always possible. First, it requires
that $M_{2}$ be orientable. If $M_{2}$ is not orientable $\Delta$
would have singularities $sing\left(\Delta\right)$ such that $M_{2}-sing\left(\Delta\right)$
is orientable. $p$-wave SF on non orientable surfaces were considered
in \cite{quelle2016edge}. The other obstruction is a mismatch in
the topology of $E^{2}$ and $TM_{2}$, and is given by $2\Phi+o\chi$,
or $\Phi-\left(g-1\right)o$ \cite{read2000paired}. If the topological
invariant $2\Phi+o\chi$ does not vanish then $\Delta$ must have
singularities. A simple way to obtain this condition is to assume
$\tilde{\omega}_{12\mu}=\omega_{12\mu}=-2A_{\mu}$, which implies
$\frac{1}{2}o\sqrt{g}\tilde{\mathcal{R}}\text{d}^{2}x=\tilde{R}_{12}=\text{d}\tilde{\omega}_{12}=-2\text{d}A=-2F$,
and use the Gauss-Bonet formula $\chi=2\left(1-g\right)=\frac{1}{4\pi}\int_{M_{2}}\tilde{\mathcal{R}}\sqrt{g}\mbox{d}^{2}x$
for the Euler characteristic. The simplest example is $M_{2}=S^{2}$
the sphere, where there must be a monopole $\Phi=o=\pm1$ for a non
singular order parameter with orientation $o$. Possible singularities
of the order parameter on the sphere without a monopole have been
studied in \cite{moroz2016chiral}. There are no obstructions to the
existence of a metric and (in the two dimensional case) of a spin
structure. 

A simple way to handle singularities of $\Delta$ is to exclude them
by working with $M_{2}-sing\left(\Delta\right)$ instead of $M_{2}$.
Then $\Delta$ defines on $M_{2}-sing\left(\Delta\right)$ and orientation,
metric, and spin structure.

The emergent geometry of space-time follows from that of space due
to the simple product structure $M_{3}=\mathbb{R}_{t}\times M_{2}$.
Thus the order parameter corresponds to the (inverse) space-time vielbein
\eqref{17}, which is globally a map $T^{*}M_{3}\rightarrow E^{2},\;v_{\mu}\mapsto e_{a}^{\;\mu}v_{\mu}$
where $E^{2}$ is now viewed as an $SO\left(1,2\right)$ vector bundle.
In other words, $e$ is globally a Solder form.

\section{Quantization of coefficients for a sum of gravitational Chern-Simons terms\label{subsec:quntization-of-coefficients}}

As stated in section \ref{subsec:quantization}, gauge invariance of 
\begin{align}
K=\alpha_{1}\int_{M_{3}}Q_{3}\left(\tilde{\omega}_{\left(1\right)}\right)+\alpha_{2}\int_{M_{3}}Q_{3}\left(\tilde{\omega}_{\left(2\right)}\right)
\end{align}
for all closed $M_{3}$ implies $\alpha_{1}+\alpha_{2}\in\frac{1}{192\pi}\mathbb{Z}$. Here we sketch the derivation, following \cite{witten2007three} (section 2.1 and the discussion leading to equation (2.27)). First, only the gauge invariance of $e^{iK}$ is required, because $K$ is a contribution to the effective action, obtained by taking the logarithm of the fermionic path integral, which is a gauge invariant object. Second, the gCS term on a general $M_{3}$ is only locally given by $\alpha\int Q_{3}\left(\tilde{\omega}\right)$,
not globally. It is convenient to globally define gCS on a given
$M_{3}$ as $\alpha\int_{M_{4}}\text{tr}\left(\tilde{R}^{2}\right)$,
where $M_{4}$ is some four manifold with $M_{3}$ as a boundary,
$\partial M_{4}=M_{3}$. This is based on the fact that locally on
$M_{4}$ we have $\text{d}Q_{3}\left(\tilde{\omega}\right)=\text{tr}\left(\tilde{R}^{2}\right)$.
With this definition, we have 
\begin{align}
e^{iK_{M_{4}}}=e^{i\left[\alpha_{1}\int_{M_{4}}\text{tr}\left(\tilde{R}_{\left(1\right)}^{2}\right)+\alpha_{2}\int_{M_{4}}\text{tr}\left(\tilde{R}_{\left(2\right)}^{2}\right)\right]},
\end{align}
which is clearly gauge invariant, but we must ensure that it is also
independent of the arbitrary choice of $M_{4}$. In fact, changing
$M_{4}$ corresponds precisely to performing a large gauge transformation
on $M_{3}$, see \cite{deser1998definition} for a more direct approach.
For $M_{4}\neq M_{4}'$ such that $\partial M_{4}=M_{3}=\partial M_{4}'$,
we have 
\begin{align}
e^{iK_{M_{4}}}/e^{iK_{M_{4}'}}=e^{i\left[\alpha_{1}\int_{X_{4}}\text{tr}\left(\tilde{R}_{\left(1\right)}^{2}\right)+\alpha_{2}\int_{X_{4}}\text{tr}\left(\tilde{R}_{\left(2\right)}^{2}\right)\right]},
\end{align}
where $X_{4}$ is a closed manifold obtained by gluing $M_{4},M_{4}'$
along their shared boundary, after reversing the orientation on $M_{4}'$.
Since we start with a spin manifold $M_{3}$, we assume that $M_{4},M_{4}'$
are also spin manifolds, and therefore so is $X_{4}$. On the closed spin manifold
$X_{4}$, the Atiah-Singer index theorem implies 
\begin{align}
 \int_{X_{4}}\text{tr}\left(\tilde{R}_{\left(1\right)}^{2}\right)=\int_{X_{4}}\text{tr}\left(\tilde{R}_{\left(2\right)}^{2}\right)\in2\pi\times192\pi\mathbb{Z}.
\end{align}
In particular, one can choose $M_{4}'$ such that the integer on the right hand side is 1, in which case 
\begin{align}
 e^{iK_{M_{4}}}/e^{iK_{M_{4}'}}=e^{2\pi i\left(\alpha_{1}+\alpha_{2}\right)192\pi}.
\end{align}
An $M_{4}$-independent $e^{iK_{M_{4}}}=e^{iK}$ then requires $\alpha_{1}+\alpha_{2}\in\frac{1}{192\pi}\mathbb{Z}$.

\section{Calculation of gravitational pseudo Chern-Simons currents\label{subsec:Calculation-of-certain}}

Here we derive the contributions \eqref{eq:92-1} to the bulk currents,
which come from the gpCS term $-\beta_{1}\int_{M_{3}}\tilde{\mathcal{R}}e^{a}De_{a}$
in the effective action. We write
\begin{align}
  \delta\int_{M_{3}}\tilde{\mathcal{R}}e^{a}De_{a}=\int_{M_{3}}\left(e^{a}De_{a}\right)\delta\tilde{\mathcal{R}}+\int_{M_{3}}\tilde{\mathcal{R}}\delta\left(e^{a}De_{a}\right).\label{eq:171}
\end{align}
It's convenient to calculate the first contribution in terms of scalars
using $e^{a}De_{a}=-oc\left|e\right|\text{d}^{3}x$. We need the formula
$\delta\tilde{\mathcal{R}}=-\delta g_{\mu\nu}\tilde{\mathcal{R}}^{\mu\nu}+\left(\nabla_{\mu}\nabla_{\nu}-g_{\mu\nu}\nabla^{2}\right)\delta g_{\mu\nu}$
relating the curvature variation to the metric variation, and $\delta g_{\mu\nu}=2e_{a(\nu}\delta e_{\;\mu)}^{a}$
relating the metric variation to the vielbein variation. We find 
\begin{align}
 \int_{M_{3}}&\left(e^{a}De_{a}\right)\delta\tilde{\mathcal{R}}\\
 =&-2o\int_{M_{3}}\left|e\right|\left[\left(\tilde{\nabla}^{\mu}\tilde{\nabla}^{\nu}-g^{\mu\nu}\tilde{\nabla}^{2}\right)c-\tilde{\mathcal{R}}^{\mu\nu}c\right]e_{a\nu}\delta e_{\;\mu}^{a}.\nonumber
\end{align}
The second contribution in \eqref{eq:171} is simpler to calculate
in terms of differential forms \cite{hughes2013torsional}, 
\begin{align}
  \delta\int\tilde{\mathcal{R}}&e_{a}De^{a}\label{eq:394-1-1}\\
 =&\int_{M}\tilde{\mathcal{R}}\left(\delta e^{a}T_{a}+e^{a}\mbox{d}\delta e_{a}+e^{a}\delta\omega_{ab}e^{b}+e^{a}\omega_{ab}\delta e^{b}\right)\nonumber \\
 =&\int_{M}\left(\tilde{\mathcal{R}}2\delta e_{a}T^{a}-\tilde{\mathcal{R}}\delta\omega_{ab}e^{a}e^{b}-\delta e_{a}e^{a}\mbox{d}\tilde{\mathcal{R}}\right)\nonumber\\
 &+\int_{\partial M}\delta e_{a}\tilde{\mathcal{R}}e^{a},\nonumber 
\end{align}
which implies 
\begin{align}
 *&\mathsf{J}^{a}=-\beta_{1}\left(2\tilde{\mathcal{R}}T^{a}-e^{a}\mbox{d}\tilde{\mathcal{R}}\right),\;*\mathsf{J}^{ab}=-\beta_{1}\left(-\tilde{\mathcal{R}}e^{a}e^{b}\right),\nonumber\\
*&\mathsf{j}^{a}=-\beta_{1}\tilde{\mathcal{R}}e^{a},\;*\mathsf{j}^{ab}=0. 
\end{align}
Here we kept track of boundary terms and calculated the contributions
to boundary currents $\mathsf{j}^{a}=\mathsf{j}_{\mu}^{\;a}\mbox{d}x^{\mu},\;\mathsf{j}^{ab}=\mathsf{j}_{\;\;\mu}^{ab}\text{d}x^{\mu}$, which are relevant for our discussion in section \ref{subsec:Additional-contributions}.
Collecting all of the bulk contributions one finds \eqref{eq:92-1}.

In section \ref{subsec:Additional-contributions} we wrote down \eqref{eq:92-1}
for a product geometry with respect to the coordinate $z$, and assumed
torsion vanishes. Here we generalize to non-zero torsion. With non
zero torsion, \eqref{eq:94-1} generalizes to 
\begin{align}
  \left\langle \mathsf{J}^{\alpha z}\right\rangle _{\text{gpCS}}=&-\beta_{1}\frac{1}{\left|e\right|}\varepsilon^{z\alpha\beta}\partial_{\beta}\tilde{\mathcal{R}},\label{eq:11-2-2}\\
 \left\langle \mathsf{J}^{z\alpha}\right\rangle _{\text{gpCS}}=&\beta_{1}\left[\frac{1}{\left|e\right|}\varepsilon^{z\alpha\beta}\partial_{\beta}\tilde{\mathcal{R}}+\frac{1}{\left|e\right|}\varepsilon^{z\beta\gamma}C_{\beta\gamma}^{\;\;\alpha}\tilde{\mathcal{R}}\right].\nonumber 
\end{align}
For $z=t$, which describes a time independent situation, we find
\begin{align}
  \left\langle J_{E}^{i}\right\rangle _{\text{gpCS}}=&\beta_{1}\varepsilon^{ij}\partial_{j}\tilde{\mathcal{R}},\label{eq:11-2-3}\\
 \left\langle P_{i}\right\rangle _{\text{gpCS}}=&-\beta_{1}\left[g_{ik}\varepsilon^{kj}\partial_{j}\tilde{\mathcal{R}}+2o\left|e\right|\tilde{\mathcal{R}}C_{12i}\right],\nonumber 
\end{align}
which generalizes \eqref{eq:92-2}. Explicit expressions for the contorsion
$C_{12i}$ are given in appendix \ref{subsec:Explicit-formulas-for}.
Equation \eqref{eq:11-2-3} is compatible with the operator equation
\eqref{eq:49}, and the density response \eqref{eq:98}.

In the case $z=y$, the inflow \eqref{eq:104-1} generalizes to
\begin{align}
  \left\langle t_{\text{cov}\;\alpha}^{y}\right\rangle _{\text{gpCS}}&=-\left|e\right|\left\langle \mathsf{J}_{\;\alpha}^{y}\right\rangle _{\text{gpCS}}\\
  &=-\beta_{1}\left[g_{\alpha\beta}\varepsilon^{\beta\gamma y}\partial_{\gamma}\tilde{\mathcal{R}}+2o\left|e\right|C_{01\alpha}\tilde{\mathcal{R}}\right].\nonumber
\end{align}
For the order parameter $\Delta=\Delta_{0}e^{i\theta\left(t,x\right)}\left(1+f\left(t,x\right),\pm i\right)$
that we consider in this case, we find using appendix \ref{subsec:Explicit-formulas-for}
that $C_{01t}=0,\;C_{01x}=e_{1}^{\;i}\frac{1}{2}\partial_{t}g_{ij}$.
The boundary current \eqref{eq:98-3} is unchanged, but the bulk+boundary
conservation equation \eqref{eq:99} is generalized to 
\begin{align}
  \tilde{\nabla}_{\alpha}\left\langle \mathsf{j}_{\;\beta}^{\alpha}\right\rangle _{\text{gpCS}}-C_{ab\beta}\left\langle \mathsf{j}^{[ab]}\right\rangle _{\text{gpCS}}=\left\langle \mathsf{J}^{y\beta}\right\rangle _{\text{gpCS}},
\end{align}
so that bulk+boundary conservation still holds for the current from
gpCS, in the presence of torsion.

\section{Perturbative calculation of the effective action \label{subsec:Perturbative-calculation-of}}

Here we present a perturbative calculation of the effective action
for the RC background fields $e,\omega$ induced by a Majorana spinor
in 2+1 dimensions. A perturbative calculation requires three types
of input: free propagators, interaction vertices, and a renormalization
scheme to handle UV divergences. In our case the propagator and vertices
are standard in the context of the coupling of relativistic fermions
to gravity, but the renormalization scheme will not be standard in
this context.

The standard renormalization schemes used in the literature are aimed
at preserving Lorentz symmetry, obtaining properly quantized coefficients
for CS terms, and obtaining finite results that do not depend on a
regulator \cite{hughes2013torsional,parrikar2014torsion}.
This is usually done as follows. First, one introduces a Lorentz invariant
regulator, such as a frequency and wave-vector cutoff $\Lambda_{\text{rel}}$,
then one introduces Pauli-Villars regulators, and tunes their masses
such that the limit $\Lambda_{\text{rel}}\rightarrow\infty$ produces
finite results and properly quantized CS coefficients. 

In contrast, we take the lattice model \eqref{eq:2-1} as a microscopic
description of the $p$-wave SC, and the relativistic continuum limit
as an approximation of it. As we obtained naturally in sections \ref{subsec:Band-structure-and}
and \ref{subsec:Coupling-the-Lattice}, this means that there are
four Majorana spinors, with different orientations and masses, and
a wave-vector cutoff $\Lambda_{UV}\sim1/a$, but no frequency cutoff,
as dictated by the lattice model. Note that these multiple Majorana
spinors are not Pauli-Villars regulators, simply because they are
all fermions. None of them has the ``wrong statistics''. The cutoff
$\Lambda_{UV}$ is a physical parameter of the model and we do not
wish to take it to infinity. Thus wave-vector integrals cannot diverge.
In contrast, since time is continuous, there is no physical frequency
cutoff, and divergences in frequency integrals do appear. These divergences
are unphysical, and can be viewed as a byproduct of the construction
of the path integral by time discretization. These divergences need
to be renormalized in the usual sense, and we do this by minimal subtraction.


To set up the perturbative calculation, we write the action $S_{\text{RC}}$
in terms of the spinor densities $\Psi=\left|e\right|^{1/2}\chi$,
and using the explicit form \eqref{eq:40}, 
\begin{align}
  S_{\text{RC}}  =\frac{1}{2}\int\mbox{d}^{3}x\overline{\Psi}&\left[\frac{1}{2}ie_{a}^{\;\mu}\gamma^{a}\overleftrightarrow{\partial_{\mu}}-\frac{1}{4}\omega_{abc}\varepsilon^{abc}-m\right]\Psi&\nonumber\\
   =\frac{1}{2}\int\mbox{d}^{3}x\overline{\Psi}&\left[ie_{a}^{\;\mu}\gamma^{a}\partial_{\mu}+\frac{i}{2}\left(\partial_{\mu}e_{a}^{\;\mu}\right)\gamma^{a}\right.\nonumber\\
 &\left.-\frac{1}{4}\omega_{abc}\varepsilon^{abc}-m\right]\Psi.\label{eq:40-1} 
\end{align}
Assuming for now that the vielbein has a positive orientation, we
insert $e_{a}^{\;\mu}=\delta_{a}^{\mu}+h_{a}^{\;\mu}$ with small
$h$, and split the action into an inverse  propagator $G^{-1}$ and vertices $V$,
\begin{align}
  S_{\text{RC}}=&\frac{1}{2}\int\mbox{d}^{3}x\Psi^{\dagger}\gamma^{0}\left[G^{-1}+V\right]\Psi,\\
  G^{-1}=&i\delta_{a}^{\;\mu}\gamma^{a}\partial_{\mu}-m,\;V=V_{1}+V_{2},\nonumber \\
  V_{1}=&i\gamma^{a}h_{a}^{\;\mu}\partial_{\mu}+\frac{i}{2}\gamma^{a}\left(\partial_{\mu}h_{a}^{\;\mu}\right),\;V_{2}=-\frac{1}{4}\omega_{abc}\varepsilon^{abc}.\nonumber 
\end{align}
The vertex $V_{1}$ is first order in the perturbation $h$. The
vertex $V_{2}$ is given explicitly by 
\begin{align}
  V_{2}=-\frac{1}{4}\omega_{ab\mu}e_{c}^{\;\mu}\varepsilon^{abc}=-\frac{1}{4}\omega_{ab\mu}\delta_{c}^{\mu}\varepsilon^{abc}-\frac{1}{4}\omega_{ab\mu}h_{c}^{\;\mu}\varepsilon^{abc},
\end{align}
and therefore contains a term of order $\omega$ and a term quadratic
in the perturbations, of order $h\omega$. Terms in vertices which
are nonlinear in perturbations are sometimes called contact terms,
and the above contribution to $V_{2}$ is the only contact term in
our scheme. Note that there is no vertex related to the volume element
$\left|e\right|$, because the fundamental fermionic degree of freedom
is the spinor density $\Psi$, see appendix \ref{subsec:Equality-of-path}.
In expressions written in terms of $h_{a}^{\;\mu}$ we use $\eta_{\mu\nu}$
to raise and lower coordinate indices and $\delta_{a}^{\mu}$ to map
internal indices to coordinate indices, so in practice there is no
difference between these indices in such expressions. 

The perturbative expansion of the effective action is given by 
\begin{align}
  2W_{\text{RC}}  =&-2i\log\mbox{Pf}\left(i\gamma^{0}\left(G^{-1}+V\right)\right)\\
  =&-i\mbox{Tr}\left(\log i\gamma^{0}G^{-1}\right)\nonumber\\
  &-i\mbox{Tr}\left(GV\right)+\frac{i}{2}\mbox{Tr}\left(GV\right)^{2}+O\left(V^{3}\right),\nonumber 
\end{align}
which, apart from the first term, is a sum over Feynman diagrams with
a fermion loop and any number of vertices $V$. We will be interested
in $W_{\text{RC}}$ to second order in the perturbations $h$ and
$\omega$ and up to third order in derivatives. Terms of first oder
in $h,\omega$ correspond to properties of the unperturbed ground
state, or vacuum, while terms of second order correspond to linear
response coefficients. The first term is independent of $h,\omega$
and corresponds to the ground state energy of the unperturbed system.
This information can also be obtained from the term linear in $h$,
and we therefore ignore $\mbox{Tr}\log i\gamma^{0}G^{-1}$ in the
following. Expanding the vertices, 
\begin{align}
  2W_{\text{RC}}=&-i\mbox{Tr}\left(GV_{1}\right)-i\mbox{Tr}\left(GV_{2}\right) +\frac{i}{2}\mbox{Tr}\left(GV_{1}\right)^{2}\label{eq:80}\\
 &+\frac{i}{2}\mbox{Tr}\left(GV_{2}\right)^{2}+i\mbox{Tr}\left(GV_{1}GV_{2}\right)+O\left(V^{3}\right).\nonumber 
\end{align}
These functional traces can now be written as integrals over Fourier
components and traces over spinor indices, 
\begin{widetext}
\begin{align}
 \mbox{Tr}\left(GV_{1}\right)=&-h_{a}^{\;\mu}\left(p=0\right)\int_{q}q_{\mu}\mbox{tr}\left(\gamma^{a}G_{q}\right),\\
  \mbox{Tr}\left(GV_{2}\right)=&\omega\left(p=0\right)\int_{q}\mbox{tr}\left(G_{q}\right),\nonumber \\
  \mbox{Tr}\left(GV_{1}\right)^{2}=&\int_{p}h_{a}^{\;\mu}\left(p\right)h_{b}^{\;\nu}\left(-p\right)\int_{q}\left(q+\frac{1}{2}p\right)_{\mu}\left(q+\frac{1}{2}p\right)_{\nu}\text{tr}\left(\gamma^{a}G_{q}\gamma^{b}G_{p+q}\right),\nonumber \\
  \mbox{Tr}\left(GV_{2}\right)^{2}=&\int_{p}\omega\left(p\right)\omega\left(-p\right)\int_{q}\text{tr}\left(G_{q}G_{p+q}\right),\nonumber\\
  \mbox{Tr}\left(GV_{1}GV_{2}\right)=&-\int_{p}h_{a}^{\;\mu}\left(p\right)\omega\left(-p\right)\int_{q}\left(q+\frac{1}{2}p\right)_{\mu}\text{tr}\left(\gamma^{a}G_{q}G_{p+q}\right),\nonumber 
\end{align}
\end{widetext}
where $\omega=-\frac{1}{4}\omega_{ab\mu}e_{c}^{\;\mu}\varepsilon^{abc}$,
and $\int_{p}=\int\frac{\text{d}^{3}p}{\left(2\pi\right)^{3}}$. Our
conventions for the Fourier transform of a function $f$ is $f\left(x\right)=\int_{q}e^{iq_{\mu}x^{\mu}}f\left(q\right)$.
The Fourier transform of the Greens function is then $G_{q}=-\frac{1}{\fsl{q}+m}=-\frac{\fsl{q}-m}{q^{2}-m^{2}}$.
The spinor traces are evaluated using the usual identities for gamma
matrices in 2+1 dimensions, 
\begin{align}
  \text{tr}&\left(\gamma^{a}\right)=0,\;\text{tr}\left(\gamma^{a}\gamma^{b}\right)=2\eta^{ab},\;\text{tr}\left(\gamma^{a}\gamma^{b}\gamma^{c}\right)=\pm2i\varepsilon^{abc},\nonumber\\
  \text{tr}&\left(\gamma^{a}\gamma^{b}\gamma^{c}\gamma^{d}\right)=2\left(\eta^{ab}\eta^{cd}-\eta^{ac}\eta^{bd}+\eta^{ad}\eta^{bc}\right).\label{eq:177} 
\end{align}
The sign $\pm$ distinguishes the two inequivalent representations
of gamma matrices in 2+1 dimensions, and with our chosen representation,
$\text{tr}\left(\gamma^{a}\gamma^{b}\gamma^{c}\right)=2i\varepsilon^{abc}$.
Using these identities yields for the single vertex diagrams 
\begin{eqnarray}
 &  & \mbox{Tr}\left(GV_{1}\right)=2\eta^{ab}h_{a}^{\;\mu}\left(p=0\right)\int_{q}\frac{q_{\mu}q_{b}}{q^{2}-m^{2}},\label{eq:83-1}\\
 &  & \mbox{Tr}\left(GV_{2}\right)=2m\omega\left(p=0\right)\int_{q}\frac{1}{q^{2}-m^{2}}.\nonumber 
\end{eqnarray}
The expressions for the diagrams with two vertices are more complicated,
so let us start by analyzing the single vertex diagrams. This will
suffice to demonstrate our renormalization scheme and compare it to
direct calculations within the lattice model and to renormalizations
which are more natural in the context of relativistic QFT.

\subsection{Single vertex diagrams }

From \eqref{eq:80} and \eqref{eq:83-1} it follows that 
\begin{align}
 W_{\text{RC}}=\Lambda_{\;\mu}^{a}\int\text{d}^{3}xh_{a}^{\;\mu}+s\int\text{d}^{3}x\omega+O\left(V^{2}\right),
\end{align}
where
\begin{align}
  \Lambda_{\;\mu}^{a}=-i\eta^{ab}\int\frac{\text{d}^{3}q}{\left(2\pi\right)^{3}}\frac{q_{\mu}q_{b}}{q^{2}-m^{2}},\;s=-i\int\frac{\text{d}^{3}q}{\left(2\pi\right)^{3}}\frac{m}{q^{2}-m^{2}},
\end{align}
can now be recognized as the energy-momentum tensor and spin density
of the unperturbed ground state, 
\begin{align}
    \left\langle \mathsf{J}_{\;\mu}^{a}\right\rangle =&-\Lambda_{\;\mu}^{a},\\
\left\langle \mathsf{J}^{ab\mu}\right\rangle =&-\frac{1}{4}\frac{1}{2}\left\langle \overline{\chi}\chi\right\rangle \delta_{c}^{\mu}\varepsilon^{abc}=-\frac{1}{4}s\delta_{c}^{\mu}\varepsilon^{abc}.\nonumber
\end{align}
 Preforming a Wick rotation $q_{0}\mapsto iq_{0}$, 
\begin{align}
  \Lambda_{\;\mu}^{a}=\delta^{ab}\int\frac{\text{d}^{3}q}{\left(2\pi\right)^{3}}\frac{q_{\mu}q_{b}}{\left|q\right|^{2}+m^{2}},\;s=-\int\frac{\text{d}^{3}q}{\left(2\pi\right)^{3}}\frac{m}{\left|q\right|^{2}+m^{2}},\label{eq:181}
\end{align}
where $\left|\cdot\right|$ is the euclidian norm. We start by calculating
$s$ in our lattice motivated renormalization scheme. In this scheme
the integral reads 
\begin{align}
  s=-\int_{\left|\boldsymbol{q}\right|<\Lambda_{UV}}\frac{\text{d}^{2}\boldsymbol{q}}{\left(2\pi\right)^{2}}\int_{-\infty}^{\infty}\frac{\text{d}q_{0}}{2\pi}\frac{m}{q_{0}^{2}+\left|\boldsymbol{q}\right|^{2}+m^{2}},
\end{align}
where $\Lambda_{UV}$ is a physical cutoff related to the lattice
spacing by $\Lambda_{UV}\sim a^{-1}$. The $q_{0}$ integral converges,
and does not require renormalization. It yields the result within
the lattice motivated scheme, 
\begin{align}
 s=-\frac{1}{2}\int_{\left|\boldsymbol{q}\right|<\Lambda_{UV}}\frac{\text{d}^{2}\boldsymbol{q}}{\left(2\pi\right)^{2}}\frac{m}{\sqrt{\left|\boldsymbol{q}\right|^{2}+m^{2}}},\label{eq:183}
\end{align}
and adding the operator ordering correction gives the ground state
charge density 
\begin{align}
  \rho=\left\langle J^{t}\right\rangle =-\frac{1}{2}\int_{\left|\boldsymbol{q}\right|<\Lambda_{UV}}\frac{\text{d}^{2}\boldsymbol{q}}{\left(2\pi\right)^{2}}\left(1-\frac{m}{\sqrt{\left|\boldsymbol{q}\right|^{2}+m^{2}}}\right).
\end{align}
After summing over low energy Majorana spinors and restoring units, this coincides
with the relativistic limit of the exact ground state charge density of the
lattice model \cite{read2000paired}, 
\begin{align}
  \rho=-\frac{1}{2}\int_{BZ}\frac{\mbox{d}^{2}\boldsymbol{q}}{\left(2\pi\right)^{2}}\left(1-\frac{h_{\boldsymbol{q}}}{\sqrt{\left|\Delta_{\boldsymbol{q}}\right|^{2}+h_{\boldsymbol{q}}^{2}}}\right),
\end{align}
where $h_{\boldsymbol{q}},\Delta_{\boldsymbol{q}}$ were defined
in section \ref{subsec:Band-structure-and}. 

For comparison we calculate the $s$ integral in a standard renormalization
scheme of relativistic QFT. In this approach the integral does not
converge. We introduce a frequency and wave-vector cutoff $\Lambda_{\text{rel}}$,
and restrict the integration to $\left|q\right|<\Lambda_{\text{rel}}$.
This yields 
\begin{align}
  s=&-\int_{\left|q\right|<\Lambda_{\text{rel}}}\frac{\text{d}^{3}q}{\left(2\pi\right)^{3}}\frac{m}{\left|q\right|^{2}+m^{2}}\\
  =&-\frac{\Lambda_{\text{rel}}m}{2\pi^{2}}+\frac{m^{2}\text{sgn}m}{4\pi}+O\left(\frac{m}{\Lambda_{\text{rel}}}\right).\nonumber
\end{align}
A simple way to proceed is to preform minimal subtraction, which means
we remove the diverging piece, and take $\Lambda_{\text{rel}}/m\rightarrow\infty$.
This gives the fully relativistic result
\begin{eqnarray}
 s=\frac{m^{2}\text{sgn}m}{4\pi}.
\end{eqnarray}
Comparing with \eqref{eq:72} we find $\zeta_{H}=s=\frac{m^{2}\text{sgn}m}{4\pi}$
for a positive orientation which is essentially the torsional Hall
viscosity of \cite{hughes2013torsional}\footnote{It is not exactly the same result because we did not use the same
relativistic renormalization scheme.}. The relativistic result can also be obtained by expanding the lattice
result \eqref{eq:183} in $\Lambda_{UV}$ and keeping the $O\left(1\right)$
piece. This is a general feature, the $O\left(1\right)$ piece of
any coefficient in the effective action is always relativistic. 

Let us now turn to the calculation of the ground state energy-momentum
tensor $\Lambda_{\;\mu}^{a}$. With a relativistic regulator $\Lambda_{\;\mu}^{a}$ is $O\left(3\right)$ invariant and must therefore be proportional to the identity,
\begin{align}
\Lambda_{\;\mu}^{a}=\delta^{ab}\int_{\left|q\right|<\Lambda_{\text{rel}}}\frac{\text{d}^{3}q}{\left(2\pi\right)^{3}}\frac{q_{\mu}q_{b}}{\left|q\right|^{2}+m^{2}}=\delta_{\mu}^{a}\frac{\Lambda}{2\kappa_{N}},
\end{align}
with the cosmological constant 
\begin{align}
  \frac{\Lambda}{2\kappa_{N}}=&\frac{1}{3}\int_{\left|q\right|<\Lambda_{\text{rel}}}\frac{\text{d}^{3}q}{\left(2\pi\right)^{3}}\frac{\left|q\right|^{2}}{\left|q\right|^{2}+m^{2}}\\
  =&\frac{1}{3}\left[\frac{\Lambda_{\text{rel}}^{3}}{6\pi^{2}}-\frac{\Lambda_{\text{rel}}m^{2}}{2\pi^{2}}+\frac{\left|m\right|^{3}}{4\pi}+O\left(\frac{m}{\Lambda_{\text{rel}}}\right)\right].\nonumber
\end{align}
Keeping the $O\left(1\right)$ piece we find the relativistic expression
\begin{align}
   \Lambda_{\;\mu}^{a}=\delta_{\mu}^{a}\frac{\Lambda}{2\kappa_{N}}=\delta_{\mu}^{a}\frac{\left|m\right|^{3}}{6\pi},
\end{align}
which again, is essentially the result of \cite{hughes2013torsional}.
With the lattice motivated renormalization scheme, 
\begin{align}
  \Lambda_{\;\mu}^{a}=\delta^{ab}\int_{\left|\boldsymbol{q}\right|<\Lambda_{UV}}\frac{\text{d}^{2}\boldsymbol{q}}{\left(2\pi\right)^{2}}\int_{-\infty}^{\infty}\frac{\text{d}q_{0}}{2\pi}\frac{q_{\mu}q_{b}}{q_{0}^{2}+\left|\boldsymbol{q}\right|^{2}+m^{2}}.
\end{align}
Here the $q_{0}$ integral for $\Lambda_{\;t}^{0}$ does not converge,
and needs to be regularized. We do this by introducing a frequency
cutoff $\Lambda_{0}$, 
\begin{align}
  \Lambda_{\;\mu}^{a}=\delta^{ab}\int_{\left|\boldsymbol{q}\right|<\Lambda_{UV}}\frac{\text{d}^{2}\boldsymbol{q}}{\left(2\pi\right)^{2}}\int_{-\Lambda_{0}}^{\Lambda_{0}}\frac{\text{d}q_{0}}{2\pi}\frac{q_{\mu}q_{b}}{q_{0}^{2}+\left|\boldsymbol{q}\right|^{2}+m^{2}}.
\end{align}
Unlike $\Lambda_{UV}$ which is a physical parameter of the model,
$\Lambda_{0}$ is a fictitious cutoff which we take to infinity at
the end of the calculation. The $q_{0}$ divergence can be interpreted
as an artifact of time discretization \cite{shankar1994renormalization}. At this point
the domain of integration is not a ball in Euclidian Fourier space
but a cylinder, so it is not invariant under $O\left(3\right)$, only
under $O\left(2\right)$\footnote{More accurately, the domain of integration for each lattice fermion
is not the disk $\left\{ \left|\boldsymbol{q}\right|<\Lambda_{UV}\right\} $
but the square $\left[-\Lambda_{UV}/2,\Lambda_{UV}/2\right]^{2}$
with $\Lambda_{UV}=\pi/a$, which is a quarter of the Brillouin
zone $BZ$, see section \ref{subsec:Coupling-the-Lattice}. The symmetry
group of this domain is not $O\left(2\right)$ but the point group
symmetry of the lattice $D_{4}\subset O\left(2\right)$. This subtlety
has no effect on the following. } and the reflection $q\mapsto-q$. This implies that the tensor $\Lambda_{\;\mu}^{a}$
takes the form 
\begin{align}
  \Lambda_{\;t}^{0}=&\int_{\left|\boldsymbol{q}\right|<\Lambda_{UV}}\frac{\text{d}^{2}\boldsymbol{q}}{\left(2\pi\right)^{2}}\int_{-\Lambda_{0}}^{\Lambda_{0}}\frac{\text{d}q_{0}}{2\pi}\frac{q_{0}^{2}}{q_{0}^{2}+\left|\boldsymbol{q}\right|^{2}+m^{2}},\\
  \Lambda_{\;j}^{A}=&\frac{1}{2}\delta_{j}^{A}\int_{\left|\boldsymbol{q}\right|<\Lambda_{UV}}\frac{\text{d}^{2}\boldsymbol{q}}{\left(2\pi\right)^{2}}\int_{-\Lambda_{0}}^{\Lambda_{0}}\frac{\text{d}q_{0}}{2\pi}\frac{\left|\boldsymbol{q}\right|^{2}}{q_{0}^{2}+\left|\boldsymbol{q}\right|^{2}+m^{2}},\nonumber
\end{align}
with all other components vanishing. The $q_{0}$ integral for
the energy density $\Lambda_{\;t}^{0}$ gives 
\begin{align}
  \Lambda_{\;t}^{0}=\frac{1}{2}\int_{\left|\boldsymbol{q}\right|<\Lambda_{UV}}\frac{\mbox{d}^{2}\boldsymbol{q}}{\left(2\pi\right)^{2}}\left[\frac{2\Lambda_{0}}{\pi}-\sqrt{\left|\boldsymbol{q}\right|^{2}+m^{2}}+O\left(\frac{m}{\Lambda_{0}}\right)\right].
\end{align}
Keeping the $O\left(1\right)$ piece we find, within the lattice motivated scheme, 
\begin{eqnarray}
  \Lambda_{\;t}^{0}=-\frac{1}{2}\int_{\left|\boldsymbol{q}\right|<\Lambda_{UV}}\frac{\mbox{d}^{2}\boldsymbol{q}}{\left(2\pi\right)^{2}}\sqrt{\left|\boldsymbol{q}\right|^{2}+m^{2}},\label{eq:196}
\end{eqnarray}
which is the familiar expression for the ground state energy of a
single Majorana fermion, which is half the energy of a filled Dirac
sea. The $q_{0}$ integral for $\Lambda_{\;j}^{A}$ converges and
gives the pressure 
\begin{align}
  \Lambda_{\;j}^{A}=\frac{1}{2}\delta_{j}^{A}\frac{1}{2}\int_{\left|\boldsymbol{q}\right|<\Lambda_{UV}}\frac{\mbox{d}^{2}\boldsymbol{q}}{\left(2\pi\right)^{2}}\frac{\left|\boldsymbol{q}\right|^{2}}{\sqrt{\left|\boldsymbol{q}\right|^{2}+m^{2}}}.
\end{align}
We see that the ground state energy density and pressure are no longer
equal. In other words the ground state energy-momentum tensor is not
Lorentz invariant, due to the lattice renormalization
scheme. It may be surprising that the expression \eqref{eq:196} for
the energy density is part of an energy momentum tensor which is not
Lorentz invariant. This has been discussed in the literature in the
context of the cosmological constant problem \cite{ossola2003considerations,koksma2011cosmological,visser2016lorentz}. 

Let us now compare the above with the lattice model. For the energy
density we need to add the operator ordering correction, 

\begin{align}
  \varepsilon=\left\langle t_{\text{cov}\;t}^{t}\right\rangle =\frac{1}{2}\int_{\left|\boldsymbol{q}\right|<\Lambda_{UV}}\frac{\mbox{d}^{2}\boldsymbol{q}}{\left(2\pi\right)^{2}}\left(m-\sqrt{\left|\boldsymbol{q}\right|^{2}+m^{2}}\right).
\end{align}
Restoring units and summing over Dirac points, we recognize the above
as the relativistic approximation of the ground state energy density
of the lattice model \cite{volovik2009universe}, 
\begin{align}
  \varepsilon=\frac{1}{2}\int_{BZ}\frac{\text{d}^{2}\boldsymbol{q}}{\left(2\pi\right)^{2}}\left(h_{\boldsymbol{q}}-E_{\boldsymbol{q}}\right).
\end{align}
The above calculations of simple ground state properties serve as
consistency checks. We have seen explicitly that these quantities
are UV sensitive. With the lattice motivated renormalization scheme
the effective action produces physical quantities that approximate
those of the lattice model, which are distinct from those obtained
with a relativistic scheme. In the following we will focus on UV insensitive
terms. In doing so we will also ignore operator ordering corrections,
because these always contain $\delta^{2}\left(0\right)\sim\int_{\left|\boldsymbol{q}\right|<\Lambda_{UV}}\frac{\text{d}^{2}\boldsymbol{q}}{\left(2\pi\right)^{2}}\sim\Lambda_{UV}^{2}$
and are therefore UV sensitive.

\subsection{Two vertex diagrams}

Let us now turn to the calculation of the more interesting second
order terms, which correspond to linear responses. After preforming
the traces over gamma matrices one finds
\begin{widetext}
\begin{align}
  \mbox{Tr}\left(GV_{1}\right)^{2}=&  -2im\varepsilon^{abc}\int_{p}h_{a}^{\;\mu}\left(p\right)h_{b}^{\;\nu}\left(-p\right)p_{c}\int_{q}\frac{\left(q+\frac{1}{2}p\right)_{\mu}\left(q+\frac{1}{2}p\right)_{\nu}}{\left(q^{2}-m^{2}\right)\left(\left(p+q\right)^{2}-m^{2}\right)}\label{eq:200}\\
  &+2m^{2}\eta^{ab}\int_{p}h_{a}^{\;\mu}\left(p\right)h_{b}^{\;\nu}\left(-p\right)\int_{q}\frac{\left(q+\frac{1}{2}p\right)_{\mu}\left(q+\frac{1}{2}p\right)_{\nu}}{\left(q^{2}-m^{2}\right)\left(\left(p+q\right)^{2}-m^{2}\right)}\nonumber \\
  &+2\left(\eta^{ac}\eta^{bd}-\eta^{ab}\eta^{cd}+\eta^{ad}\eta^{cb}\right)\int_{p}h_{a}^{\;\mu}\left(p\right)h_{b}^{\;\nu}\left(-p\right)\int_{q}\frac{\left(q+\frac{1}{2}p\right)_{\mu}\left(q+\frac{1}{2}p\right)_{\nu}q_{c}\left(p+q\right)_{d}}{\left(q^{2}-m^{2}\right)\left(\left(p+q\right)^{2}-m^{2}\right)},\nonumber\\ 
 \mbox{Tr}\left(GV_{2}\right)^{2}=&\int_{p}\omega\left(p\right)\omega\left(-p\right)\int_{q}\frac{2\eta^{ab}q_{a}\left(p+q\right)_{b}+2m^{2}}{\left(q^{2}-m^{2}\right)\left(\left(p+q\right)^{2}-m^{2}\right)},\\
 \mbox{Tr}\left(GV_{1}GV_{2}\right)= & -2i\varepsilon^{abc}\int_{p}h_{a}^{\;\mu}\left(p\right)\omega\left(-p\right)p_{c}\int_{q}\frac{\left(q+\frac{1}{2}p\right)_{\mu}q_{b}}{\left(q^{2}-m^{2}\right)\left(\left(p+q\right)^{2}-m^{2}\right)}\label{eq:202}\\
 &+2m\eta^{ab}\int_{p}h_{a}^{\;\mu}\left(p\right)\omega\left(-p\right)\int_{q}\frac{\left(q+\frac{1}{2}p\right)_{\mu}\left(2q+p\right)_{b}}{\left(q^{2}-m^{2}\right)\left(\left(p+q\right)^{2}-m^{2}\right)}.\nonumber 
\end{align}
\end{widetext}
One is then left with the calculation of the integrals over the loop
momenta $q$ in the above equations. The first step in doing so is
Wick rotating to euclidian signature by changing $q_{0}\mapsto iq_{0},\;p_{0}\mapsto ip_{0}$
in the $q$ integrals. 

At this point one can use Feynman parameters to simplify the form
of the integrands, but since we are only interested in the effective
action to low orders in derivatives of the background fields, we find
it simpler to expand the integrands in powers of $p/m$. 

We start with the first integral in \eqref{eq:200}, which contains
the gCS term. Expanding the integrand in $p/m$ we find
\begin{widetext}
\begin{align}
  \frac{\left(q+\frac{1}{2}p\right)_{\mu}\left(q+\frac{1}{2}p\right)_{\nu}}{\left(q^{2}+m^{2}\right)\left(\left(p+q\right)^{2}+m^{2}\right)}=&
 \frac{q_{\mu}q_{\nu}}{\left(m^{2}+q^{2}\right)^{2}} 
+\left[\frac{p_{(\mu}q_{\nu)}}{\left(m^{2}+q^{2}\right)^{2}}-2\frac{q_{\mu}q_{\nu}p\cdot q}{\left(m^{2}+q^{2}\right)^{3}}\right] \label{eq:H33}\\
  &+\left[\frac{p_{\mu}p_{\nu}}{4\left(m^{2}+q^{2}\right)^{2}}-\frac{p^{2}q_{\mu}q_{\nu}}{\left(m^{2}+q^{2}\right)^{3}}-\frac{2p_{(\mu}q_{\nu)}p\cdot q}{\left(m^{2}+q^{2}\right)^{3}}+\frac{4q_{\mu}q_{\nu}\left(p\cdot q\right)^{2}}{\left(m^{2}+q^{2}\right)^{4}}\right]
 +O\left(p^{3}\right)\nonumber 
\end{align}
\end{widetext}
where terms are grouped according to their order in $p/m$.The $q$ integral over
the $O\left(1\right)$ terms diverges, and therefore produces
a UV sensitive term in the effective action. With a relativistic renormalization
we find 
\begin{align}
  2W_{\text{RC}}=\frac{m^{2}\text{sgn}\left(m\right)}{4\pi}\int\text{d}^{3}xh_{a}^{\;\mu}\eta_{\mu\nu}\varepsilon^{abc}\partial_{b}h_{c}^{\;\nu}+\cdots
\end{align}
Comparing with \eqref{eq:72} and using $e_{a}De^{a}=\varepsilon^{abc}h_{a\nu}\partial_{b}h_{c}^{\;\nu}\text{d}^{3}x+\cdots$
we find again the torsional Hall viscosity $\zeta_{H}=\frac{m^{2}\text{sgn}\left(m\right)}{4\pi}$,
for positive orientation. With the lattice renormalization the $\eta_{\mu\nu}$
in the above is replaced by a non Lorentz invariant tensor, but in
this work we are only interested in UV insensitive responses and we
will not discuss it further. 

The $q$ integral over the $O\left(p/m\right)$ terms vanishes because
it is odd under the reflection $q\mapsto-q$.

The $O\left(p^{2}/m^{2}\right)$ contributions are most interesting for
us. The $q$ integral over these converges, and therefore produces
UV insensitive terms in the effective action. Instead of calculating
the integral with the finite physical $\Lambda_{UV}$, we can calculate
it with $\Lambda_{UV}/m=\infty$ at the expense of producing small $O\left(m/\Lambda_{UV}\right)$
corrections. Then the calculation
reduces to a standard calculation within relativistic QFT which has
appeared a few times in the literature with slightly different conventions  \cite{goni1986massless,van1986topological,vuorio1986parity,vuorio1986parityErr}, and which is done below for completeness. See also \cite{kurkov2018gravitational} for a recent heat kernel calculation and review of the literature, and \cite{bonetti2013one} for similar computations in 4+1 dimensions. 
With $\Lambda_{UV}/m=\infty$ the integral is Lorentz invariant and
this implies the standard reductions to radial functions such as $q_{\mu}q_{\nu}\mapsto\frac{1}{3}\eta_{\mu\nu}q^{2}$.
The $O\left(p^{2}/m^{2}\right)$ contributions in \eqref{eq:H33} then reduce to
\begin{align}
  \frac{1}{4}&\frac{p_{\mu}p_{\nu}}{\left(m^{2}+q^{2}\right)^{2}}-\frac{1}{3}\frac{p^{2}\eta_{\mu\nu}q^{2}}{\left(m^{2}+q^{2}\right)^{3}}-\frac{2}{3}\frac{p_{\mu}p_{\nu}q^{2}}{\left(m^{2}+q^{2}\right)^{3}}\nonumber\\
  &+\frac{4}{15}\frac{p^{2}\eta_{\mu\nu}+2p_{\mu}p_{\nu}}{\left(m^{2}+q^{2}\right)^{4}}q^{4},
\end{align}
and preforming the $q$ integral yields 
\begin{align}
  \frac{1}{96\pi\left|m\right|}\left(p_{\mu}p_{\nu}-p^{2}\eta_{\mu\nu}\right).
\end{align}
This corresponds to the following term in the effective action
\begin{align}
  2W_{\text{RC}}=&\frac{\text{sgn}\left(m\right)}{2}\frac{1}{96\pi}2\int\text{d}^{3}xh_{a}^{\;\mu}\varepsilon^{abc}\partial_{c}\left(\partial_{\mu}\partial_{\nu}-\partial^{2}\eta_{\mu\nu}\right)h_{b}^{\;\nu}\nonumber\\
  &+\cdots
\end{align}
To identify this term it is easiest to fix a Lorentz gauge where
$h_{\left[\mu\nu\right]}=0$. In terms of the $p$-wave SC this corresponds
to $U\left(1\right)$ gauge fixing the phase $\theta$ of the order
parameter to $0$, along with an additional boost which is only a symmetry 
in the relativistic limit. Then $h$ corresponds also to the first
order metric perturbation, $g_{\mu\nu}=\eta_{\mu\nu}-2h_{(\mu\nu)}=\eta_{\mu\nu}-2h_{\mu\nu}$,
and the above corresponds to the expansion of the gCS term 
\begin{eqnarray}
 &  & 2W_{\text{RC}}=\frac{\text{sgn}\left(m\right)}{2}\frac{1}{96\pi}\int Q_{3}\left(\tilde{\Gamma}\right)+\cdots\label{eq:H38}
\end{eqnarray}
to second order in $h$. In preforming such expansions we found the
Mathematica package xAct very useful \cite{brizuela2009xpert,xAct}.
Equation \eqref{eq:H38} corresponds to $\kappa_{H}=\frac{1}{48\pi}\frac{\text{sgn}\left(m\right)}{2}$. We note that within the perturbative calculation there is no difference between $\int Q_{3}\left(\tilde{\Gamma}\right)$
and $\int Q_{3}\left(\tilde{\omega}\right)$, see \eqref{eq:70}.

The above result is valid for a vielbein $e_{a}^{\;\mu}=\delta_{a}^{\mu}+h_{a}^{\;\mu}$
which has a positive orientation. A vielbein with a negative orientation
can be written as $e_{a}^{\;\mu}=L_{\;a}^{b}\left(\delta_{b}^{\mu}+h_{b}^{\;\mu}\right)$
where $L$ is a Lorentz transformation with $\text{det}L=-1$. We
can deal with such vielbeins by absorbing $L$ into the gamma matrices,
$\gamma^{a}\mapsto L_{\;a}^{b}\gamma^{a}$. The only effect that this
change has on the traces \eqref{eq:177} is changing $\text{tr}\left(\gamma^{a}\gamma^{b}\gamma^{c}\right)=2i\varepsilon^{abc}$
to $\text{tr}\left(\gamma^{a}\gamma^{b}\gamma^{c}\right)=-2i\varepsilon^{abc}$.
The metric is independent of the orientation and so is $\tilde{\Gamma}$,
so the result valid for both orientations is 
\begin{eqnarray}
 &  & 2W_{\text{RC}}=\frac{\text{sgn}\left(m\right)o}{2}\frac{1}{96\pi}\int Q_{3}\left(\tilde{\Gamma}\right)+\cdots
\end{eqnarray}
where $o=\text{sgn}\left(\text{det}e\right)$ is the orientation.
The second and third lines in \eqref{eq:200} correspond, with a relativistic
regulator, to $O\left(h^{2}\right)$ contributions to the cosmological
constant and E-H term which are UV sensitive. 

One can compute the other traces in the same manner. The only additional
UV insensitive contribution comes from the second integral in \eqref{eq:202}.
It is given by 
\begin{align}
  2W_{\text{RC}}=&\frac{\text{sgn}\left(m\right)}{2}\frac{1}{96\pi}\int\text{d}^{3}x4\omega\left(\partial_{\mu}\partial_{\nu}-\partial^{2}\eta_{\mu\nu}\right)2h^{\mu\nu}\nonumber\\
  &+\cdots
\end{align}
This corresponds to the expansion of the gpCS term to second
order in the vertices,
\begin{align}
  2W_{\text{RC}}  =\frac{\text{sgn}\left(m\right)}{2}\frac{1}{96\pi}\int\text{d}^{3}x\left|e\right|\tilde{\mathcal{R}}c+\cdots
\end{align}
where we have used \eqref{eq:76}, the expansion of the curvature
$\tilde{\mathcal{R}}=-2\left(\partial_{a}\partial_{b}-\partial^{2}\eta_{ab}\right)h^{ab}+O\left(h^{2}\right)$,
the definition $c=\varepsilon^{abc}\left(\omega_{abc}-\tilde{\omega}_{abc}\right)$,
and the expansion $\tilde{\omega}_{abc}\varepsilon^{abc}=-\varepsilon^{abc}\partial_{a}h_{bc}+O\left(h^{2}\right)$
of the LC spin connection. Note that in the Lorentz gauge $h_{\left[ab\right]}=0$,
$\tilde{\omega}_{abc}\varepsilon^{abc}$ vanishes to first order.
This completes the calculation of the UV insensitive terms in the
effective action which we have studied in this paper. 


\bibliographystyle{apsrev4-1}

\end{document}